\documentclass[11pt]{report}
\usepackage{csthesis}
\usepackage{makeidx}  
\makeindex  

\begin{document}
\figurespagetrue
\dedicationpagetrue



%
%


\title{Chaos and Quantum-Classical Correspondence for Two Coupled Spins}
\author{Joseph V. Emerson}
\qualification{M.Sc., Simon Fraser University, 1995}
\qualification{B.Sc., McGill University, 1993}
\submitdate{May 2001}
\copyrightyear{2001}


\chair{Dr.~Mike Hayden} 
\signatory{Dr.~Leslie E.~ Ballentine, Senior Supervisor
}
\signatory{Dr.~Howard Trottier, Supervisor
}
\signatory{Dr.~Michael Wortis, Supervisor
}
\signatory{Dr.~Igor Herbut,
        Internal Examiner
}
\signatory{Dr.~Jorge V. Jose, External Examiner \\
        Department of Physics,
        Northeastern University
}

\beforepreface

\prefacesection{Abstract}


Two approaches to quantum-classical correspondence 
are distinguished according to the classical dynamical theory 
with which quantum theory is compared. The first 
of these,  Ehrenfest correspondence, defines a dynamical regime 
in which the quantum expectation values follow approximately 
a classical trajectory. The second of these, Liouville correspondence, 
applies when the quantum probability distributions remain well 
approximated by a density in the classical phase space. 
The former applies only for narrow states,  
whereas the latter may remain valid 
even for quantum states that have spread to the system size. 

A spin model is adopted for this correspondence 
study because the quantum state is 
discrete and finite-dimensional, and thus no articifical truncation 
of the Hilbert space is required. 
The quantum time-evolution is 
given by a discrete unitary mapping.  The corresponding classical model 
is volume-preserving (non-dissipative) and the time-evolution is 
given by a symplectic map.  

In classically chaotic regimes, the widths of initially narrow 
quantum states grow, on average, exponentially with 
time, until saturation at the system size. 
This initial 
spreading rate is well approximated by the classical Lyapunov 
exponent when the accessible classical phase space is 
predominantly chaotic.  
Because of the exponential growth rate of the quantum variance,  
the Ehrenfest regime is delimited by a break-time that grows 
logarithmically with increasing quantum numbers. 

The small differences between quantum expectation values and  
corresponding Liouville averages of dynamical variables 
also grow exponentially, 
initially, if the classical behaviour is chaotic. This exponential 
rate is independent of $\hbar$ and consistently larger than the 
classical Lyapunov exponent by at least a factor of two. 
Interestingly,  
this exponential growth rate does not continue until these differences 
approach magnitudes of order the system size, but crosses over to 
power-law growth (or simply saturates), when the differences have 
reached magnitudes no larger than order $\hbar$. 
Because of the exponential growth of these quantum-classical differences, 
the Liouville regime is also delimited by a break-time 
that grows logarithmicly with increasing quantum numbers. 
However, due to the early saturation 
of the exponential growth of the differences, 
the Liouville log break-time applies only 
in a restricted domain. 

After spreading to the system size, 
the quantum states relax to an approximately time-independent 
state.  This is characterised by equilibrium distributions that are 
subject to rapidly oscillating fluctuations about the coarse-grained 
classical steady-state.  If the accessible classical 
phase space is predominantly chaotic, then the equilibrium quantum  
probability distributions are nearly microcanonical. 
Otherwise the steady-state 
quantum probability distributions accurately reflect the details of the  
KAM surfaces in the classical phase space. 
The equilibrium regime quantum fluctuations, which are 
quantum-classical differences, are shown to approach zero as a 
negative power of the quantum numbers. 

Since the differences between quantum expectation values 
and {\em individual} classical trajectories grow exponentially in time to 
a magnitude that scales with the system size, 
the Ehrenfest regime is delimited by an {\em unrestricted}  
log break-time.  
The log break-time may be short compared to 
experimental time-scales for some chaotic macroscopic bodies, and,   
as a result, an observable {\em breakdown} of Ehrenfest 
correspondence may arise for some macroscopic systems. Although the 
Liouville regime is also delimited by a log break-time, 
an observable {\em breakdown} of correspondence between quantum mechanics 
and Liouville mechanics for macroscopic bodies is not likely since their  
differences, expressed relative to the dimensions of 
the observable, approach zero in the limit of large quantum numbers.

\dedication{\begin{center}\vspace{-2.5in}For Coyote\end{center}}

\dedicquotation

\prefacesection{Acknowledgments}

I would like to thank my thesis advisor, Dr.\ Leslie Ballentine, 
and the Faculty of Science for financial support 
provided throughout my doctoral studies. I would also like 
to thank the Centre for Experimental and Constructive Mathematics 
for access to their computational resources. 

I am indebted to several individuals for their intellectual 
support: Dr.\ Leslie Ballentine, for his critical 
insights and numerous clarifications, and several careful readings of the 
manuscript; Karn Kallio, my office mate and friend, for sharing 
his vast knowledge of mathematics and, more importantly, 
his infectious enthusiasm for learning the art of physics; and 
my old friend Rob Spekkens, for the many thought-provoking discussions 
that stimulated my interest in 
the physics and philosophy of quantum mechanics.   

Finally, I wish to thank my family and 
friends for their encouragement, the Bookman for 
his advice, and, especially, Julia Lianne Bevin Shugarman - 
for seeing me through.


\lists


\beforetext

\chapter{Introduction}


Though it is widely accepted that classical 
mechanics should emerge from quantum theory in an appropriate limit, 
this ``correspondence principle'' meets with severe difficulties 
when the classical dynamics exhibit chaos. 
Research into the physical conditions required for the 
emergence of classical chaos from the underlying quantum description 
has become a topic of active investigation, 
and has led to a number of differing perspectives. 

Ford and coworkers \cite{Ford91,Ford92a,Ford92b} have  
considered the criterion of {\it algorithmic complexity}, which 
has led them to the conclusion 
that classical chaos can not emerge from quantum 
theory in the macroscopic limit, and, therefore, that quantum-classical 
correspondence must break down for chaotic motion.      
While Zurek and Paz \cite{ZP94,ZP95a} and 
others \cite{Kol96,Zurek98a,Zurek98b} 
agree that a breakdown of correspondence does arise, they 
have argued that the required correspondence is restored  
when the effect of a weak-coupling  
to an environment, a process sometimes 
called {\em decoherence} \cite{Joos85,Dec96},  
is taken into account. Alternatively, 
Ballentine \cite{Ball95}, Takahashi \cite{Tak89}, and   
Casati and Chirikov \cite{CC95} have emphasized  
that correspondence should emerge in the macroscopic limit 
given an appropriate degree of coarse-graining which 
may be considered to represent the finite  
resolution of macroscopic measurements. 
The variety of perspectives on this issue is due 
in part to differing views on which observable properties  
must be taken into consideration in order to demonstrate a formal 
correspondence in the macroscopic limit \cite{Ball94,Ball95,Ball98,Zurek98a}. 

The details of the quantum-classical transition and 
the quantum manifestations of classical chaos are of considerable 
interest also for the accurate and practical description of 
mesoscopic systems. The onset of widespread chaos 
in the {\em classical} Hamiltonian model of the highly excited states 
of hydrogen has been shown to accurately predict 
the unexpected onset of ionization observed in these states 
under increasing microwave fields \cite{BayKoch74,LP78}.      
In the classical model, the microwave 
field itself provides the non-integrable 
perturbation to the otherwise integrable classical 
dynamics.  Under strong magnetic fields, the energy level statistics 
of these Rydberg atoms exhibit the characterisitic distribution of 
spacings predicted from random matrix theory precisely when the 
classical model exhibits chaotic behaviour \cite{Jensen}.  

A better understanding of the quantum-classical transition is required 
also for practical computation in the mesoscopic regime. 
On the one hand, the {\it exact} theoretical description of 
even simple microscopic and mesoscopic systems in 
quantum theory demands computational resources (specifically, 
{\em classical} computational resources!) that increase as a power of 
the quantum numbers and exponentially with the number of interacting 
subsystems.  On the other hand, 
the semiclassical techniques developed by Gutzwiller \cite{Gutzwiller} 
and others become computationally impractical due to the proliferation
of (unstable) periodic orbits in the case of chaotic motion.  
Moreover, the accuracy of the semiclassical approximation 
is not well understood in either the time-domain \cite{TH93} or the 
energy domain \cite{Haake96}, and remains a topic of active investigation.  
The results in this thesis suggest that purely classical 
techniques may be applied to accurately describe the time-domain behaviour 
of chaotic systems even in the case of surprisingly small quantum numbers. 

The theoretical identification of the characteristic 
properties of quantum systems with chaotic classical counterparts 
has been an interesting and illuminating topic of research in its own right. 
This nascent area of research, sometimes called ``quantum chaos,'' 
is not a unified field but encompasses a wide variety of approaches.  
In analogy with the manner in which classical chaos came to be understood, 
much of the theoretical progress in quantum chaos has proceeded 
through the study of idealized model systems, usually taking the form 
of stroboscopic mappings. Some of this research will be reviewed 
further below. I should also mention here that, whereas a fairly 
complete understanding of the microscopic  
underpinnings of irreversible thermodynamics has recently emerged  
from the perspective of classical dynamical systems 
theory \cite{Gaspard,Dorfman}, 
very little is known about how these characteristic properties emerge 
from the more fundamental quantum mechanical treatment of these models. 
The results presented in this thesis will clarify how these statistical  
properties emerge from the underlying quantum dynamics specifically when 
the classical model exhibits chaotic behaviour. 

Certainly the most intellectually intriguing aspect  
of this research pertains to the possibility of demonstrating 
that {\it experimentally distinct} predictions result 
from two very different interpretations of the quantum state. 
This pedagogical issue is tied to the following practical question 
about quantum-classical correspondence: 
which classical theory of mechanics actually emerges 
from the quantum description of  macroscopic systems?  
One alternative is Hamilton's equations of motion 
(Newton's laws); another is the classical   
Liouville equation that describes the dynamics of probability 
densities with non-vanishing support. 
The former describes the exact, deterministic evolution of single  
macroscopic systems, whereas the latter provides only probabilistic 
predictions for ensembles of similarly prepared systems. 

The central focus of the work in this thesis is to demonstrate that, 
for the chaotic motion of some macroscopic systems, standard 
quantum mechanics may be unable to reproduce Newton's laws of motion 
over experimentally relevant time-scales. Moreover, contrary 
to the conclusion of a recent argument by Zurek \cite{Zurek98b}, I will 
provide evidence that this specific {\it breakdown} of 
quantum-classical correspondence 
is not circumvented by taking into account the decoherence effects 
of a stochastic environment. This striking conclusion does not imply 
a failure of the correspondence principle, however,  
since the statistical predictions of Liouville mechanics are shown to 
be compatible with the predictions of quantum theory for such systems.  
The decisive implications of this scenario for the interpretation of 
the correspondence principle, as well as the interpretation 
of quantum theory itself, will be drawn out and related to some 
of the historical debate on this topic.  

In the remainder of this 
Introduction I will review the relevant literature 
on quantum chaos and quantum-classical correspondence, before 
undertaking, in the body of this thesis, 
a detailed comparison of the quantum 
and classical dynamics of a non-integrable model system 
comprised of two coupled spins. 
The perspective developed from the results of this analysis 
will generalize and extend much of the earlier work 
on the dynamical properties of 
quantum chaos and quantum-classical correspondence 
that has been developed through the study of 
different model systems. 


\section{Classical Trajectories and Quantum Expectation Values}

A convenient starting point for this subject 
follows from considering an old theorem due to Ehrenfest \cite{Ehrenfest} 
and the oft-repeated correspondence argument that derives from 
it \cite{Liboff,Schiff,Messiah,Merzbacher}.  For a simple Hamiltonian 
system of the form $H= p^2/2m + V(q)$, the 
time-dependence of the expectation values 
$\langle q(t) \rangle = \langle \psi(t) | q | \psi(t) \rangle$ 
and $\langle p(t) \rangle= \langle \psi(t) | p | \psi(t) \rangle$, 
in an arbitrary quantum state $\psi(t)$, is prescribed 
by the differential equations, 
\begin{eqnarray}
\label{eqn:qdiff}
\frac{d \langle q(t) \rangle }{d t} & = &
\frac{\langle p(t) \rangle}{m} \nonumber \\
\frac{d \langle p(t) \rangle }{d t} & = & 
\langle F(q) \rangle. 
\end{eqnarray}
These equations do not form a closed set since the 
time-dependence of the force function, 
\begin{equation}
 \langle F(q) \rangle = \langle \psi(t)| F(q) |\psi(t) \rangle 
\neq F(\langle q \rangle) 
\end{equation}
has not been specified. However, if the operator $F(q)= d V /d q$ 
can be expanded in a Taylor series about the position centroid of 
the quantum state, $\langle q(t) \rangle$, then, 
\begin{equation}
\label{eqn:taylor} 
	\langle F(q) \rangle = F(\langle q \rangle) 
	+ \frac{(\Delta q)^2} {2} 
	\left. \frac{d^2 F}{dq^2} \right|_{q =\langle q \rangle}  + \dots 
\end{equation}
where $ (\Delta q)^2 
= \langle q^2 \rangle - \langle q \rangle^2$ is the state variance.  
When the quadratic and higher-order terms of 
this expansion are negligible, the 
system of differential equations (\ref{eqn:qdiff}) becomes  
closed and the expectation values of the dynamical variables 
reduce to Hamilton's canonical equations of motion, 
\begin{eqnarray}
\label{eqn:cdiff}
\frac{d q_c(t) }{d t} &= &
\frac{p_c(t)}{m} \nonumber \\
\frac{d p_c(t) }{d t} &= & \{H,p_c \} = F(q_c),  
\end{eqnarray}
In the above, $\{\cdot ,\cdot \}$ is a Poisson bracket
and $q_c(t)$ and $p_c(t)$ are classical dynamical variables. 

At this point it is natural to ask: under what conditions may we 
expect this approximation to apply? Clearly, if the potential $V(q)$  
is either linear or quadratic in $q$, as for a harmonic oscillator, 
then the quantum equations are identical to the classical ones. 
More generally, this correspondence follows provided the state variance 
is narrow compared to significant non-linear variation in the prescribed 
force,  
\begin{equation}
\label{eqn:loc1}
(\Delta q)^2 \ll L^2 \sim 
\frac{F(\langle q \rangle)}{ \left. \frac{d^2 F}{dq^2} \right|_{q =\langle q \rangle}}, 
\end{equation}
and higher-order contributions to the expansion (\ref{eqn:taylor}) 
are also negligible. 
The dynamical regime in which 
(\ref{eqn:loc1}) holds is sometimes called 
the {\it Ehrenfest} regime \cite{Ball94,Ball98,EB00a}. 

A common textbook argument \cite{Liboff,Messiah,Merzbacher} maintains that 
the approximation (\ref{eqn:loc1}) should be valid, 
quite generally, for {\em macroscopic} systems. 
This arguments holds that, for a macroscopic body, 
the width of a well-localised state 
may be extremely narrow compared to characteristic variation in the   
macroscopic potential (or some other relevant system dimension).  
Consequently, the approximation, 
\begin{equation}
\label{eqn:loc2}
	\langle F(q) \rangle = F(\langle q \rangle),  
\end{equation}
should hold for a generic macroscopic system described by such a 
localised state. 
The important conclusion is that the differential equations governing 
the time-evolution of the expectation values (\ref{eqn:qdiff}) then 
accurately approximate Hamilton's equations of motion (\ref{eqn:cdiff}). 
Provided it is possible to match the classical initial conditions, 
quantum theory can then describe the {\it deterministic} motion 
of a single macroscopic body by {\em associating} 
expectation values with the classical dynamical variables.  
The requirement on the initial conditions is simply that the 
quantum centroids must be set equal to the initial phase 
space coordinates of the classical trajectory; that is,  
$ \langle q(0) \rangle = q_c(0)$ and $ \langle p(0) \rangle = p_c(0)$. 

These observations may be taken as means of justifying the  
``correspondence principle'', that is, the view that 
the predictions of quantum mechanics must coincide 
with those of classical mechanics in cases where the latter 
is known to be valid \cite{Merzbacher,Messiah}. 
In the present context, ``classical mechanics''  
is taken to refer specifically to Hamilton's equations of motion. 
I will call this interpretation of the correspondence principle 
the {\it Ehrenfest} correspondence principle. 
This conjecture may be stated heuristically as,  
\begin{equation}
\label{eqn:Ehrcorr}
	\left. 
	\begin{array}{c} | \langle q \rangle - q_c | \to 0,  \\ 
	| \langle p \rangle - p_c | \to 0,  
	\end{array} \right\} \;
	\;\;\; \mathrm{for} \;\;\; \frac{{\mathcal J}}
	{\hbar} \rightarrow \infty, 
\end{equation}
where ${\mathcal J}$ is a characteristic action and $\hbar$ is Planck's 
constant. I will provide a more useful statement of this 
conjecture in Chapter 5.

Of course, the ratio ${\mathcal J}/\hbar$ is actually finite 
for real physical systems. Moreover, initially 
localised quantum states will generally diffuse with time, 
and the accuracy of the approximation (\ref{eqn:loc2}) eventually 
breaks down. Therefore, to confirm the validity of the conjecture 
(\ref{eqn:Ehrcorr}) as a principle of correspondence,  
it is important to determine the time-scale 
on which significant deviations between the 
quantum and classical predictions arise.  
Let ${\bf x}_c$ denote the generalised classical coordinates, 
{\it e.g}.\ ${\bf x}_c=(q_c,p_c)$,  and $\langle {\bf x} \rangle$  
denote the corresponding quantum expectation values. Then, the 
magnitude of their difference  
\begin{equation}
  {\bf \epsilon }(t)  = | \langle {\bf x} \rangle - {\bf x}_c| 
\end{equation}
is a time-dependent measure of the degree of correspondence. Given some 
criterion defining {\em adequate} 
correspondence, {\it i.e}., a threshold $\chi$, one may 
define \cite{Chi88,HB94} a time-scale $t_{Ehr}$  
over which Ehrenfest correspondence holds from the condition, 
\begin{equation}
\label{eqn:Ehrcond}
	{\bf \epsilon }(t) \le \chi  \; \; \; \; \mathrm{for} 
\;\;\;\ t < t_{Ehr}. 
\end{equation}
This threshold-dependent condition delineates the Ehrenfest 
regime more precisely than the approximation (\ref{eqn:loc1}). 
This condition differs slightly from 
the ``narrow state'' condition (\ref{eqn:loc1}), though 
both definitions will coincide, approximately, as a result 
of Eqs.\ (\ref{eqn:taylor}) and (\ref{eqn:qdiff}), 
if  the tolerance threshold $\chi$ for the position is set at the 
characteristic system size, $L$, given by the RHS of (\ref{eqn:loc1}) 
and the explicit degree of correspondence for the momentum is 
ignored. 

The time-scale of Ehrenfest correspondence will generally depend on the 
details of the potential and other system parameters, but 
it is worthwhile to get some sense of the duration of this regime 
by considering the simplest case of a free-particle. 
For a particle of mass $m$, 
starting from a Gaussian state with 
initial standard deviation $\Delta q(0) = a_o $ and momentum 
$\hbar k_o$, the time-dependent probability 
distribution is given by, 
\begin{equation}
	P(q,t) = | \psi(q,t) |^2 = \frac{1}{a\sqrt{2\pi}(1+ t^2/\tau^2)^{1/2}}
\exp\left[- \frac{(q - (\hbar k_o t /m )^2}{2 a_o^2 (1+t^2/\tau^2 )}\right].
\end{equation}
The width $\Delta q$ of this Gaussian doubles on the 
time-scale $\tau = 2 m a_o^2/\hbar$. 
The time it takes 
this width to grow to a size $\Delta q >> a_o$ is given by 
\begin{equation}
	t \sim {\Delta q \; a_o \; m \over \hbar}. 
\end{equation}
Therefore, in the case of a quantum state describing a macroscopic 
object, such as a baseball, initially localised to within 1 Angstrom, 
the time it takes this state to 
diffuse to a width of 1 cm is of order $10^{14}$ 
years \cite{Ehrenfest,Liboff}.  
This is much longer than the estimated lifetime of the universe!

Thus it appears plausible to conclude that quantum 
expectation values can adequately describe the deterministic motion 
of individual macroscopic objects over time-scales much longer than 
the duration of any conceivable experiment. In the following 
section I will explain how the presence of chaos in 
the classical dynamics challenges this simple picture, 
as well as the adequacy of the conjecture (\ref{eqn:Ehrcorr}) as a 
principle of quantum-classical correspondence.  


\section{Classical Chaos and Quantum Chaos}

The task of identifying criteria for chaos in quantum mechanics 
in a manner that extends or generalizes the criteria for chaos 
in classical mechanics meets with several immediate difficulties. 
Classical chaos is characterized by ``extreme sensitivity to initial 
conditions.'' This property is normally identified with  
an exponential divergence, on average, in the separation  
between initially nearby classical trajectories.  
The rate of this divergence 
is characterized by a set of exponents, called Lyapunov exponents. 
In the simplest case of a mapping, ${\bf x}(n+1) = {\bf F}({\bf x}(n))$, 
the largest such exponent, $\lambda_L$, may be defined in the following way. 
Let ${\bf d}(n)= {\bf x}_a(n) - {\bf x}_b(n) $ designate the 
difference between the 
phase space coordinates, ${\bf x}_a$ and ${\bf x}_b$, of 
two nearby classical trajectories. 
For small enough $|{\bf d}|$, the growth of this 
difference vector is governed by the tangent matrix, 
\begin{equation}
	M_{ij}(x) = \left. \frac{\partial F_i({\bf x})} {\partial x_j} 
\right|_{x(n)}.
\end{equation}
The eigenvalues  of this matrix characterize the local stability 
of the mapping 
about the trajectory ${\bf x}(n)$ \cite{LL,Arnold}. 
(If the mapping describes 
a Hamiltonian system, then 
these eigenvalues must sum to zero.) In particular, 
the Lyapunov exponents are obtained from 
the geometric mean of these matrices 
evaluated along some reference trajectory. The largest of these 
exponents may be defined from the double limit \cite{Tabor}, 
\begin{equation}
	\lambda_L = 
\lim_{n \to \infty} \; 
\lim_{|{\bf d}(0)| \to 0} \; \frac{1}{n} 
\ln \frac{|{\bf d}(n)|}{|{\bf d}(0)|}. 
\end{equation}
A more general definition (one that may be applied 
for practical computation) is provided in Chapter 2.

However, there is no definite 
concept of ``trajectory'' in quantum theory.    
Each quantum state, if pure, is 
specified by a vector in Hilbert space, and not a point 
in phase space. Moreover, 
the time-dependent Schrodinger equation is linear (more generally, 
the quantum time-development is given by a unitary transformation), 
whereas classical 
chaos requires  a nonlinear tranformation of the coordinates. 
For bounded systems, the eigenvalue spectrum is discrete, and the 
quantum dynamics are at most quasi-periodic. 

The significance of these differences may be illustrated  
by considering some   
measure of proximity that applies to quantum states, and then 
determining how this measure evolves in time. 
One definition of proximity that may be constructed in Hilbert space 
is the degree of overlap between two quantum states, 
$ | \langle \psi | \phi \rangle |$. 
Nearby states are then those that the satisfy,  
$ | \langle \psi | \phi \rangle | = 1 - \delta$ 
with $\delta \ll 1$.  However, this scalar quantity 
is invariant under unitary time-evolution,  
\begin{equation}
 | \langle \psi(t) | \phi(t) \rangle | = 
 | \langle \psi(0) | \; U^{-1}(t) \; U(t) \; | \phi(0) \rangle | 
 = | \langle \psi(0) | \phi(0) \rangle |.  
\end{equation}
Therefore, {\em any} two quantum states 
do not separate at all with time, for any system! 

This attempt at characterizing chaos in quantum mechanics, though 
unsuccessful, draws attention to two of the central questions 
that have motivated research in the field of ``quantum chaos.'' 
The first of these pertains to identifying signatures of 
chaos that arise in quantum mechanics:  
which criteria, if any, distinguish 
chaotic systems from regular ones in quantum theory? 
The second pertains to quantum-classical correspondence: 
how does classical 
chaos emerge from the underlying quantum description 
in the macroscopic limit? 
In particular, how can quantum theory  
reproduce the exponential divergence 
exhibited by the classical {\em trajectories}?   

Much of the work that attempts to identify manifestations of 
chaos in quantum theory has been obtained 
from numerical computation of the properties of specific Hamiltonian 
models. These results have suggested a few, possibly universal,  
signatures of chaos in quantum systems. 
Before proceeding to introduce some of these approaches, 
it is important to clarify that 
there is currently no {\em definitive} criterion for 
chaos in quantum theory.   
Consequently, in this thesis, when the term ``chaotic'' is invoked to 
describe a quantum system, this should be understood as a 
shorthand expression denoting that the classical counterpart 
to the quantum system exhibits chaotic behaviour. 

Perhaps the most well-known signature of ``quantum chaos'' 
arises in the distribution of 
spacings between consecutive energy eigenvalues.
Quantum systems with chaotic classical counterparts exhibit the same 
characteristic distributions for these spacings as those of 
a random matrix with the same symmetry properties as the Hamiltonian.  
These distributions exhibit {\em level repulsion} relative to their 
integrable counterparts:  small spacings are suppressed (leading to 
``avoided crossings'') and most  
spacings are clustered about an average value \cite{Haake91}.  
This signature exhibited in the eigenvalue spectrum 
has no obvious classical analogue.

Although there is no sensitivity to changes in the 
initial conditions in quantum mechanics (given the same unitary evolution 
for both initial conditions), 
the presence of classical chaos is evident in the quantum state's 
sensitivity to small perturbations in the Hamiltonian.  
This distinctive feature of the chaotic quantum dynamics  
has been demonstrated in several model systems, although  
the perturbations have been applied using very different techniques 
\cite{Peres84,Peres93,SC93,SC94,SC96,BZ96}.  
The basic idea is to evolve a given initial state, 
$|\psi(0) \rangle$, according to a unitary transformation,  
$| \psi(t) \rangle = U |\psi(0) \rangle$, 
where $U = \exp(-i H t / \hbar)$ for some Hamiltonian $H$,  and 
then examine the overlap, $|\langle \psi'(t)|\psi(t) \rangle|$, 
defined between this final state with another 
{\em perturbed} final state,   
$| \psi'(t) \rangle = U'|\psi(0) \rangle$. Here $U'$ may differ from $U$ 
according to some small perturbation in the Hamiltonian, or by some 
small coordinate transformation applied at the end of the 
unpertubed evolution \cite{BZ96}. 
If the classical Hamiltonian is chaotic, then the overlap 
decreases much more rapidly, and to a lower mean asymptotic value, than 
if the classical Hamiltonian is regular \cite{Peres93}. 
As noted by several authors \cite{SC92,BZ96}, 
this sensitivity criterion arises 
also in the case of a classical 
density that is evolved according to the Liouville equation. 
As will be demonstrated further below, 
much can be learned about quantum  mechanics, and quantum chaos, 
from the connection between quantum mechanics and the 
classical Liouville equation.

\section{Chaos and Ehrenfest Correspondence}

Several authors have addressed the question of quantum-classical 
correspondence for chaotic systems by comparing the 
dynamical behaviour of the quantum expectation values with that of   
the classical dynamical variables satisfying Hamilton's 
equations of motion \cite{BZ78,Frahm85,Haake87}. This approach 
is clearly motivated by the Ehrenfest correspondence conjecture 
explained in the previous section (\ref{eqn:Ehrcorr}). 
On this view,  the quantum dynamics may be identified as chaotic 
provided the quantum expectation values 
approximately follow a chaotic classical trajectory.  
If nearby classical trajectories are diverging, 
on average, exponentially, then the quantum expectation values will 
exhibit a similar exponential divergence, at least while the 
quantum state remains well-localised, that is, in the 
initial Ehrenfest regime, $t < t_{Ehr}$. 

Accordingly, one of 
the main interests has been to determine how the correspondence 
time, $t_{Ehr}$, increases with increasing system size. In the 
literature this limit is often denoted using the short-hand 
expression ``$\hbar \rightarrow 0$'', where in this context $\hbar$ 
always denotes the dimensionless ratio $(\hbar/{\cal J})$, 
and the magnitude of ${\cal J}$ is given by the quantum numbers 
or some combination of the system parameters. 
It is also useful to note that $(\hbar / {\cal J})$ does not actually 
denote a mathematical limit in quantum theory, but refers to a sequence 
of quantum models with increasing values of the ratio $({\cal J}/\hbar)$, 
but with other parameters adjusted so that each quantum model is associated 
with the same classical model. 

Haake {\it et al}.\ have stuided this time-scale in the special case of 
a simple non-autonomous 
spin model which they call the kicked top \cite{Haake87}.
The length of the spin is conserved, so the quantum number 
$j$ is a measure of the system dimension 
$j \hbar \sim {\cal J}$. 
For initial SU(2) coherent states, they obtain  
an approximate mapping relation for the  {\em expectation values} 
of the angular momentum components, $\langle J_k \rangle$, 
that consists of the corresponding classical mapping 
modifed by correction terms in powers of $(1/j)$. They are 
able to relate the growth of the first-order corrections to  
the tangent matrix of the classical mapping. The eigenvalues of 
this matrix govern the local stability of the classical motion. 
For mappings, the definition of the Lyapunov exponent is 
obtained from the geometric mean of these matrices evaluated 
along the classical trajectory (see Chapter 2). 
If the classical motion is regular, then 
the perturbation should grow, at most, as some 
power of the time. 
Consequently, for large $j$,  
significant deviations from the classical map 
occur on a very late time-scale, 
\begin{equation}
\label{eqn:powerbreaktime}
 t \sim (\frac{{\cal J}}{\hbar})^{1/\alpha}, 
\end{equation}
where $\alpha \simeq 1$.
This scaling is consistent 
with the example of the baseball considered above. 
On the other hand, if the classical mapping is chaotic,   
then the first-order quantum corrections terms should grow, on average, 
exponentially with time. For chaotic dynamics, 
the quantum mapping therefore deviates from the classical one 
on the time-scale, 
\begin{equation}
\label{eqn:logbreaktime}
 t  \sim \lambda_L^{-1} \ln( {\cal J}/\hbar).   
\end{equation}

The above correspondence scaling results for 
regular (\ref{eqn:powerbreaktime}) 
and chaotic (\ref{eqn:logbreaktime}) 
motion are consistent with estimates 
derived by other authors in several different 
model systems \cite{BZ78,Berry79a,Zas81,CC81,Frahm85}, though 
under a variety of different approximation techniques. 
In Chapter 5, I will provide a simpler and much more general argument 
than the one outlined above, which demonstrates that 
Eqs.\ (\ref{eqn:powerbreaktime}) 
and (\ref{eqn:logbreaktime}) should hold (for regular and 
chaotic systems respectively) in the general case of 
a system of particles interacting only through position 
dependent potentials.   

There are a few features of the above break-time rules 
that are worth stressing immediately. 
First, Eqs.\ (\ref{eqn:powerbreaktime}) and (\ref{eqn:logbreaktime})
apply specifically in the case of initial  
minimum uncertainty states, and in this sense 
they {\it maximize} the time-scale of correspondence. 
Second, these scaling results are guaranteed to apply 
only when considering the onset of very small 
differences, since the exponential growth of the differences 
is derived on the basis of the local stability of the classical flow. 
The scaling for large  
differences, that is, differences given by a significant 
fraction of the system dimension, is left undetermined. 
Finally, under the restrictive conditions noted just above, 
these expressions may be taken as an estimate of how 
the Ehrenfest break-time, defined by (\ref{eqn:Ehrcond}), 
scales with increasing system size for chaotic systems. 
This follows from the fact that the coordinate centroids of the  
quantum states are associated with 
the phase space coordinates of a classical trajectory. 
Consequently, although 
the time-scale (\ref{eqn:logbreaktime}) is usually called 
simply the ``break-time'' of quantum-classical correspondence for a chaotic 
system, in this thesis I will refer to this result more specifically as    
the {\em Ehrenfest} break-time, for reasons that will become clear shortly.
As a result of the time-scale 
(\ref{eqn:logbreaktime}), several authors have argued that 
deterministic chaos can exist 
only as a transient phenomena in quantum mechanics \cite{CC,Chi88}. 



Another distinct feature of quantum chaos
was first identified by Fox and coworkers in numerical studies of the 
kicked top and the kicked rotor (which they call the kicked pendululm) 
\cite{Fox94a,Fox94b}.  These authors showed that, 
when the classical dynamics are chaotic, the widths of 
initially localised quantum states grow exponentially with time,    
until saturation at the system size.  Specifically, their 
numerical studies showed that,   
\begin{equation}
\label{eqn:fox}
 \Delta q(t)  
\simeq \Delta q(0) \exp \lambda_w t, 
\end{equation}
where $\Delta q^2 
= \langle q^2 \rangle - \langle q\rangle^2$ 
is the variance of the quantum state and $\lambda_w$ is a characteristic 
exponent of order the largest Lyapunov exponent. 

A separate, though equally important, 
contribution of their work was to recognize 
that the exponential spreading of the quantum state mimicks that of a 
classical distribution of trajectories concentrated around the initial 
quantum centroids.  More precisely, they 
constructed initial classical densities which matched 
the moments of the Husimi phase space functions for 
the initial coherent states \cite{Fox94b}.  
This close correspondence in the growth rates of quantum states 
and classical densities has been characterized more carefully 
in recent work by Ballentine and McRae \cite{Ball98,Ball00}, 
who have studied this problem in the Henon-Heiles model.  
The exponential growth of quantum and classical variances 
is demonstrated also for the chaotic states of 
the coupled spin model examined in this thesis (see Chapter 3). 

This exponential spreading of quantum states  
is closely related to the earlier estimate (\ref{eqn:logbreaktime}) 
that Ehrenfest correspondence 
is governed by a log(${\cal J}/\hbar$) break-time. 
As noted above, the width of the quantum state gives rise to 
the first correction term in the expansion (\ref{eqn:taylor}), 
and the precision of the correspondence between 
the quantum expectation values and the classical trajectory 
is limited by the magnitude of this correction term. 
In particular, for an initial minimum-uncertainty state,  
$\Delta q(0) \simeq \hbar / \Delta p(0) $, 
the time at which 
the width in position space has grown sufficiently large that 
(\ref{eqn:loc1}) no longer holds 
is obtained by setting the RHS of (\ref{eqn:fox})
equal to the RHS of (\ref{eqn:loc1}) and solving for the 
Ehrenfest break-time. This gives, 
\begin{equation}
\label{eqn:tehr}
t \sim \lambda_w^{-1} \ln(\frac{L \Delta p(0)} {\hbar}),  
\end{equation}
where $L$ is a characteristic length 
defined {\em relative} to the system size. 
The magnitude $L$ may be interpreted as a threshold indicating a ``break'' 
between the quantum and classical predictions. 
It is important to stress that the validity of 
(\ref{eqn:tehr}) is based on a numerical result (\ref{eqn:fox}), 
rather than derived, but is more general than 
the earlier estimate of the break-time scaling (\ref{eqn:logbreaktime}),  
since the derivation of Eq.\ (\ref{eqn:logbreaktime}) 
applied strictly in the limit of vanishingly small differences. 
As noted earlier, the results of Fox {\it et al}.\
indicate that this exponential growth persists until 
the state width reaches the system size; therefore  
the scaling rule (\ref{eqn:tehr}) remains valid even for estimating 
the onset of very large quantum-classical differences.  


Clearly, for a macroscopic system, 
the ratio ${\cal J}/\hbar$ is  astronomically large. 
The incredibly large magnitude of this ratio is a crucial 
ingredient in Ehrenfest's approach to extracting 
Newton's Laws from quantum mechanics \cite{Ehrenfest}. 
However, Zurek and Paz \cite{ZP95a} have observed that a 
$\log({\cal J}/\hbar)$ time-scale may be  short even for 
systems of macroscopic dimension, {\it i.e}., systems with  
${\cal J}/\hbar >> 1$. Specifically, they have estimated 
that the time-scale $t \simeq \lambda_L^{-1} \ln({\cal J}/\hbar)$
is comparable to experimentally relevant time-scales in the case of 
Hyperion, one of the moons of Saturn, which is known to exhibit 
a chaotic tumble \cite{Hyperion84,Hyperion94}. Therefore the 
Ehrenfest correspondence principle is at risk of {\it breaking 
down} in the case of chaotic motion. I will examine 
this possibility in detail in Chapter 5, and conclude that 
such a breakdown actually may be subject to experimental confirmation!  
First, however, in view of the far-reaching 
implications of this result, it is worthwhile to 
reconsider the assumptions about quantum-classical correspondence 
that are reflected in the Ehrenfest correspondence principle  
and explore more carefully 
the alternatives to this correspondence conjecture. 

\section{Quantum Correspondence with Liouville Mechanics} 

In the preceding discussion I have drawn attention to 
some similarities between the 
properties of a time-evolved quantum state and those of 
a corresponding ensemble of classical trajectories.  
The general description of the time-evolution of 
this ensemble is prescribed by the classical Liouville equation. 
This connection may provide further insight 
into the problem of identifying the distinctive features of the 
quantum dynamics that arise under conditions of 
chaos in the classical Hamiltonian dynamics. 
Moreover, in view of the practical and conceptual 
difficulties that challenge the validity of the Ehrenfest 
correspondence conjecture (that is, the view  
that the predictions of quantum mechanics coincide with 
Newton's laws for macroscpic objects), it is important  
to determine whether the correspondence between quantum 
mechanics and classical Liouville mechanics is constrained 
by similar difficulties in the case of  chaotic dynamics.  

A useful starting point for investigating the 
correspondence between these two theories 
starts with a description of the classical Liouville equation.
Since one typically does not have {\em exact} 
knowledge of the initial coordinates 
that describe a physical system, Liouville was interested in describing 
how this cloud, or ensemble, of {\em possible} 
initial coordinates evolves in time. 
The time-dependence is given by the following partial 
differential equation \cite{Goldstein},   
\begin{equation}
\label{int:liouville}
	\frac{\partial \rho({\bf x},t) }{\partial t}
		= \{ \rho({\bf x}), H({\bf x})\} 
\end{equation}
where $\rho({\bf x},t)$ is a non-negative distribution function 
(a probability density), ${\bf x}$ are the phase space 
coordinates, and 
the symbols $\{ \cdot , \cdot \}$ denote the Poisson bracket. 
A unique solution to the time-dependence requires the specification 
of an initial boundary condition for the density, {\it i.e}., 
some initial state $\rho(q,p,0)$.  
A classical dynamical system is also associated with 
an invariant (Liouville) measure $ \mu({\bf x})$ on the  
phase space manifold ${\cal P}$. Then, at each point in time, 
one can calculate the ensemble average for 
an observable $A({\bf x})$ from the prescription, 
\begin{equation}
\label{eqn:lexp}
	\langle A({\bf x})\rangle_c = \int_{\cal P} d \mu({\bf x}) 
			\; \rho_c({\bf x}) \; A({\bf x}). 
\end{equation}

Ballentine {\it et al}.\  \cite{Ball94} have shown that,  
in the generic case of Hamiltonians of the form, $H = p^2/2m + V(q)$, 
the time-dependence of the classical ensemble averages 
exhibit a striking similarity to the quantum ones. 
The Liouville description even includes 
some of the ostensibly ``quantum'' corrections to the classical 
canonical equations of motion.
From (\ref{int:liouville}) and (\ref{eqn:lexp}) 
and integrating by parts, one gets,  
\begin{eqnarray}
\label{eqn:Ldiff}
\frac{d \langle q \rangle_c }{d t} &= &
\frac{\langle p \rangle_c}{m} \nonumber \\
\frac{d \langle p \rangle_c }{d t} &= & 
\langle F(q) \rangle_c.  
\end{eqnarray}
As before, it is useful to expand the force function 
about the position centroid (of the classical state), which gives,
\begin{eqnarray}
\label{eqn:Ldiff2}
\frac{d \langle q \rangle_c }{d t} &= &
\frac{\langle p \rangle_c}{m} \nonumber \\
\frac{d \langle p \rangle_c }{d t} &= & 
F(\langle q \rangle_c) + 
\frac {\Delta q_c^2}{2} 
\left. \frac{d^2 F(q)}{ dq^2} \right|_{q=\langle q \rangle_c}
+ \dots  
\end{eqnarray}
where $\Delta q_c^2 = \langle q^2 \rangle_c - \langle q \rangle_c^2$.
As noted above for the time-development of 
the quantum expectation values (\ref{eqn:qdiff}), the 
classical ensmble averages also exibit a 
first-order ``correction'' term which is proportional to the 
state variance and which will lead to deviations from 
the canonical equations of motion (\ref{eqn:cdiff}).

Of course, the quantum and classical Liouville theories 
do not provide identical theoretical descriptions 
of the same physical system. 
There are two independent sources of discrepency: 
dynamical differences and differences in the initial conditions. 
The dynamical differences become apparent only when the 
time-development of the higher-order moments (that is, the additional  
terms in the expansion (\ref{eqn:taylor})) are calculated explicitly.  
These calculations 
are carried out most easily in the Wigner-Weyl representation of 
quantum mechanics \cite{Wigner}. 
The relevent features of this 
representation are introduced in Appendix B. 

The time-dependence of the Wigner quasi-distribution $\rho_W(q,p)$ 
can be expressed as \cite{Moyal}, 
\begin{equation}
\label{eqn:wm}
	\frac{d \rho_w(q,p,t) } {d t} 
= \{ H , \rho_w \} + \sum_{n=1}^{\infty} 
\left(\frac{\hbar}{2}\right)^{2n} \frac{(-1)^n}{(2n+1)!}
\frac {\partial^{2n+1} \rho_w} {\partial p^{2n+1}} 
\frac {\partial^{2n+1} H} {\partial q^{2n+1}}
\end{equation}
where $\{\cdot,\cdot \}$ denotes the classical Poisson bracket. 
As explained in Appendix B, although the ``Moyal terms'' contain 
an explicit proportionality to 
powers of $\hbar$, these terms are not guaranteed to become vanishingly  
small in the limit ``$\hbar \rightarrow 0$'' since the momentum 
derivatives will generally produce factors proportional to 
$(q/\hbar)$. 

The Wigner quasi-distribution is not a unique quantum phase 
space distribution, and more importantly, may take on negative values.  
Consequently, it may not be interpreted as a probability density and 
does not have direct experimental significance.  
The origin of the dynamical differences between the quantum and classical 
theories is clearer if we consider differences arising explicitly 
at the level of observable quantities. 
As shown in Appendix B, for a general time-independent 
Weyl-ordered operator, $A(q,p)$, 
the time-development of the corresponding quantum expectation value 
$\langle A(q,p) \rangle$, is given by, 
\begin{equation}
	\frac{d \langle A_w(q,p) \rangle } {d t} 
= \{ H_w , A_w \} + \sum_{n=1}^{\infty} 
\left(\frac{\hbar}{2}\right)^{2n} \frac{(-1)^{n+1}}{(2n+1)!}
\frac {\partial^{2n+1} A} {\partial p^{2n+1}} 
\frac {\partial^{2n+1} H} {\partial q^{2n+1}}. 
\end{equation}
The Moyal corrections to the dynamics for Liouville averages do not  
arise {\em explicitly} unless the operator $A(q,p)$ contains at least a cubic 
power of the momentum. Differences arise {\em implicitly}, however, 
even for the low-order moments, since the 
time-development of the low-order moments will eventually become  
sensitive to the Moyal corrections through their dependence on 
the time-development of the higher moments.   
The quantum and Liouville time-evolution equations evidently exhibit 
a much greater degree of similarity than that observed between 
the quantum equations and Hamilton's equations of motion. 

Similary, the Liouville picture provides a much more versatile 
framework for matching the details of the initial quantum state. 
Using a Liouville density with non-vanishing support, it is possible to 
match not only the initial quantum expectation values (associated  
with the classical dynamical variables), but also the   
higher-order quantum moments.  The ``best-case'' scenario 
arises for the coherent states, which are minimum uncertainty 
states, $ \Delta q \Delta p = \hbar/2$. These states are the unique 
quantum states that are associated with non-negative Wigner functions.  
Consequently all of the quantum moments (given by Hermitian Weyl-ordered 
operators) can be matched exactly by those of a classical density. 
The appropriate classical density is a Gaussian with support 
concentrated over an area $ \Delta q_c \Delta p_c \simeq \hbar$, 
as expected from the uncertainty principle. 
Perhaps not surprisingly, this 
matching may not be carried out to arbitrary precision 
for a general quantum state. An intuitive way 
of understanding this is to note that 
the Wigner quasi-distribution will generally 
take on negative values,  
and therefore some of  the quantum moments will take on 
numerical values that cannot 
be reproduced by a non-negative Liouville density.  


These formal observations motivate 
an alternative interpretation of the requirements of the  
correspondence principle. This interpretation consists of 
the conjecture that the 
the predictions of quantum mechanics should coincide with those 
of classical Liouville mechanics when the predictions of 
the latter are experimentally valid. 
It will sometimes be convenient to refer to this conjecture as 
the {\it Liouville} correspondence principle.  
This is a weaker conjecture than the Ehrenfest correspondence principle,  
which was stated in the discussion 
preceding the heuristic formula (\ref{eqn:Ehrcorr}).   

Ballentine and coworkers 
have studied the details of this quantum-Liouville 
correspondence for a few non-integrable systems.
Ballentine, Yang, and Zibin \cite{Ball94}  
compared quantum expectation values and classical 
ensemble averages for the low-order moments, {\it e.g}.\ 
the dynamical variables and their variances, primarily 
at early times, that is, in the initial Ehrenfest regime 
of a few low-dimensional model systems. 
For fixed system size, the correspondence with 
Liouville mechanics was shown to be much more accurate, and 
to last much longer, than the correspondence with a single 
classical trajectory. They demonstrated a remarkable degree of 
improved correspondence  
under conditions of classical chaos. 

Ballentine and McRae \cite{Ball98} and Ballentine \cite{Ball00} 
examined the degree of Liouville  
correspondence for the Henon-Heiles model. 
They demonstrated 
that both the quantum and classical 
variances grow at an approximately exponential rate, 
a result obtained previously by Fox and 
coworkers for different model systems \cite{Fox94a,Fox94b}. 
They also characterised the dynamical 
behaviour of the quantum-Liouville differences 
for the dynamical variables and their variances. 
They determined that, for states initiated from a chaotic 
region, the quantum-Liouville differences 
grow exponentially, in the initial 
Ehrefenst regime, with a characteristic exponent that is 
larger than the largest Lyapunov exponent. 




The duration of the correspondence between quantum 
mechanics and Liouville mechanics has been considered 
by Zurek and Paz \cite{ZP94,ZP95b} in the ``optimal''
case of minimum uncertainty states.  
Using very general arguments about the nature of the 
chaotic classical dynamics and the growth of the 
Moyal corrections in Eq.\ (\ref{eqn:wm}), they have estimated  
that the  quantum-Liouville differences should be 
governed by a break time that scales as, 
\begin{equation}
\label{int:tb}
	t_b \simeq \frac{1}{\lambda_L} \ln( L \; \Delta p(0)/\hbar)
\end{equation}
where $L$ and $\Delta p(0)$ are defined as in Eqs.\ 
(\ref{eqn:loc1}) and (\ref{eqn:tehr}). 
Casati and Chirikov \cite{CC95} have argued that 
(\ref{int:tb}) is not a distinct result, 
but is essentially the 
same as the (Ehrenfest) break-time, given by (\ref{eqn:logbreaktime}),   
that was first developed in Refs.\ \cite{BZ78,CC81}. 
Although the estimates (\ref{int:tb}) 
and (\ref{eqn:logbreaktime}) bear a formal  
similarity, this Introduction makes clear that they 
are based on distinct correspondence criteria.  
(The conditions for Liouville and Ehrenfest correspondence 
are often not clearly distinguished in the literature.)  

However, the formal similarity between $t_b$ and $t_{Ehr}$ suggests 
that the Liouville correspondence 
principle may also be at risk of breaking down in the case of  
chaotic macroscopic motion, for systems such as Hyperion, 
since the break-time scaling 
exhibited by Eq.\ (\ref{int:tb}) suggests the possibility  
of observable quantum-classical differences on an experimentally 
accessible time-scale \cite{ZP95a}.  While Zurek has 
raised the possibility of a {\em breakdown} of the 
correspondence principle \cite{Zurek98b} in this context, 
he has also suggested  
that this breakdown is not actually observed because of the 
``decoherence'' effects that result from weak coupling to the ubiquitous 
environment.  


Habib {\it et al}.\ \cite{Zurek98a} have 
considered the effects of decoherence on the 
degree of Liouville correspondence at the 
level of expectation values for low-order moments and also  
by comparing Wigner quasi-distributions directly 
with the classical density in study of the Henon-Heiles model.   
They demonstrated  
that the degree of correspondence (for fixed $\hbar/{\cal J}$) 
improved when the effects of a stochastic enviroment (or ``decoherence'') 
are taken into account. 
The positive effects of decoherence on the degree of quantum-Liouville  
correspondence have been demonstrated in several recent theoretical 
studies of classically chaotic models \cite{Kol96,HB96,BGLR96,Gong99}. 
In the case of the kicked rotor model,  
a suppression of quantum localisation 
due to environmental noise has been demonstrated {\it experimentally} 
in a simulation of the model 
using trapped cesium atoms subjected to a pulsed 
laser beam \cite{Ammann98}. The ``environment'' in this case consists  
of spontaneous emission due to the ``vacuum fluctuations.'' 
However, Farini {\it et al}.\ \cite{Farini} have shown  
that the magnitude of the Moyal terms in Eq.\ (\ref{eqn:wm})  
can remain small even for 
an isolated system, and therefore the effects of decoherence 
may not be necessary to recover classical properties. 

The scaling of the Liouville break-time estimate (\ref{int:tb}) has been 
examined by Roncaglia {\it et al}.\ \cite{RBWG95} in numerical 
studies of the {\it anomolous diffusion regime} of the kicked rotor. 
These authors provided some numerical evidence in support 
of the scaling with 
increasing system size expressed in Eq.\ (\ref{int:tb}) by 
considering the differences between 
the quantum expectation values and Liouville ensemble 
averages for low-order moments, such as $\langle p^2 \rangle$. 
However, it is important to emphasize that the $\log({\cal J}/\hbar)$ 
scaling for the Liouville break-time 
does not hold in the general case. In particular, 
the quantum-Liouville break-time that arises in the  
{\it localisation regime} of the 
kicked rotor is believed to grow as a small power 
of the ratio $({\cal J}/\hbar)$ \cite{Haake91}, which is more than adequate 
for correspondence in the macroscopic limit.

These considerations have been outlined in order to motivate 
the following questions: (1) Under what conditions does the 
Liouville break-time estimate (\ref{int:tb}) hold? 
(2) Although environmental 
noise may improve the degree of quantum-Liouville correspondence, 
is decoherence necessary for quantum-Liouville 
correspondence in the macroscopic limit? 
(3) Can decoherence help restore Ehrenfest correspondence?  


The main focus of the work presented in this thesis 
is to characterize the properties of Liouville and Ehrenfest 
correspondence, particularly under conditions of 
classical chaos, for an isolated 
few degree-of-freedom system. 
The central challenge of this analysis consists of 
extrapolating the predictions of quantum theory (and therefore 
the degree of correspondence) 
into the macroscopic regime, where direct calculation is 
of course not feasible in the generic 
case of a non-integrable system. 

This thesis is organised as follows.
In Chapter 2 I will describe 
the quantum and classical versions of the model examined 
in this thesis. 
Since the model is novel I will examine the behaviours of 
the chaotic classical dynamics in some detail.  
I will also describe the initial 
quantum states, which are SU(2) coherent states, 
and then define a corresponding classical density on the 
2-sphere which is a good analog for these  
states. I show in Appendix A that a perfect match is 
impossible: no distribution on ${\mathcal S}^2$ can reproduce 
the moments of the SU(2) coherent states exactly. 
At the end of Chapter 2 I will describe some of the 
numerical techniques that were 
required for the quantum and classical Liouville 
descriptions of the dynamics.

In Chapter 3 I will examine the characteristics of quantum-Liouville 
correspondence at the level of the low-order moments. 
First I will demonstrate that a  
close correspondence with Liouville mechanics 
persists well after the Ehrenfest correspondence has broken down. 
The widths of both the quantum states and the Liouville densities 
grow exponentially until saturation at the system size. 
I will then confirm that the quantum-Liouville differences exhibit 
an initial phase of exponential growth 
for chaotic states, but also demonstrate that this growth 
terminates well before the differences reach the system size. 
I will also show that the Liouville 
break-time scales according to the estimate Eq.\ (\ref{int:tb}), but only 
for the onset of very small quantum-Liouville differences. 
By ``very small'' I mean differences that remain a factor $(\hbar/{\cal J})$
smaller than the system size.  As a result, I argue that decoherence 
is not necessary for the emergence of these classical properties 
in the macroscopic limit. 

This demonstrated 
correspondence at the level of the low-order moments still 
leaves room for significant quantum-Liouville deviations. 
In Chapter 4 I will examine the degree of correspondence 
between the quantum and classical probability distributions, emphasizing 
the details of this correspondence after the states have spread 
to the system size. 
In particular, I will show that the initially localised 
quantum states approach the classical equilibrium 
on the same relaxation time-scale as the corresponding Liouville density.  
When the classical Hamiltonian exhibits widespread chaos, 
the quantum probability distributions 
approach microcanonical equilibrium configurations. 
I also demonstrate that 
the quantum fluctuations away from the classical equilibrium 
become vanishingly small in the macroscopic limit. 
These results reinforce the conclusion 
that quantum-Liouville differences for classical observables become 
increasingly difficult to detect as the system size increases.  

In Chapter 5 I will examine the properties of Ehrenfest correspondence. 
I will demonstrate that the Ehrenfest  
differences, in contrast with the quantum-Liouville differences,
actually grow to a macroscopic size on the short time-scale  
(\ref{eqn:tehr}) in the case of chaotic classical dynamics.  
Moreover, decoherence is unable to restore 
the {\it Ehrenfest} correspondence for these observables.  
I will conclude that Ehrenfest correspondence may be subject 
to experimental refutation for some chaotic macroscopic bodies.
At the end of Chapter 5 I will address the 
implications of this result for the correspondence principle 
and the interpretation of the quantum state. 


\chapter{Description of the Model} 


This chapter is taken from Emerson and Ballentine \cite{EB00a}.

\section{The Model}
\label{sec2}

We consider the quantum and classical dynamics 
generated by a non-integrable model of two interacting spins, 
\begin{equation}
        H = a (S_z + L_z) + c S_x L_x \sum_{n = -\infty}^{\infty}\delta(t-n) 
\label{eqn:ham}
\end{equation}
where ${\bf S} = (S_x,S_y,S_z)$ and 
${\bf L} = (L_x,L_y,L_z)$.
The first two terms in (\ref{eqn:ham}) 
correspond to simple rotation of both spins 
about the $z$-axis. 
The sum over coupling terms describes an infinite 
sequence of $\delta$-function interactions 
at times $t=n$ for integer n. 
Each interaction term corresponds to an impulsive rotation 
of each spin about the $x$-axis by an angle proportional 
to the $x$-component of the other spin. 

\subsection{The Quantum Dynamics}


To obtain the quantum dynamics we interpret 
the Cartesian components of the spins 
as operators satisfying the usual angular momentum 
commutation relations,  
\begin{eqnarray*}
  & [ S_i,S_j ] = & i \epsilon_{ijk} S_k \\
  & [ L_i,L_j ] = & i \epsilon_{ijk} L_k \\ 
  & [ J_i,J_j ] = & i \epsilon_{ijk} J_k .
\end{eqnarray*}
In the above we have set $\hbar =1$ and introduced the total 
angular momentum vector ${\bf J} = {\bf S} + {\bf L}$.

The Hamiltonian (\ref{eqn:ham}) possesses kinematic 
constants of the motion, $ [ {\bf S}^2,H] = 0$ and $[ {\bf L}^2,H] =0 $,  
and the total state vector $|\psi\rangle$ 
can be represented in a finite Hilbert space of  
dimension $ N =(2s+1) \times (2l+1)$. 
This space  is spanned by the orthonormal vectors 
$|s,l,m_s,m_l \rangle = |s,m_s \rangle \otimes |l,m_l \rangle$ 
with $m_s \in \{ s, s-1, \dots, -s \}$ 
and $m_l \in \{ l, l-1, \dots, -l \}$.  
These are the joint eigenvectors of the four spin operators 
\begin{eqnarray}
\label{eqn:basis}
	{\bf S}^2 |s, l, m_s,m_l \rangle & =& s(s+1) 
| s,l,m_s,m_l \rangle,  \nonumber \\
	S_z | s, l,m_s,m_l \rangle & =& m_s 
| s,l,m_s,m_l \rangle,  \\
	{\bf L}^2 | s,l,m_s,m_l \rangle &= & l(l+1) 
| s,l,m_s,m_l \rangle,  \nonumber \\
	L_z | s,l,m_s,m_l \rangle &= & m_l 
| s,l,m_s,m_l \rangle.  \nonumber 
\end{eqnarray}

The periodic sequence of interactions introduced by the 
$\delta$-function produces a quantum mapping. 
The time-evolution 
for a single iteration, from just before a kick to just before 
the next, is produced by the unitary transformation, 
\begin{equation}
\label{eqn:qmmap}
	| \psi(n+1) \rangle = F \; | \psi(n) \rangle,  
\end{equation}
where $F$ is the single-step Floquet operator, 
\begin{equation}
\label{eqn:floquet}
        F = \exp \left[ - i a (S_z + L_z) \right] 
                \exp \left[ - i c S_x L_x \right].  
\end{equation}
Since $a$ is a rotation its range is $2\pi$ radians. 
The quantum dynamics are thus specified by two 
parameters, $a$ and $c$, 
and two quantum numbers, $s$ and $l$. 

An explicit representation of the 
single-step Floquet operator can be obtained in the basis 
(\ref{eqn:basis}) by first re-expressing the 
interaction operator in (\ref{eqn:floquet})  
in terms of rotation operators, 
\begin{eqnarray}
 \exp \left[ - i c S_x \otimes L_x \right] & =  & 
[R^{(s)}(\theta,\phi) \otimes  R^{(l)}(\theta,\phi)] \;  
\exp \left[ - i c S_z \otimes L_z \right]	\nonumber \\
& &  \times [R^{(s)}(\theta,\phi) \otimes R^{(l)}(\theta,\phi)]^{-1}, 
\end{eqnarray}
using polar angle $\theta=\pi/2$ and azimuthal angle $\phi=0$. 
Then non-zero off-diagonal terms arise only in the expressions  
for the rotation matrices, which take the form,   
\begin{equation}
	\langle j,m'| R^{(j)}(\theta,\phi) | j,m \rangle = \exp(-i m' \phi )  
		d^{(j)}_{m',m}(\theta).
\end{equation}
The matrix elements, 
\begin{equation}
d^{(j)}_{m',m}(\theta) = \langle j,m'| \exp( - i \theta J_y ) | j,m \rangle
\end{equation}
are given explicitly by Wigner's formula \cite{Sakurai}.

We are interested in studying the different time-domain 
characteristics of quantum observables when the corresponding 
classical system exhibits either regular or chaotic dynamics. 
In order to compare quantum systems with different quantum numbers   
it is convenient to normalize subsystem observables 
by the subsystem magnitude 
 $\sqrt{\langle {\bf L}^2\rangle} = \sqrt{l(l+1)}$. 
We denote such normalized observables with a tilde, where
\begin{equation}  
	\langle \tilde{L}_z (n) \rangle  =  
{ \langle \psi(n)| L_z  |\psi(n) \rangle  \over \sqrt{l(l+1)} }
\end{equation}
and the normalized variance at time $n$ is defined as, 
\begin{equation}  
	\Delta {\tilde {\bf L}  }^2 (n)   
	= {\langle  {\bf L}^2 \rangle - 
\langle  {\bf L }(n) \rangle^2 \over l(l+1) } .  
\end{equation}

We are also interested in evaluating the properties 
of the quantum probability distributions. The 
probability distribution 
corresponding to the observable $L_z$ is given by the trace, 
\begin{equation}
P_z(m_l) = {\mathrm T} {\mathrm r} \left[ \rho^{(l)}(n) | l, m_l \rangle 
\langle l, m_l | \right] = \langle l,m_l |  \rho^{(l)}(n) |  l,m_l\rangle , 
\end{equation}
where $\rho^{(l)}(n)= {\mathrm T}{\mathrm r}^{(s)} \left[ \; | \psi(n) \rangle 
\langle \psi(n) | \; | s, m_s \rangle 
\langle s, m_s | \; \right]$ is the reduced state operator for 
the spin ${\bf L}$ at time $n$ and ${\mathrm T}{\mathrm r}^{(s)}$ denotes 
a trace over the factor space corresponding to the spin ${\bf S}$. 


\subsection{Classical Map} 

For the Hamiltonian (\ref{eqn:ham}) the corresponding 
classical equations of motion 
are obtained by interpreting the angular momentum components 
as dynamical variables satisfying,   
\begin{eqnarray*}
  & \{ S_i,S_j \} = & \epsilon_{ijk} S_k \\
  & \{ L_i,L_j \} = & \epsilon_{ijk} L_k \\ 
  & \{ J_i,J_j \} = & \epsilon_{ijk} J_k ,
\end{eqnarray*}
with $\{ \cdot,\cdot \}$ denoting the Poisson bracket. 
The periodic $\delta$-function in the coupling term 
can be used to define surfaces at $t=n$, for integer $n$, 
on which the time-evolution reduces to a stroboscopic mapping,  
\begin{eqnarray}
\label{eqn:map}
        \tilde{S}_x^{n+1} & = & \tilde{S}_x^n \cos( a) - 
        \left[ \tilde{S}_y^n \cos( \gamma r \tilde{L}_x^n ) - 
        \tilde{S}_z^n \sin( \gamma r 
        \tilde{L}_x^n)\right] \sin( a), \nonumber \\ 
        \tilde{S}_y^{n+1} & = & \left[ \tilde{S}_y^n \cos( 
        \gamma r \tilde{L}_x^n) - \tilde{S}_z^n \sin( \gamma r
        \tilde{L}_x^n ) \right] \cos( a) + 
        \tilde{S}_x^n \sin( a), \nonumber \\
        \tilde{S}_z^{n+1} & = &  \tilde{S}_z^n \cos( 
        \gamma r \tilde{L}_x^n) + \tilde{S}_y^n \sin( 
        \gamma r \tilde{L}_x^n), \\
        \tilde{L}_x^{n+1} & = & \tilde{L}_x^n \cos( a) - 
        \left[\tilde{L}_y^n\cos(\gamma \tilde{S}_x^n) - \tilde{L}_z^n 
        \sin(\gamma\tilde{S}_x^n )\right]\sin(a), \nonumber \\ 
        \tilde{L}_y^{n+1} & = & \left[ \tilde{L}_y^n \cos( \gamma 
        \tilde{S}_x^n) - \tilde{L}_z^n \sin( \gamma
        \tilde{S}_x^n) \right] \cos( a) + 
        \tilde{L}_x^n \sin( a), \nonumber \\
        \tilde{L}_z^{n+1} & = &  \tilde{L}_z^n \cos( 
        \gamma \tilde{S}_x^n )+ \tilde{L}_y^n \sin( \gamma 
        \tilde{S}_x^n), \nonumber 
\end{eqnarray}
where ${\tilde {\bf L}} = {\bf L} / |{\bf L}|$ , 
${\tilde {\bf S}} = {\bf S} / |{\bf S}|$   
and we have introduced the parameters 
$ \gamma = c |{\bf S}| $ and $ r = | {\bf L}| / | {\bf S}| $. 
The mapping equations (\ref{eqn:map}) describe 
the time-evolution of (\ref{eqn:ham}) 
from just before one kick to just before 
the next. 

Since the magnitudes of both spins are conserved, 
$ \{ {\bf S}^2,H \} = \{ {\bf L}^2,H \} =0$, 
the motion is actually confined to the four-dimensional 
manifold ${\mathcal P} ={\mathcal S}^2 \times {\mathcal S}^2$, 
which corresponds to the surfaces of two spheres. 
This is manifest when the mapping (\ref{eqn:map}) 
is expressed in terms of the four {\it canonical} coordinates 
${\bf x} = (S_z, \phi_s, L_z , \phi_l )$, where 
$\phi_s = \tan (S_y /S_x)$ and $\phi_l = \tan(L_y/L_x) $. 
We will refer to the mapping (\ref{eqn:map}) in canonical form 
using the shorthand notation ${\bf x}^{n+1} = {\bf F}({\bf x}^n)$. 
It is also useful to introduce a complete set of spherical coordinates 
$ \vec{\theta} = (\theta_s,\phi_s,\theta_l,\phi_l) $ where 
$\theta_s = \cos^{-1} (S_z / |{\bf S}|) $ and 
$\theta_l = \cos^{-1} (L_z / |{\bf L}|) $.

The mapping (\ref{eqn:map}) on the reduced surface  
${\mathcal P}$ enjoys a rather large 
parameter space; the dynamics are  
determined from three independent dimensionless 
parameters: $a \in [0,2\pi)$, $\gamma \in (-\infty,\infty)$, 
and $r \ge 1$. 
The first of these, $a$, controls the angle of 
free-field rotation about the $z$-axis. 
The parameter $\gamma= c |{ \bf S }|$ 
is a dimensionless coupling strength 
and $r = |{\bf L}|/ |{ \bf S }| $ corresponds 
to the relative magnitude of the two spins.

\begin{figure}[!t]
\begin{center}
\input{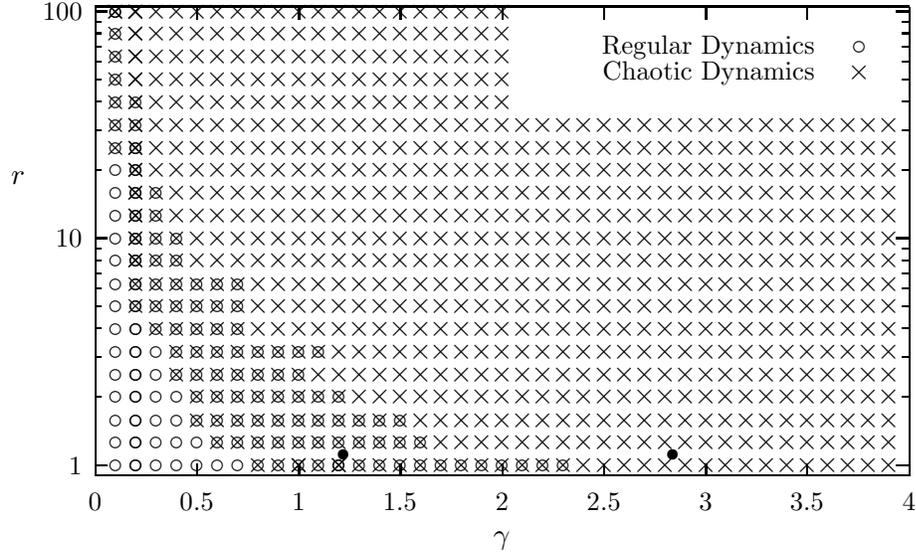}
\caption{
Behaviour of the classical mapping for 
different values of $r=|{\bf L}|/ |{\bf S}|$ 
and $\gamma =c |{\bf S}|$ with $ a =5$.   
Circles correspond to parameter values for which at least 99\% 
of the surface area ${\mathcal P}$ produces regular dynamics and 
crosses correspond to parameter values for which 
the dynamics are at least 99\% chaotic.  
Superpositions of circles and crosses correspond to parameter values which 
produce a mixed phase space. We investigate quantum-classical correspondence 
for the parameter values $\gamma = 1.215$ (mixed regime) 
and $\gamma= 2.835$ (global chaos), with $r=1.1$, 
which are indicated by filled circles.  
}
\label{regimes}
\end{center}
\end{figure}

We are particularly interested in the effect of increasing  
the coupling strength $\gamma$ for different fixed values of $r$. 
In Fig.\ \ref{regimes} 
we plot the dependence of the classical behaviour 
on these two parameters for the case $a=5$, which produces typical results. 
The data in this figure 
were generated by randomly sampling initial conditions on
${\mathcal P}$, using the invariant measure,
\begin{equation}
\label{eqn:measure}
d \mu( {\bf x} ) = d \tilde{S}_z d \phi_s d \tilde{L}_z d \phi_l ,
\end{equation}
and then calculating the largest Lyapunov exponent associated 
with each trajectory. Open circles 
correspond to regimes where at least $99\%$ of the 
initial conditions were found to exhibit regular behaviour 
and crosses correspond to regimes where at least $99\%$ of 
these randomly sampled initial conditions were found to exhibit 
chaotic behaviour.
Circles with crosses through them (the superposition of 
both symbols) correspond to regimes with a mixed phase space. 
For the case $a=5$ and with $r$ held constant, 
the scaled coupling strength $\gamma$ plays the role of a   
perturbation parameter: 
the classical behaviour varies from regular, to mixed, 
to predominantly chaotic as $|\gamma|$ is increased from zero.

The fixed points of the classical map (\ref{eqn:map}) provide 
useful information about the parameter dependence of the classical 
behaviour and, more importantly, in the case of mixed regimes, help   
locate the zones of regular behaviour in the 4-dimensional phase space. 
We find it sufficient to consider only 
the four trivial (parameter-independent) fixed points 
which lie at the poles along the $z$-axis: 
two of these points correspond to parallel spins,  
$(S_z,L_z) = \pm (|{\bf S}|,|{\bf L}|)$,  
and the remaining two points correspond to anti-parallel spins, 
$(S_z,L_z) = (\pm |{\bf S}|, \mp|{\bf L}|)$. 

The stability around these fixed points can be determined from the 
eigenvalues of the tangent map matrix,
${\bf M} = \partial {\bf F} / \partial {\bf x}$, 
where all derivatives are evaluated at the fixed point of interest. 
(It is easiest to derive $M$  
using the six {\it non-canonical} mapping equations (\ref{eqn:map})
since the tangent map for the {\it canonical} mapping equations 
exhibits a coordinate system singularity at these fixed points.) 
The eigenvalues corresponding to  
the four trivial fixed points are obtained from  
the characteristic equation, 
\begin{equation}
	[ \xi^2 - 2 \xi \cos a  + 1 ]^2 
	\pm \xi^2 \gamma^2 r \sin^2 a = 0, 
\end{equation}
with the minus (plus) sign corresponding to the parallel 
(anti-parallel) cases and I have suppressed the trivial 
factor $(1 -\xi)^2$ which arises since the six equations (\ref{eqn:map})  
are not independent.  
For the parallel fixed points the four eigenvalues are  
\begin{eqnarray}
\label{eqn:parallel} 
	\xi_{1,2}^P  & = & \cos a  \pm {1 \over 2} \sqrt{r \gamma^2 \sin^2 a} 
	+ {1 \over 2} 
	\sqrt{ \pm 4 \cos a  \sqrt{ \gamma^2 r \sin^2 a}  
	- (\sin^2 a) (4 - \gamma^2 r) }, \nonumber \\ 
	\xi_{3,4}^P  & = & \cos a  \pm {1 \over 2} \sqrt{r \gamma^2 \sin^2 a} 
	- {1 \over 2} 
	\sqrt{ \pm 4 \cos a  \sqrt{ \gamma^2 r \sin^2 a}  
	- (\sin^2 a) (4 - \gamma^2 r) },  
\end{eqnarray}
and the eigenvalues for the anti-parallel cases, $\xi^{AP}$, are obtained 
from (\ref{eqn:parallel}) 
through the substitution $r \rightarrow -r $. 
A fixed point becomes unstable if and only if $|\xi| > 1$ for 
at least one of the four eigenvalues. 

\subsubsection{Mixed Phase Space: $\gamma = 1.215$}

\begin{figure}[!t]
\begingroup%
  \makeatletter%
  \newcommand{\GNUPLOTspecial}{%
    \@sanitize\catcode`\%=14\relax\special}%
  \setlength{\unitlength}{0.1bp}%
\begin{picture}(3600,2160)(0,0)%
\special{psfile=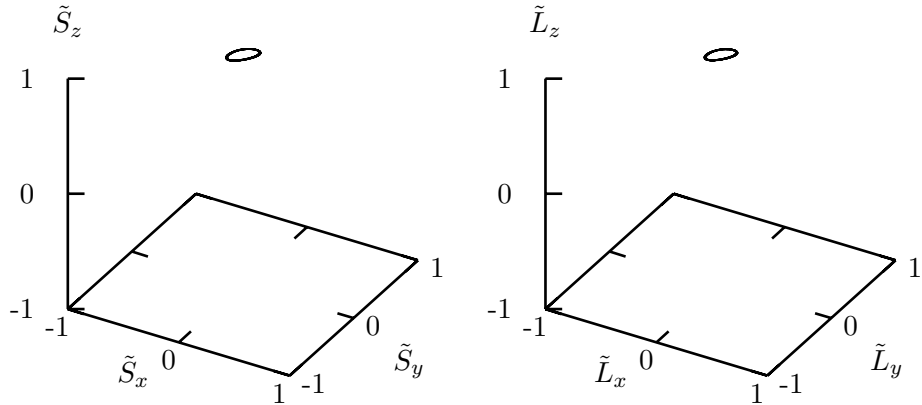 llx=0 lly=0 urx=720 ury=504 rwi=7200}
\put(2508,1906){\makebox(0,0){$\tilde{L}_z$}}%
\put(2382,1688){\makebox(0,0)[r]{1}}%
\put(2382,1254){\makebox(0,0)[r]{0}}%
\put(2382,819){\makebox(0,0)[r]{-1}}%
\put(3794,623){\makebox(0,0){$\tilde{L}_y$}}%
\put(3874,974){\makebox(0,0)[l]{1}}%
\put(3632,756){\makebox(0,0)[l]{0}}%
\put(3391,539){\makebox(0,0)[l]{-1}}%
\put(2756,585){\makebox(0,0){$\tilde{L}_x$}}%
\put(3306,501){\makebox(0,0){1}}%
\put(2888,626){\makebox(0,0){0}}%
\put(2470,752){\makebox(0,0){-1}}%
\put(708,1906){\makebox(0,0){$\tilde{S}_z$}}%
\put(582,1688){\makebox(0,0)[r]{1}}%
\put(582,1254){\makebox(0,0)[r]{0}}%
\put(582,819){\makebox(0,0)[r]{-1}}%
\put(1994,623){\makebox(0,0){$\tilde{S}_y$}}%
\put(2074,974){\makebox(0,0)[l]{1}}%
\put(1832,756){\makebox(0,0)[l]{0}}%
\put(1591,539){\makebox(0,0)[l]{-1}}%
\put(956,585){\makebox(0,0){$\tilde{S}_x$}}%
\put(1506,501){\makebox(0,0){1}}%
\put(1088,626){\makebox(0,0){0}}%
\put(670,752){\makebox(0,0){-1}}%
\end{picture}%
\endgroup
 
\vspace{-0.3in}
\caption{Stroboscopic trajectories on the unit sphere launched from 
a regular zone of the mixed regime with   $\gamma=1.215$, 
$r=1.1$, $a=5$ and $\vec{\theta}(0) = (5^o,5^o,5^o,5^o)$.
}
\label{r1.1.ric2}
\end{figure}

\begin{figure}[!t]
\begingroup%
  \makeatletter%
  \newcommand{\GNUPLOTspecial}{%
    \@sanitize\catcode`\%=14\relax\special}%
  \setlength{\unitlength}{0.1bp}%
\begin{picture}(3600,2160)(0,0)%
\special{psfile=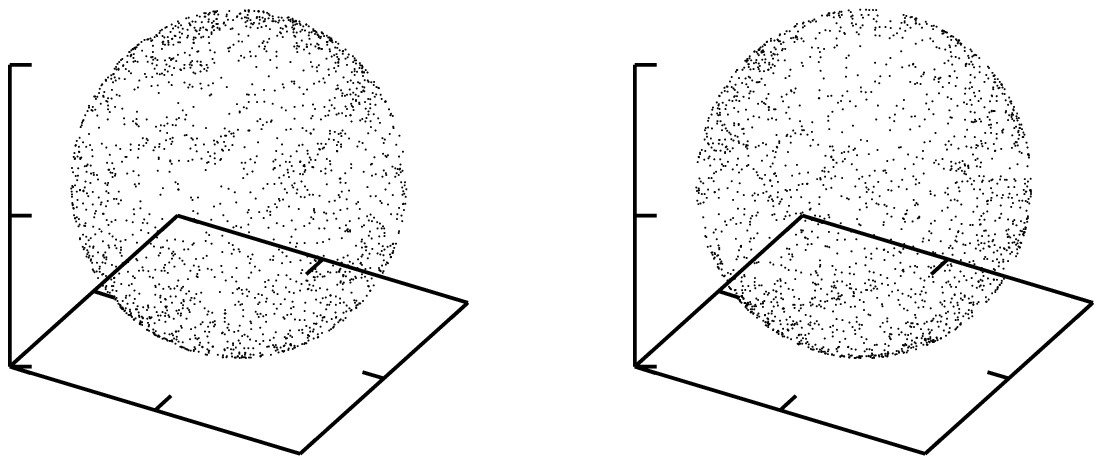 llx=0 lly=0 urx=720 ury=504 rwi=7200}
\put(2508,1906){\makebox(0,0){$\tilde{L}_z$}}%
\put(2382,1688){\makebox(0,0)[r]{1}}%
\put(2382,1254){\makebox(0,0)[r]{0}}%
\put(2382,819){\makebox(0,0)[r]{-1}}%
\put(3794,623){\makebox(0,0){$\tilde{L}_y$}}%
\put(3874,974){\makebox(0,0)[l]{1}}%
\put(3632,756){\makebox(0,0)[l]{0}}%
\put(3391,539){\makebox(0,0)[l]{-1}}%
\put(2756,585){\makebox(0,0){$\tilde{L}_x$}}%
\put(3306,501){\makebox(0,0){1}}%
\put(2888,626){\makebox(0,0){0}}%
\put(2470,752){\makebox(0,0){-1}}%
\put(708,1906){\makebox(0,0){$\tilde{S}_z$}}%
\put(582,1688){\makebox(0,0)[r]{1}}%
\put(582,1254){\makebox(0,0)[r]{0}}%
\put(582,819){\makebox(0,0)[r]{-1}}%
\put(1994,623){\makebox(0,0){$\tilde{S}_y$}}%
\put(2074,974){\makebox(0,0)[l]{1}}%
\put(1832,756){\makebox(0,0)[l]{0}}%
\put(1591,539){\makebox(0,0)[l]{-1}}%
\put(956,585){\makebox(0,0){$\tilde{S}_x$}}%
\put(1506,501){\makebox(0,0){1}}%
\put(1088,626){\makebox(0,0){0}}%
\put(670,752){\makebox(0,0){-1}}%
\end{picture}%
\endgroup
 
\vspace{-0.3in}
\caption{
Same parameters as Fig.\ \ref{r1.1.ric2} 
but the trajectory is launched from a chaotic zone of the mixed 
regime with 
initial condition $\vec{\theta}(0) = (20^o,40^o,160^o,130^o)$.
}
\label{r1.1.cic2}
\end{figure}

We are particularly interested in the behaviour of this model 
when the two spins are comparable in magnitude.
Choosing the value $r=1.1$ (with $a=5$ as before),  
we determined by 
numerical evaluation that the anti-parallel 
fixed points are unstable for $|\gamma| > 0$.  
In the case of the parallel fixed points, 
all four eigenvalues remain on the unit circle, $|\xi^{P}| =  1$, 
for $|\gamma| < 1.42$. This stability condition guarantees the presence 
of regular islands about the parallel fixed points \cite{LL}.  
In Fig.\ \ref{r1.1.ric2} we plot  
the trajectory corresponding to the parameters $a=5$, $r=1.1$, 
$\gamma=1.215$ and with initial 
condition $\vec{\theta}(0) =(5^o,5^o,5^o,5^o)$
which locates the trajectory  
near a stable fixed point 
of a mixed phase space (see Fig.\ \ref{regimes}.)
This trajectory clearly exhibits a periodic pattern 
which we have confirmed to be regular by computing 
the associated Lyapunov exponent ($\lambda_L=0$). In contrast, 
the trajectory plotted in Fig.\ \ref{r1.1.cic2} is launched 
with the same parameters but with initial condition 
$\vec{\theta}(0) = (20^o,40^o,160^o,130^o)$, which is close to one  
of the unstable anti-parallel fixed points.  
This trajectory explores a much larger portion of the 
surface of the two spheres in a seemingly random manner. 
As expected, a computation of the largest 
associated Lyapunov exponent yields a positive number ($\lambda_L = 0.04$). 

\subsubsection{Global Chaos: $\gamma = 2.835$}

If we increase the coupling strength to the value $\gamma =2.835 $, 
with $a=5$ and $r=1.1$ as before, then 
all four trivial fixed points become unstable. 
By randomly sampling ${\mathcal P}$ with $3 \times 10^4$ 
initial conditions we find that less than $0.1$\% of 
the kinematically accessible surface ${\mathcal P}$ is covered with 
regular islands (see Fig.\ \ref{regimes}). 
This set of parameters produces 
a connected chaotic zone with 
largest Lyapunov exponent $\lambda_L=0.45$. 
We will refer to this type of regime as one of ``global chaos'' 
although the reader should note that our usage of this expression 
differs slightly from that in Ref.\ \cite{LL}. 

\subsubsection{The Limit $r \gg 1$}

Another interesting limit of our model arises when one of the spins 
is much larger than the other, $r \gg 1$. We expect that 
in this limit the larger spin (${\bf L}$) will act 
as a source of essentially external ``driving'' for the 
smaller spin (${\bf S}$). 
Referring to the coupling terms in the 
mapping (\ref{eqn:map}), 
the ``driving'' strength, or perturbation 
upon ${\bf S}$ from ${\bf L}$, is 
determined from the product $\gamma r = c |{\bf L}|$, 
which can be quite large, whereas 
the ``back-reaction'' strength, or perturbation upon ${\bf L}$ from ${\bf S}$, 
is governed only by the 
scaled coupling strength $\gamma = c|{\bf S}|$, which can be quite small. 
It is interesting to examine whether a dynamical regime exists 
where the larger system might approach regular behaviour while 
the smaller `driven' system is still subject to chaotic motion. 

\begin{figure}[!t]
\begingroup%
  \makeatletter%
  \newcommand{\GNUPLOTspecial}{%
    \@sanitize\catcode`\%=14\relax\special}%
  \setlength{\unitlength}{0.1bp}%
\begin{picture}(3600,2160)(0,0)%
\special{psfile=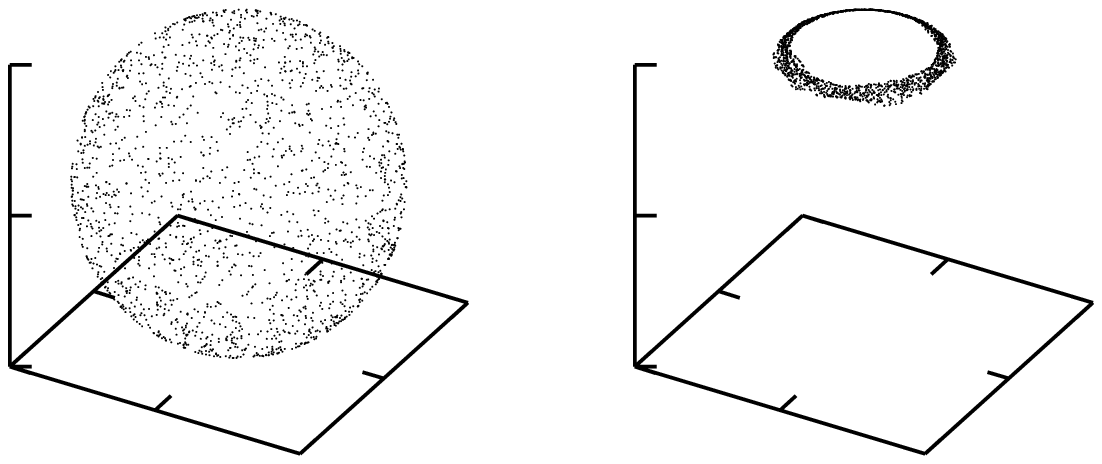 llx=0 lly=0 urx=720 ury=504 rwi=7200}
\put(2508,1906){\makebox(0,0){$\tilde{L}_z$}}%
\put(2382,1688){\makebox(0,0)[r]{1}}%
\put(2382,1254){\makebox(0,0)[r]{0}}%
\put(2382,819){\makebox(0,0)[r]{-1}}%
\put(3794,623){\makebox(0,0){$\tilde{L}_y$}}%
\put(3874,974){\makebox(0,0)[l]{1}}%
\put(3632,756){\makebox(0,0)[l]{0}}%
\put(3391,539){\makebox(0,0)[l]{-1}}%
\put(2756,585){\makebox(0,0){$\tilde{L}_x$}}%
\put(3306,501){\makebox(0,0){1}}%
\put(2888,626){\makebox(0,0){0}}%
\put(2470,752){\makebox(0,0){-1}}%
\put(708,1906){\makebox(0,0){$\tilde{S}_z$}}%
\put(582,1688){\makebox(0,0)[r]{1}}%
\put(582,1254){\makebox(0,0)[r]{0}}%
\put(582,819){\makebox(0,0)[r]{-1}}%
\put(1994,623){\makebox(0,0){$\tilde{S}_y$}}%
\put(2074,974){\makebox(0,0)[l]{1}}%
\put(1832,756){\makebox(0,0)[l]{0}}%
\put(1591,539){\makebox(0,0)[l]{-1}}%
\put(956,585){\makebox(0,0){$\tilde{S}_x$}}%
\put(1506,501){\makebox(0,0){1}}%
\put(1088,626){\makebox(0,0){0}}%
\put(670,752){\makebox(0,0){-1}}%
\end{picture}%
\endgroup
 
\vspace{-0.3in}
\caption{A chaotic trajectory for mixed regime parameters  
 $\gamma=0.06$, $r=100$, and $a=5$ with  
$\vec{\theta}(0) = (27^o,27^o,27^o,27^o)$.
The motion of the larger spin appears to remain 
confined to a narrow band on the surface of the 
sphere. 
}
\label{r100.cic3}
\end{figure}

In Fig.\ \ref{r100.cic3} we plot a 
chaotic trajectory for $r=100$ with initial 
condition $\vec{\theta}(0) = (27^o,27^o,27^o,27^o)$
which is located in a chaotic zone ($\lambda_L= 0.026$) 
of a mixed phase space (with $a=5$ and $\gamma=0.06$). 
Although the small spin wanders 
chaotically over a large portion of its kinematically 
accessible shell ${\mathcal S}^2$, the motion of the large spin remains 
confined to a `narrow' band.  Although the 
band is narrow relative to the 
large spin's length, it is not small relative to the smaller spin's 
length. The trajectories are both plotted on 
the unit sphere, so the effective area explored by the 
large spin (relative to the effective area covered 
by the small spin) scales in proportion to $r^2$.

\subsection{The Liouville Dynamics}

We are interested in comparing the 
quantum dynamics generated by (\ref{eqn:qmmap}) with 
the corresponding Liouville dynamics of a classical distribution. 
The time-evolution of a classical phase space distribution 
is generated by the partial differential equation,    
\begin{equation}
\label{eqn:liouville}
{ \partial \rho_c({\bf x},t) \over \partial t } = - \{ \rho_c , H \},   
\end{equation}
where $H$ stands for the Hamiltonian (\ref{eqn:ham})  
and ${\bf x} = (S_z, \phi_s,L_z,\phi_l)$. 

The solution to (\ref{eqn:liouville}) can be expressed in 
the compact form, 
\begin{equation}
\label{eqn:soln}
	\rho_c({\bf x},t) =   
\int_{\mathcal P} {\mathrm d} \mu({\bf y}) \;
\delta({\bf x} - {\bf x}(t,{\bf y})) \; 
\rho_c({\bf y},0), 
\end{equation}
with measure $d \mu({\bf y})$ given by (\ref{eqn:measure}).  
Each time-dependent function ${\bf x}(t,{\bf y}) \in {\mathcal P}$ is  
a solution of the equations of motion for (\ref{eqn:ham}) 
with initial condition $ {\bf y} \in {\mathcal P}$. 
The solution (\ref{eqn:soln}) simply expresses that 
Liouville's equation (\ref{eqn:liouville}) describes the dynamics 
of a classical density $\rho_c({\bf x},t)$ of points  
evolving in phase space under the Hamiltonian flow. 
We exploit this fact to numerically solve 
(\ref{eqn:liouville}) by 
randomly generating initial conditions  
consistent with an initial phase space 
distribution $\rho_c({\bf x},0)$ and 
then time-evolving each of these initial conditions 
using the equations of motion (\ref{eqn:map}). 
We then calculate the ensemble averages 
of dynamical variables,  
\begin{equation}
        \langle \tilde{L}_z(n) \rangle_c = 
\int_{\mathcal P} {\mathrm d} \mu({\bf x}) { L_z \over |{\bf L}| } \rho_c({\bf x},n).
\label{eqn:Lave}
\end{equation}
by summing over this distribution of trajectories at each time step.

\subsection{Correspondence Between Quantum and Classical Models}

For a quantum system specified by the four numbers $\{a,c,s,l\}$,  
the corresponding classical parameters $\{a,\gamma,r\}$ 
may be determined by first defining the classical magnitudes 
in terms of the quantum magnitudes, 
\begin{eqnarray}
		|{\bf S}|_c & = & \sqrt{s(s+1)} \nonumber \\
		| {\bf L}|_c & = & \sqrt{l(l+1)},
\end{eqnarray}
where the quantities on the left hand side are the lengths of the 
classical spins and those on the right are the quantum 
numbers.  If we set the Hamiltonian coefficients $a$ and $c$ 
numerically equal for both models, then the remaining two 
dimensionless classical parameters are determined, 
\begin{eqnarray}
		r & = & \sqrt{l(l+1) \over s(s+1)} \nonumber \\  
		\gamma & = & c \sqrt{s(s+1)}.
\end{eqnarray}

We are interested in extrapolating the behaviour of 
the quantum dynamics in the limit $s \rightarrow \infty$ 
and $l \rightarrow \infty$. This is accomplished by 
studying sequences of quantum models with increasing $s$ and $l$ 
chosen such that $r$ and $\gamma$ are held fixed. 
Since $s$ and $l$ 
are restricted to integer (or half-integer) values, the corresponding 
classical $r$ will actually vary slightly for each member of this sequence, 
although $\gamma$ can be matched exactly by varying the quantum parameter 
$c$.  
In the limit $s \rightarrow \infty$ and $l \rightarrow \infty$ 
this variation becomes increasingly small since 
 $r= \sqrt{l(l+1)/s(s+1)} \rightarrow l/s$. 
For convenience, the classical $r$ corresponding to 
each member of the 
sequence of quantum models is identified by its value in this limit. 
We have examined the effect of the small variations in the value of $r$ 
on the classical behaviour and found the variation to be negligible. 

\section{Initial States}

\subsection{Initial Quantum State}

We consider {\em initial} quantum states   
which are pure and separable, 
\begin{equation}
        | \psi(0) \rangle = | \psi_s(0) \rangle \otimes |\psi_l(0) \rangle. 
\end{equation}
For the initial state of each subsystem we use one of the directed 
angular momentum states,  
\begin{equation} 
\label{eqn:cs}
| \theta,\phi \rangle 
	= R^{(j)}(\theta,\phi) | j,j \rangle, 
\end{equation} 
which correspond to states of maximum polarization in the 
direction $(\theta,\phi)$. It has the properties:
\begin{eqnarray}
        \langle \theta, \phi | J_z | \theta, \phi \rangle & = & j \cos
        \theta \nonumber \\
        \langle \theta, 
\phi |J_x \pm i J_y | \theta, \phi \rangle & = &
        j e^{\pm i\phi} \sin \theta,
\end{eqnarray}
where $j$ in this section refers to either $l$ or $s$.

The  states (\ref{eqn:cs}) are the SU(2) coherent states, 
which, like their counterparts in the Euclidean 
phase space, are minimum uncertainty states \cite{cs}; 
the normalized variance of the quadratic operator,  
\begin{equation}    
\Delta {\bf \tilde J}^2  = { 
 \langle \theta, \phi | {\bf J}^2 | \theta, \phi \rangle 
- \langle \theta, \phi | {\bf J} | \theta, \phi \rangle^2 
\over  j(j+1) }  = {1 \over  (j+1)},
\end{equation}
is minimised for given $j$ and vanishes 
in the limit $j \rightarrow \infty$. 
The coherent states polarized along the $z$-axis, 
$| j,j\rangle $  and $| j,-j\rangle $,  
also saturate the inequality of the uncertainty relation,  
\begin{equation}
	\langle J_x^2 \rangle 
	\langle J_y^2 \rangle \ge { \langle J_z \rangle^2 \over 4 }, 
\end{equation}
although this inequality is not saturated 
for coherent states polarized along other axes.

\subsection{Initial Classical State 
and Correspondence in the Macroscopic Limit}

We compare the quantum dynamics 
with that of a classical Liouville density which is chosen to 
match the initial probability distributions of the 
quantum coherent state.  
For quantum systems with a Euclidean phase space it is always possible 
to construct a classical density with 
marginal probability distributions that match exactly 
the corresponding moments of the quantum coherent state.  
This follows from the fact that 
the marginal distributions for a coherent state are positive 
definite Gaussians, and therefore all of the moments can be matched 
{\it exactly} by choosing a Gaussian classical density. 
For the SU(2) coherent state, however, 
we show in Appendix A that no classical density has 
marginal distributions that can reproduce 
even the low order moments of the 
quantum probability distributions (except in the limit of infinite $j$). 
Thus from the outset it is clear that 
any choice of an initial classical state will exhibit residual 
discrepancy in matching some of the initial quantum moments. 

We have examined the initial state and dynamical 
quantum-classical correspondence using several different 
classical distributions. These included the vector model 
distribution described in Appendix A and the Gaussian distribution 
used by Fox and Elston in correspondence studies of the 
kicked top \cite{Fox94b}. For a state polarized along the $z$-axis we 
chose the density,  
\begin{eqnarray}
\label{eqn:rho}
        \rho_c (\theta,\phi) \;  \sin \theta d \theta d \phi & = & 
\; C
\exp \left[ - { 2 \sin^2({\theta\over 2}) ) 
\over \sigma^2}\right] \sin \theta d \theta d\phi \\
& = & \; C \exp \left[ - {(1 -\tilde{J}_z)\over 
 \sigma^2 } \right] \; d \tilde{J}_z  d\phi, \nonumber 
\end{eqnarray}
with $ C = \left[ 2 \pi \sigma^2 \left( 1 - \exp( -2 \sigma^{-2})
 \right) \right]^{-1} $, 
instead of those previously considered,  
because it is periodic under $2\pi$ rotation. 
An initial state directed along  
$(\theta_o,\phi_o)$ is then produced by a rigid body 
rotation of (\ref{eqn:rho}) 
by an angle $\theta_o$ about the $y$-axis followed 
by rotation with angle $\phi_o$ about the $z$-axis. 

The variance $\sigma^2$ and the magnitude $|{\bf J}|_c$ 
are free parameters of the 
classical distribution that
should be chosen to fit the quantum probabilities as well as possible.  
It is shown in Appendix A that no classical density has marginal 
distributions which can match all of the quantum moments, 
so we concentrate only on matching the lowest order moments.   
Since the magnitude of the spin is a kinematic constant  
both classically and quantum mechanically, we choose 
the squared length of the classical spin to have the correct 
quantum value, 
\begin{equation}
\label{eqn:mag}
|{\bf J}|_c^2  =  \langle J_x^2 \rangle_c + 
\langle J_y^2 \rangle_c + \langle J_z^2 \rangle_c = j(j+1). 
\end{equation}

For a state polarized along the $z$-axis, we have 
$\langle J_x \rangle  = \langle J_y \rangle  = 0 $ 
and $\langle J_y^2 \rangle = \langle J_x^2 \rangle $ 
for both distributions as a consequence of the axial symmetry.  
Furthermore, as a consequence of (\ref{eqn:mag}), we will 
automatically satisfy the condition, 
\begin{equation}
	2 \langle J_x^2 \rangle_c + \langle J_z^2 \rangle_c =  j(j+1).
\end{equation}
Therefore we only need to consider the classical moments,  
\begin{eqnarray}
\label{eqn:cm}
	\langle J_z \rangle_c =  |{\bf J}| \; G(\sigma^2)  \\
	\langle J_x^2 \rangle_c 
=  |{\bf J}|^2 \sigma^2 \; G(\sigma^2), 
\end{eqnarray}
calculated from the density (\ref{eqn:rho}) 
in terms of the remaining free parameter, $\sigma^2$, where, 
\begin{equation}
	G(\sigma^2) = \left[ {1 + \exp(-2 \sigma^{-2}) 
	\over 1 - \exp (-2 \sigma^{-2}) }\right] - \sigma^2.
\end{equation}
We would like to match both of these classical moments with 
the corresponding quantum values,  
\begin{eqnarray}
\label{eqn:qmz}
	\langle J_z \rangle =  j, \\
	\langle J_x^2 \rangle =  j/2, 
\label{eqn:qmx2}
\end{eqnarray}
calculated for the coherent state (\ref{eqn:cs}). 
However, no choice of $\sigma^2$ will satisfy both 
constraints. 

If we choose $\sigma^2$ to satisfy (\ref{eqn:qmz}) exactly 
then we would obtain, 
\begin{equation}
\sigma^2 = \frac{1}{2j} - \frac{3}{8j^2}+{\mathcal O}(j^{-3}).
\end{equation}
If we choose $\sigma^2$ to satisfy (\ref{eqn:qmx2}) exactly 
then we would obtain, 
\begin{equation}
\sigma^2 = \frac{1}{2j} + \frac{1}{4j^2}+{\mathcal O}(j^{-3}).
\end{equation}
(These expansions are most easily derived from the approximation 
$G(\sigma^2) \simeq 1 - \sigma^2$, which has an exponentially 
small error for large $j$.)

We have chosen to compromise between these values by fixing 
$\sigma^2$ so that the ratio 
$\langle J_z \rangle_c / \langle J_x^2 \rangle_c $ 
has the correct quantum value. 
This leads to the choice, 
\begin{equation}
\label{eqn:sigma}
\sigma^2 = \frac{1}{2 \sqrt{j(j+1)}} =
\frac{1}{2j} - \frac{1}{4j^2}+{\mathcal O}(j^{-3}).
\end{equation}

These unavoidable initial differences between the classical 
and quantum moments will vanish in the ``classical'' 
limit. To see this explicitly it is convenient to 
introduce a measure of the quantum-classical differences, 
\begin{equation} 
	\delta J_z(n) = | \langle J_z(n) \rangle - \langle J_z(n) \rangle_c|, 
\end{equation}
defined at time $n$. 
For an initial state polarised in direction $(\theta,\phi)$, 
the choice (\ref{eqn:sigma}) produces the initial difference, 
\begin{equation} 
\label{eqn:initdiff}
	\delta J_z(0) 
= {  \cos (\theta) \over 8j } + {\mathcal O}(j^{-2}), 
\end{equation}
which vanishes as $j \rightarrow \infty$.

\section{Numerical Methods}
\label{sect:nummethods}


The time-periodic spin 
Hamiltonian (\ref{eqn:ham}) is convenient for numerical study 
because the time-dependence reduces to a simple mapping and  
the quantum state vector is confined to a finite dimensional 
Hilbert space. Consequently we can solve the exact time-evolution 
equations (\ref{eqn:qmmap}) numerically without introducing any 
artificial truncation of the Hilbert space. 
The principal source of numerical inaccuracy arises 
from the numerical evaluation of the matrix elements
of the rotation operator 
$ \langle j, m' | R(\theta,\phi) | j, m \rangle = \exp( - i \phi m' ) 
d_{m'm}^{(j)} (\theta)$.  The rotation operator is required 
both for the 
calculation of the initial quantum coherent state,  
$ |\theta, \phi \rangle = R(\theta,\phi) | j,m=j \rangle $, 
and for the evaluation of the unitary Floquet operator. 
In order to maximise the precision of our results  
we calculated the matrix elements 
$d_{m'm}^{(j)}(\theta) = \langle j, m' | \exp(-i \theta J_y) | j,m \rangle$  
using the recursion algorithm of Ref.\ \cite{Haake96} and 
then tested the accuracy of our results 
by introducing controlled numerical errors. 
For small quantum numbers ($j<50$) 
we are able to confirm the correctness of our coded algorithm 
by comparing these results with those 
obtained by direct evaluation of 
Wigner's formula for the matrix elements $d_{m'm}^{(j)}(\theta)$.

The time evolution of the Liouville density was simulated 
by numerically evaluating between $10^8$ and $10^9$ classical trajectories 
with randomly selected initial conditions weighted according to 
the initial distribution (\ref{eqn:rho}). 
Such a large number of trajectories was required in order 
to keep Monte Carlo errors small enough to  
resolve the initial normalized quantum-classical differences, which 
scale as $1/8j^2$, over the range of $j$ values we have examined.  

We identified initial conditions of the classical map as chaotic 
by numerically calculating the largest Lyapunov 
exponent, $\lambda_L$, using the formula, 
\begin{equation}
\label{eqn:liap}
\lambda_L =  { 1 \over N } \sum_{n=1}^N \ln d(n) 
\end{equation}
where $d(n) = \sum_i | \delta x_i(n) | $ , with $d(0) = 1$. 
The differential ${\bf \delta x}(n)$ is a difference 
vector between adjacent trajectories and thus evolves under the 
action of the tangent map,
\begin{equation} 
{\bf \delta x}(n+1) = {\bf M} \cdot {\bf \delta x}(n), 
\end{equation}
where ${\bf M}$ is evaluated along some fiducial trajectory \cite{LL}. 

Since we are interested in studying quantum states, and 
corresponding classical distributions which have non-vanishing support 
on the sphere, it is also important to get an idea of the size of 
these regular and chaotic zones. By comparing the size of a given 
regular or chaotic zone to the variance of an  
initial state located within it, we can determine 
whether most of the state is contained within this zone. 
However, we can not perform 
this comparison by direct visual 
inspection since the relevant phase space 
is 4-dimensional.  
One strategy which we used to overcome this difficulty 
was to calculate the Lyapunov exponent for a large number of  
randomly sampled initial conditions and then 
project only those points which are regular (or chaotic) 
onto the plane spanned by $\tilde{S}_z = \cos \theta_s$ 
and $\tilde{L}_z = \cos \theta_l$. 
If the variance of the initial quantum state is located within,  
and several times smaller 
than, the dimensions of a zone devoid of any of these points, 
then the state in question 
can be safely identified as chaotic (or regular). 

\chapter{Correspondence for Low Order Moments} 

This chapter is taken from Emerson and Ballentine \cite{EB00a}.

\section{Introduction}
\label{sec1}

There is considerable interest in the interface between quantum and 
classical mechanics and the conditions that lead to the emergence 
of classical behaviour.  In order to characterize these conditions, 
it is important to differentiate 
two dynamical regimes of 
quantum-classical correspondence \cite{Ball94}:   
\newline
\noindent
(i) Ehrenfest correspondence, in which 
the centroid of the wave packet
approximately follows a classical trajectory; and   
\newline
\noindent
(ii) Liouville correspondence, in which the quantum probability
distributions are in approximate agreement with those of an 
appropriately constructed classical ensemble satisfying 
Liouville's equation. 

Regime (i) is relevant only when 
the width of the quantum state is small 
compared to the dimensions of the system; 
if the initial state is not narrow, this regime may be absent.
Regime (ii), which generally includes (i), applies to a much broader 
class of states, and this regime of correspondence may persist well 
after the Ehrenfest correspondence has broken down. 
The distinction between regimes (i) and (ii) has not always been 
made clear in the literature, though  
the conditions that delimit these two regimes, and in particular 
their scaling with system 
parameters, may be quite different. 

The theoretical study of quantum chaos 
has raised the question of whether the 
quantum-classical break occurs differently in chaotic states, in
states of regular motion, and in mixed phase-space systems.  
This is well understood only in the case of regime (i).  
There it is well-known \cite{BZ78,Haake87,Chi88}
that the time for a minimum-uncertainty wave packet to 
expand beyond the Ehrenfest regime scales as $\log({\mathcal J} / \hbar )$ 
for chaotic states, and as a power of ${\mathcal J} / \hbar $  
for regular states, where ${\mathcal J}$ denotes 
a characteristic system action.  

The breakdown of quantum-classical correspondence, 
in the case of regime (ii),  
is less well understood, though it has been argued that this regime 
may also be delimited by a $\log({\mathcal J}/ \hbar)$ break-time 
in classically chaotic states \cite{ZP94,Zurek98a}. 
Some numerical evidence in support of this conjecture has been 
reported in a study 
of the kicked rotor in the {\em anomolous diffusion} regime 
\cite{RBWG95}.
(On the other hand, in the regime of {\em quantum localization}, 
the break-time for the kicked rotor 
seems to scale as $({\mathcal J}/ \hbar)^2$ \cite{Haake91}.)
Since the $\log({\mathcal J}/ \hbar)$ time scale is rather short, 
it has been suggested 
that certain macroscopic objects would be predicted to exhibit 
non-classical behaviour 
on observable time scales \cite{ZP95a,Zurek98b}. 
These results highlight the importance of investigating  
the characteristics of quantum-classical correspondence in more detail. 

In this paper we study the classical and quantum dynamics of two
interacting spins.  This model is convenient because the Hilbert space of
the quantum system is finite-dimensional, and hence tractable for
computations.  
Spin models have been useful in the past for
exploring classical and quantum chaos \cite{Haake87,FP83,B91a,B91b,B93,RR98} 
and our model belongs to a class of spin models which show promise of 
experimental realization in the near future \cite{Mil99}. 
The classical limit is approached by taking the magnitude of
both spins to be very large relative to $\hbar$, while keeping their ratio 
fixed.  For our model 
a characteristic system action is given by ${\mathcal J} 
\simeq \hbar l$, where $l$ is a quantum number, 
and the classical limit is simply the limit of large 
quantum numbers, {\it i.e.} the limit $l \rightarrow \infty$. 

In the case of the chaotic dynamics for our model, we first show that 
the widths of both the quantum and classical states 
grow exponentially at a rate 
given approximately by the largest Lyapunov exponent (until saturation 
at the system dimension).  
We then show that the initially small quantum-classical differences 
also grow at an exponential rate, with an exponent $\lambda_{qc}$   
that is independent of the quantum numbers and at least twice 
as large as the largest Lyapunov exponent.  
We demonstrate how this exponential growth of differences 
leads to a log break-time rule,  
$t_b \simeq \lambda_{qc}^{-1} \ln( l p / \hbar )$, delimiting the regime 
of Liouville correspondence. The factor $p$, measured in units of $\hbar$,  
is some preset tolerance that defines a {\em break} between 
the quantum and classical expectation values, but does not scale with 
the system dimension.  
However, we also show that this logarithmic 
rule holds {\em only if} the tolerance $p$ for quantum-classical 
differences is chosen extremely small, 
in particular $p < {\mathcal O}(\hbar)$. 
For larger values of the tolerance, the break-time does not 
occur on this log time-scale and may not occur until the recurrence time.  
In this sense, log break-time rules describing Liouville correspondence 
are not robust.
These results 
demonstrate that, for chaotic states in the classical limit, 
quantum observables 
are described approximately 
by Liouville ensemble averages well beyond the 
Ehrenfest time-scale, after which both quantum and classical 
states relax towards equilibrium configurations. 
This demonstration of correspondence  
is obtained for 
a few degree-of-freedom quantum system of coupled spins 
that is described by a 
pure state and subject only to unitary evolution.

\section{Characteristics of the Quantum and Liouville Dynamics}
\label{sec:qcc}



\subsection{Mixed Phase Space}

\begin{figure}[!t]
\begin{center}
\input{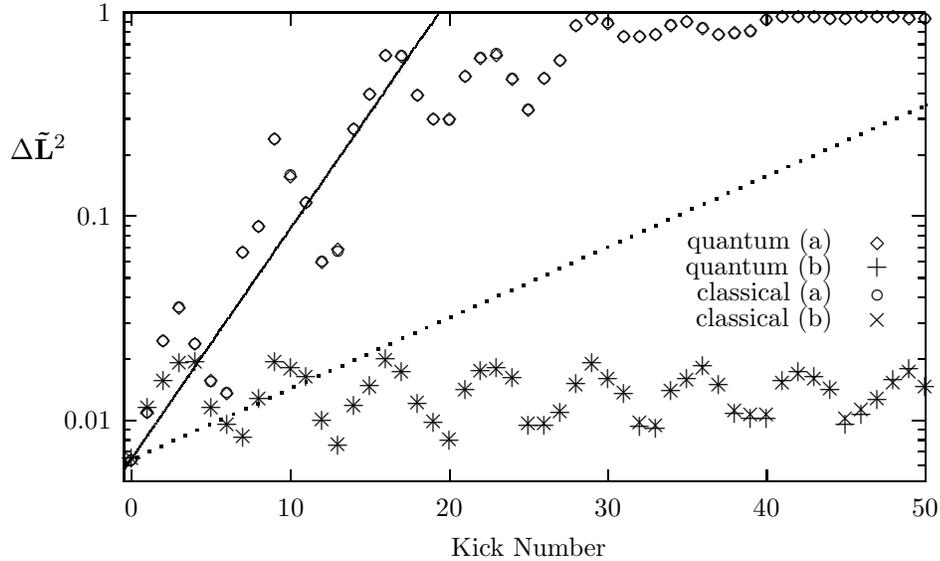}
\caption{
Growth of normalized quantum and classical variances  
in a chaotic zone (a) 
and a regular zone (b) of the mixed phase space regime 
$\gamma = 1.215$ and $r\simeq 1.1$ with $l=154$.  
Quantum and classical results are nearly indistinguishable on this scale.
In the chaotic case, the approximately exponential growth 
of both variances  
is governed by a much larger rate, $\lambda_w =0.13$ (solid line), 
than that predicted from 
the largest Lyapunov exponent, $\lambda_L=0.04$ (dotted line).
}
\label{vargrowth.1.215.ric2}
\end{center}
\end{figure}

We consider the 
time-development of initial quantum coherent states (\ref{eqn:cs})
evolved according to the mapping (\ref{eqn:qmmap})
using quantum numbers $s=140$ and $l=154$   
and associated classical  parameters 
$\gamma=1.215$, $r \simeq 1.1$, and $a=5$,  
which produce a mixed phase space (see Fig.\ \ref{regimes}). 
The classical results are generated by evolving the 
initial ensemble  (\ref{eqn:rho}) 
using the mapping (\ref{eqn:map}). 
In Fig.\ \ref{vargrowth.1.215.ric2} we compare  
the time-dependence of the normalized quantum variance, 
$\Delta {\tilde {\bf L}}^2  
= [\langle {\bf L}^2 \rangle - \langle {\bf L} \rangle^2 ] / l(l+1) $, 
with its classical counterpart,  
$\Delta {\tilde {\bf L}}^2_c 
= [\langle {\bf L}^2 \rangle_c - 
\langle {\bf L} \rangle_c^2 ] / |{\bf L}|^2 $. 
Circles (diamonds) correspond to the dynamics of 
an initial quantum (classical) 
state centered at $\vec{\theta}(0) = (20^o,40^o,160^o,130^o)$, 
which is located in the connected chaotic zone
near one of the unstable fixed points 
of the classical map. Crosses (plus signs) 
correspond to an initial quantum (classical) 
state centered on the initial condition 
$\vec{\theta}(0) =(5^o,5^o,5^o,5^o)$, which is located in 
the regular zone near one of the stable fixed points. 
For both initial conditions the quantum and classical 
results are nearly indistinguishable on the scale of the figure. 
In the case of the regular initial condition, the quantum 
variance remains narrow over long times and, 
like its classical counterpart, exhibits a regular oscillation. 
In the case of the chaotic initial condition 
the quantum variance also exhibits a periodic 
oscillation but this oscillation is superposed on a very rapid, approximately 
exponential, growth rate.  
This exponential growth persists until the variance approaches 
the system size, that is, when $\Delta {\tilde {\bf L}}^2  \simeq 1$ . 
The initial exponential growth of the quantum variance in classically chaotic 
regimes has been observed previously in several models and appears 
to be a generic feature of the quantum dynamics; this behaviour of 
the quantum variance is mimicked very accurately 
by the variance of an initially  well-matched 
classical distribution \cite{Ball98,Fox94b,Fox94a}. 

For well-localized states, in the classical case, the exponential growth 
of the distribution variance in chaotic zones is certainly related to 
the exponential divergence of the underlying trajectories, a property 
that characterizes classical chaos. To examine this connection  
we compare the observed exponential rate of growth of the widths of 
the classical (and quantum) state with the exponential rate predicted 
from the classical Lyapunov exponent. For the coherent states the 
initial variance can be calculated exactly, 
$\Delta {\tilde {\bf L}}^2(0) = 1/(l+1)$. 
Then, assuming exponential growth of this initial variance we get, 
\begin{equation}
\label{eqn:expvar} 
	\Delta {\tilde {\bf L}}^2(n)  
\simeq { 1 \over l  } \exp( 2 \lambda_w n)
\; \;\;\;\;\;\;\; {\mathrm f}{\mathrm o}{\mathrm r}  
\;\;\;  n < t_{sat},
\end{equation}
where a factor of $2$ is included in the exponent since  
$\Delta {\tilde {\bf L}}^2 $ corresponds to a 
squared length. The dotted line in Fig. \ref{vargrowth.1.215.ric2} 
corresponds to the prediction (\ref{eqn:expvar}) 
with $\lambda_w = \lambda_L = 0.04$, the value of the 
largest classical Lyapunov exponent.  As can be seen from the figure, 
the actual growth rate of the classical (and quantum) variance   
of the chaotic initial state is significantly larger 
than that predicted using the largest Lyapunov exponent.   
For comparison purposes we also plot a  
solid line in Fig.\ \ref{vargrowth.1.215.ric2} corresponding 
to (\ref{eqn:expvar}) using $\lambda_w = 0.13 $, which provides 
a much closer approximation to the actual growth rate. 
We find, for a variety of initial conditions in the chaotic zone of 
this mixed regime, that the actual 
classical (and quantum) variance growth rate is consistently larger 
than the simple prediction (\ref{eqn:expvar}) using $\lambda_L$ for the 
growth rate.  This systematic bias requires some explanation. 

As pointed out in Ref.\ \cite{Fox94b}, the presence of 
some discrepancy between $\lambda_w$ 
and $\lambda_L$ can be expected from the fact that the 
Lyapunov exponent is defined  
as a geometric mean of the tangent map eigenvalues 
sampled over the entire connected chaotic zone (corresponding 
to the infinite time limit $n\rightarrow \infty$) whereas 
the {\it actual} growth rate of a given 
distribution over a small number of time-steps will be 
determined largely by a few eigenvalues 
of the local tangent map. In mixed regimes these local eigenvalues 
will vary considerably over the phase space manifold and the 
product of a few of these eigenvalues 
can be quite different from the geometric 
mean over the entire connected zone.  


However, we find that the actual growth rate is consistently {\it larger}  
than the Lyapunov exponent prediction. 
It is well known that in mixed regimes the remnant KAM tori 
can be `sticky'; these sticky regions can have a significant decreasing  
effect on a calculation of the Lyapunov exponent. 
In order to identify an initial condition as 
chaotic, we specifically choose initial states 
that are concentrated away from these KAM surfaces (regular islands).  
Such initial states will then be exposed mainly to the larger 
local expansion rates found away from these surfaces. 
This explanation is supported by our observations that, when  
we choose initial conditions closer to these remnant tori, we 
find that the growth rate of the variance is significantly reduced. 
These variance growth rates are still slightly larger than the 
Lyapunov rate, but this is not surprising since 
our initial distributions are concentrated over a 
significant fraction of the phase space and 
the growth of the distribution  
is probably more sensitive to contributions 
from those trajectories subject to large eigenvalues away from the 
KAM boundary than those stuck near the boundary. 
These explanations are further 
supported by the results of the following section, where  
we examine a phase space regime that is nearly devoid of regular islands.  
In these regimes we find that the Lyapunov exponent serves as a much 
better approximation to the variance growth rate. 

\subsection{Regime of Global Chaos}

If we increase the dimensionless coupling strength to $\gamma=2.835$, 
with $a=5$ and $r \simeq 1.1$ as before, then the classical flow 
is predominantly chaotic on the surface ${\mathcal P}$
(see Fig.\ \ref{regimes}). 
Under these conditions 
we expect that generic initial classical distributions 
(with non-zero support) will spread to cover the full surface  
${\mathcal P}$ and then quickly relax close to microcanonical equilibrium. 
We find that the initially localised quantum states also exhibit  
these generic features when the quantum map is governed by 
parameters that produce these conditions classically.  

For the non-autonomous Hamiltonian system (\ref{eqn:map}) 
the total energy is not conserved, but 
the two invariants of motion, ${\bf L}^2$ and ${\bf S}^2$, confine 
the dynamics to the 4-dimensional 
manifold ${\mathcal P} = {\mathcal S}^2 \times {\mathcal S}^2 $, which is 
the surface of two spheres.  
The corresponding microcanonical distribution
is a constant on this surface, 
with measure (\ref{eqn:measure}),  
and zero elsewhere.  
From this distribution we can calculate 
microcanonical equilibrium values 
for low order moments, where, for example,     
$\{ L_z \} = ( 4 \pi)^{-2} \int_{\mathcal P} L_z d \mu 
= 0$ and $\{ \Delta {\bf L}^2 \} = \{ {\bf L}^2 \}-\{ {\bf L} \}^2 = |{\bf L}|^2$. 
The symbols $\{ \cdot \}$ denote a microcanonical average. 

\begin{figure}[!t]
\begin{center}
\input{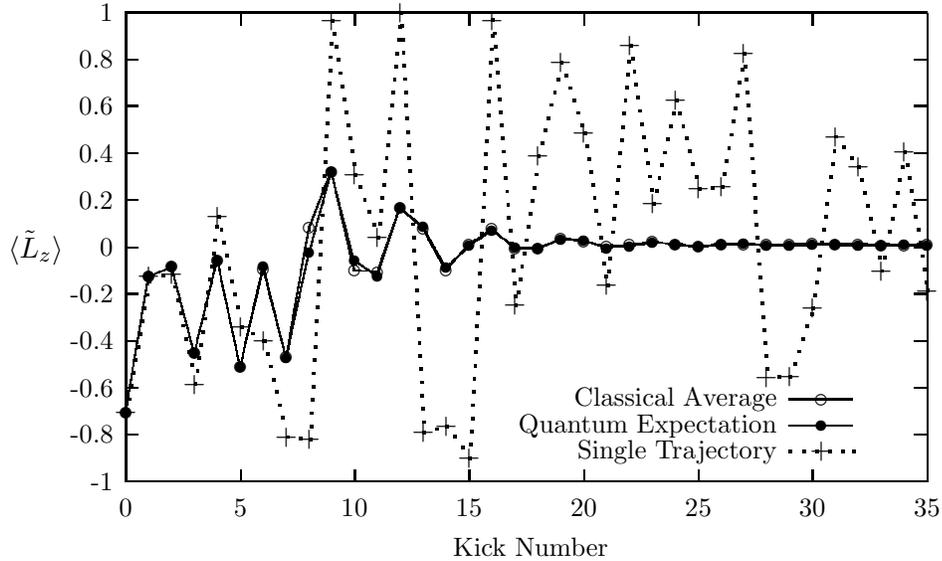}
\caption{Comparison of quantum expectation value and 
corresponding classical average $\langle L_z \rangle_c$
in the regime of global chaos $\gamma=2.835$ and $r\simeq 1.1$
with  $l=154$  
and initial condition $\vec{\theta}_o = (45^o,70^o,135^o,70^o) $. 
The points of the stroboscopic map 
are connected with lines to guide the eye.
The quantum expectation 
value and the Liouville average 
exhibit esentially the same rate of relaxation to 
microcanonical equilibrium, 
a behaviour which is qualitatively distinct from 
that of the single trajectory. 
}
\label{qmlmnm}
\end{center}
\end{figure}

To give a sense of the accuracy of the correspondence between 
the classical ensemble and the quantum dynamics 
in Fig.\ \ref{qmlmnm}, we show a direct comparison 
of the dynamics of the quantum expectation value $
 \langle \tilde{L}_z \rangle $ 
with $l=154$ 
and the classical distribution average $ \langle \tilde{L}_z \rangle_c $ 
for an initial coherent state and corresponding 
classical distribution centered at 
$\vec{\theta} = (45^o,70^o,135^o,70^o)$. 
To guide the eye in this figure we have drawn lines connecting 
the stroboscopic points of the mapping equations.   
The quantum expectation value exhibits essentially 
the same dynamics as the classical Liouville average, not only at 
early times, that is, in the 
initial Ehrenfest regime \cite{Ball94,HB94},   
but for times well into the equilibrium regime where the classical 
moment $ \langle L_z \rangle$ has relaxed close to the microcanonical 
equilibrium value $\{ L_z \} = 0 $. 
We have also provided results for a single trajectory launched 
from the same initial condition in order to emphasize the 
qualitatively distinct behaviour it exhibits.

\begin{figure}[!t]
\begin{center}
\input{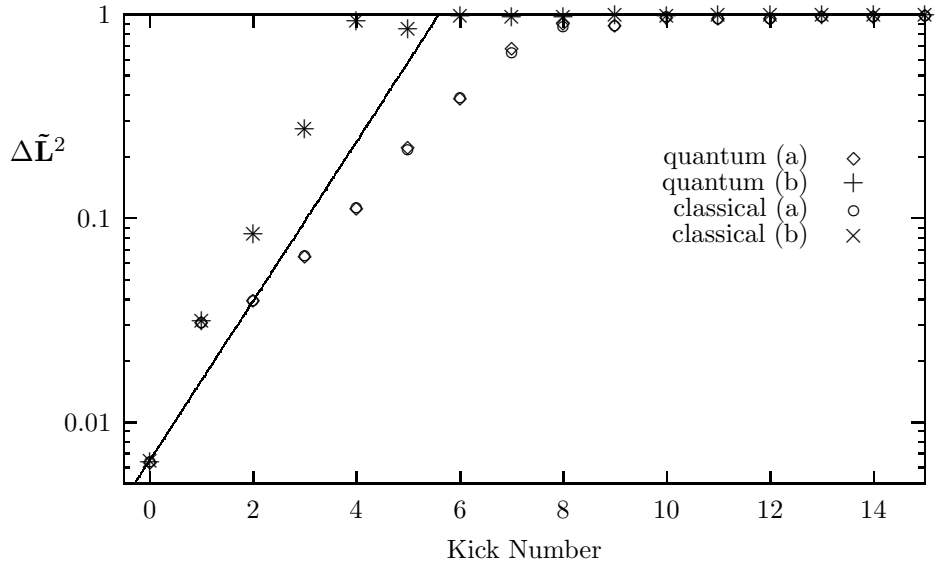}
\caption{
Growth of normalized quantum and classical variances 
in the regime of global chaos, $\gamma =2.835$ and $r \simeq 1.1$ with l=154,
for the two initial conditions cited in the text.  
Quantum-classical differences are nearly imperceptible on this scale. 
In this regime the largest Lyapunov exponent $\lambda_L=0.45$ 
provides a much better estimate of the initial variance 
growth rate. 
}
\label{vargrowth.2.835.140}
\end{center}
\end{figure}

In Fig.\ \ref{vargrowth.2.835.140} we show the 
exponential growth of the normalized quantum and 
classical variances on a semilog plot for 
the same set of parameters and quantum numbers.
Numerical data for (a) correspond to initial condition
$\vec{\theta}(0) = (20^o,40^o,160^o,130^o)$ 
and those for (b) correspond to $\vec{\theta}(0) = (45^o,70^o,135^o,70^o)$. 
As in the mixed regime case,  
the quantum-classical differences are nearly 
imperceptible on the scale of the figure,  
and the differences between the quantum and classical 
variance growth rates are many orders of magnitude 
smaller than the small differences in the growth rate arising 
from the different initial conditions. 

In contrast with the mixed regime case, 
in this regime of global chaos the prediction (\ref{eqn:expvar})  
with  $\lambda_w= \lambda_L=0.45$ 
now serves as a much better approximation 
to the exponential growth rate of the quantum variance, and associated 
relaxation rate of the quantum and classical states. 
In this regime the exponent $\lambda_w$ is also much larger than in 
the mixed regime case 
due to the stronger degree of classical chaos. As a result,   
the initially localised quantum and classical distributions 
saturate at system size much sooner. 

It is useful to apply (\ref{eqn:expvar}) 
to estimate the time-scale at which the quantum (and classical) 
distributions saturate at system size.   
From the condition 
$\Delta {\tilde {\bf L}}^2(t_{sat}) \simeq 1$ and 
using (\ref{eqn:expvar}) we obtain,  
\begin{equation}
\label{eqn:nsat}
	t_{sat} \simeq (2 \lambda_w)^{-1} \ln( l) 
\end{equation}
which serves as an estimate of this characteristic time-scale. 
In the  regimes for which the full surface ${\mathcal P}$ 
is predominately chaotic, 
we find that the actual 
exponential growth rate of the width of the quantum state, $\lambda_w$,   
is well approximated by the largest Lyapunov exponent $\lambda_L$.  
For $a=5$ and $r=1.1$, the approximation $\lambda_w \simeq \lambda_L$  
holds for coupling strengths $\gamma > 2$, for which 
more than 99\% of the surface ${\mathcal P}$ is covered by one connected 
chaotic zone (see Fig.\ \ref{regimes}).  

By comparing the quantum probability distribution to its classical 
counterpart, we can learn much more about the relaxation properties of the 
quantum dynamics. 
In order to compare each $m_l$ value of the quantum distribution, $P_z(m_l)$,  
with a corresponding piece of the continuous classical marginal 
probability distribution, 
\begin{equation}
P_c(L_z) = \int \! \! \int \! \! \int \! 
d \tilde{S}_z d \phi_s d \phi_l \; \rho_c(\theta_s,\phi_s, \theta_l, \phi_l), 
\end{equation}
we discretize the latter into $2j+1$ bins of width $\hbar=1$. 
This procedure produces a discrete classical 
probability distribution $P_z^c(m_l)$ 
that prescribes the probability 
of finding the spin component $L_z$ in the interval $[m_l+1/2,m_l-1/2]$ 
along the $z$-axis. 

\begin{figure}[!t]
\begin{center}
\input{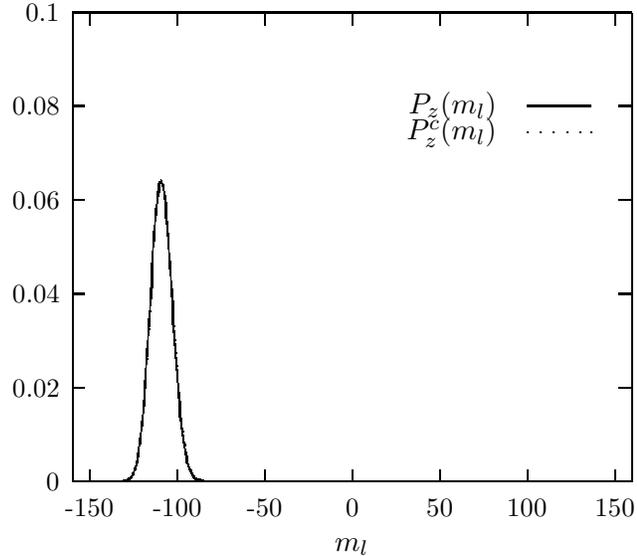}
\caption{
Initial probability distributions for $L_z$ 
for $\vec{\theta}(0) = (45^o,70^o,135^o,70^o) $ 
with $l=154$. 
The quantum and classical distributions are 
indistinguishable on the scale of the figure.
}
\label{probdist0}
\end{center}
\end{figure}

\begin{figure}[!t]
\begin{center}
\input{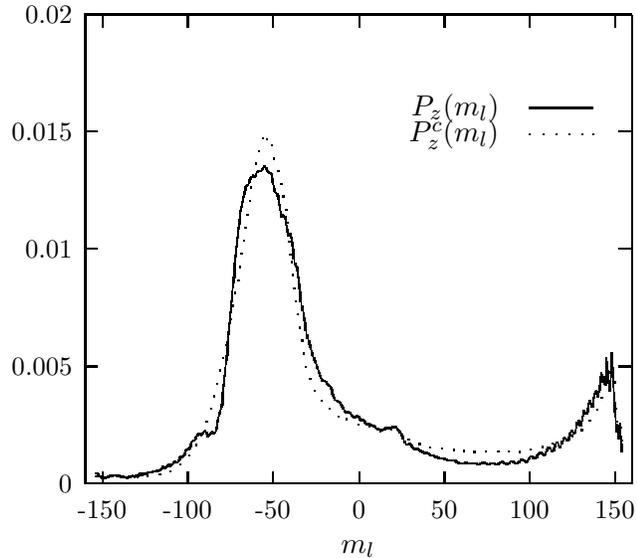}
\caption{
Same as Fig.\ \ref{probdist0} but the states have evolved to $n=6$ 
in the regime of global chaos $\gamma=2.835$ and $r\simeq 1.1$. 
Both the quantum and classical distribution have spread to system dimension 
and exhibit their largest differences on this saturation time-scale. 
}
\label{probdist6}
\end{center}
\end{figure}

To illustrate the time-development of these distributions 
we compare the quantum and classical probability distributions 
for three successive values of the kick number $n$,    
using the same quantum numbers and 
initial condition as in Fig.\ \ref{qmlmnm}. 
In Fig.\ \ref{probdist0} 
the initial quantum and classical states
are both well-localised and nearly 
indistinguishable on the scale of the figure. 
At time $n=6 \simeq t_{sat}$, shown in Fig.\ \ref{probdist6}, 
both distributions have grown to fill the accessible 
phase space. 
It is at this time 
that the most significant quantum-classical 
discrepancies appear. 

\begin{figure}[!t]
\begin{center}
\input{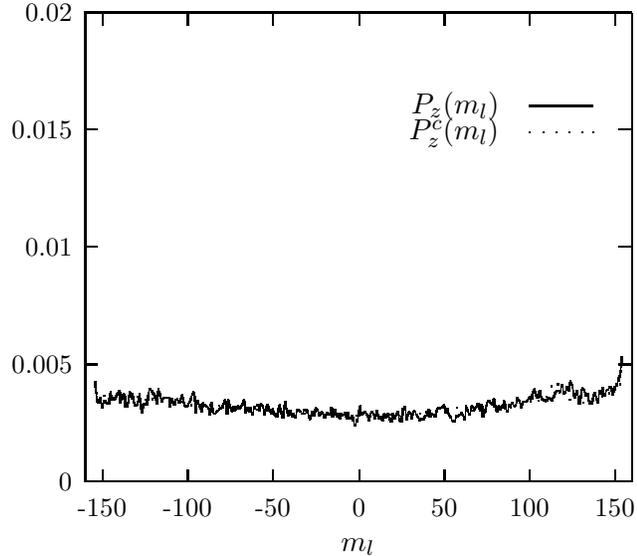}
\caption{
Same as Fig.\ \ref{probdist6}, but for $n=15$. Both quantum and classical 
distributions have relaxed close to the microcanonical equilibrium. 
}
\label{probdist15}
\end{center}
\end{figure}

For times greater than $t_{sat}$, however, these emergent 
quantum-classical discrepencies do not continue to grow,  
since both distributions begin relaxing towards equilibrium distributions. 
Since the dynamics are confined to a {\it compact} phase space,  
and in this parameter regime the remnant KAM tori 
fill a negligibly small fraction of the kinematicaly accessible phase space, 
we might expect the classical equilibrium distribution to be very 
close to the microcanonical distribution. 
Indeed such relaxation close to microcanonical equilibrium is 
apparent for both the quantum and the classical distribution 
at very early times, as demonstrated in Fig. \ref{probdist15}, 
corresponding to $n=15$. 

Thus the signature of a classically hyperbolic flow,  
namely, the exponential relaxation of an 
arbitrary distribution (with non-zero measure) 
to microcanonical equilibrium \cite{Dorfman}, holds to good approximation  
in this model in a regime of global chaos.
More suprisingly, this classical signature  
is manifest also in the dynamics of the quantum distribution.
In the quantum case, however, as can be seen 
in Fig. \ref{probdist15}, the probability distribution 
is subject to small irreducible time-dependent fluctuations about the 
classical equilibrium.  We examine 
these quantum fluctuations in detail Chapter 4. 


\section[Time-Domain Characteristics of Quantum-Classical Differences]{Time-Domain Characteristics of Quantum-Classical \newline Differences}

\begin{figure}[!t]
\begin{center}
\input{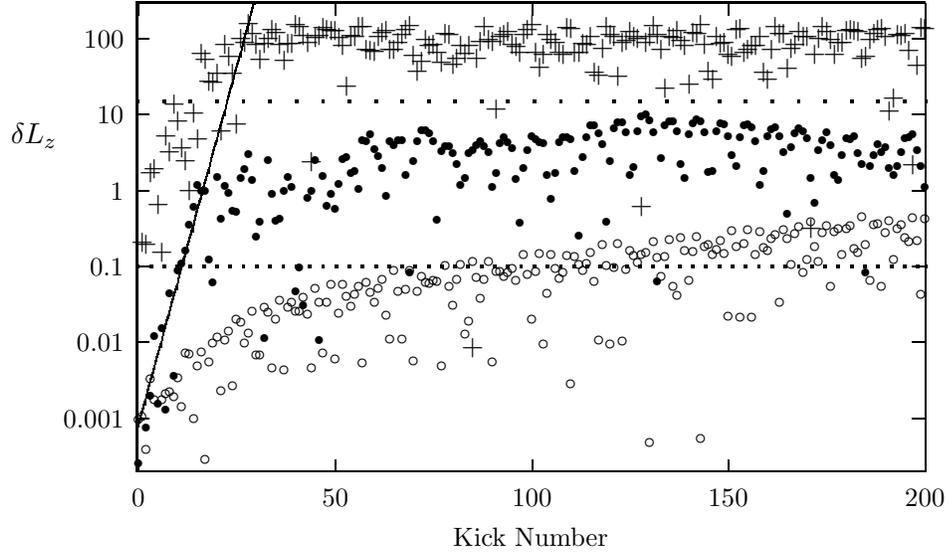}
\caption{
Time-dependence of quantum-classical differences in a regular 
zone (open circles) and a chaotic zone (filled circles) of mixed 
regime ($\gamma=1.215$ and $r \simeq 1.1$) 
with $l=154$. 
For the chaotic state,  
$\delta L_z = | \langle L_z \rangle - \langle L_z \rangle_c| $ is 
contrasted with the Ehrenfest difference $ | \langle L_z \rangle - L_z| $ 
between the quantum expectation value and a single trajectory (plus signs), 
which grows until saturation at system dimension. 
The solid line corresponds to (\ref{eqn:expansatz}) 
using $\lambda_{qc} = 0.43$. 
The horizontal lines indicate two different values of the difference 
tolerance $p$ which may be used to determine the break-time;  
for $p=0.1$ (dotted line) $t_b$ occurs on a logarithmic time-scale,  
but for $p=15.4$ (sparse dotted line) 
$t_b$ is not defined over numerically accessible time-scales.  
}
\label{delta.1.215.140.n200}
\end{center}
\end{figure}

We consider the time dependence of quantum-classical differences 
defined along the $z$-axis of the spin ${\bf L}$,
\begin{equation}
\label{eqn:diff}
        \delta L_z(n) = 
| \langle L_z(n)  \rangle - \langle L_z(n) \rangle_c  |, 
\end{equation}
at the stroboscopic times $t=n$. 
In Fig.\ \ref{delta.1.215.140.n200}
we compare the time-dependence of 
$\delta L_z(n)$ on a semi-log plot 
for a chaotic state (filled circles), with 
$\vec{\theta}(0) = (20^o,40^o,160^o,130^o)$, 
and a regular state (open circles), 
 $\vec{\theta}(0) = (5^o,5^o,5^o,5^o)$, 
evolved using the same mixed regime parameters 
($\gamma=1.215$ and $r\simeq1.1$) and quantum numbers
($l=154$) as in Fig.\ \ref{vargrowth.1.215.ric2}. 

We are interested in the behaviour of the upper envelope of the data 
in Fig.\ \ref{delta.1.215.140.n200}. For the regular case, 
the upper envelope of the quantum-classical differences grows very 
slowly, as some polynomial function of time. 
For the chaotic case, on the other hand, at early times 
the difference measure (\ref{eqn:diff}) grows 
exponentially until saturation around 
$n=15$, which is 
well before reaching system dimension, $|{\bf L}| \simeq l = 154$.  
After this time, which we denote $t^*$, 
the quantum-classical differences exhibit no definite growth, 
and fluctuate about the equilibrium value $\delta L_z \sim 1 \ll |{\bf L}|$. 
In  Fig.\ \ref{delta.1.215.140.n200} we also include data for the 
time-dependence of the Ehrenfest difference 
$| \langle L_z \rangle - L_z| $, which is 
defined as the difference between 
the quantum expectation value and the dynamical 
variable of a single trajectory initially centered on the quantum state. 
In contrast to $\delta L_z$, the rapid growth of the Ehrenfest difference 
continues until saturation at the system dimension. 

\begin{figure}[!t]
\begin{center}
\input{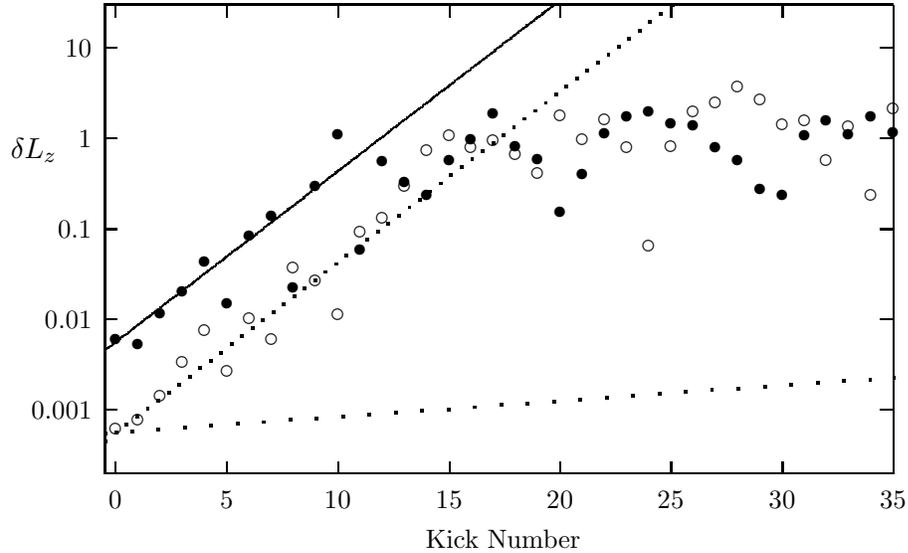}
\caption{
Growth of the quantum-classical difference 
$\delta L_z$ 
in the chaotic zone 
of a mixed regime, $\gamma =1.215$ and $r\simeq 1.1$,    
with $l=22$ (filled circles) 
and $l=220$ (open circles). 
For $l=220$ the exponential growth rate 
(\ref{eqn:expansatz}) is plotted using the classical 
Lyapunov exponent, $\lambda_L =0.04$ (sparse dotted line),  
and for both $l$ values (\ref{eqn:expansatz}) is plotted using 
the exponent $\lambda_{qc}=0.43$ (solid line for $l=22$, 
dotted line for $l=220$), 
which is obtained from a fit of (\ref{eqn:tb}) to the 
corresponding break-time data in Fig.\ \ref{break_time_p0.1}.
}
\label{delta.1.215.dvsl}
\end{center}
\end{figure}

In Fig.\ \ref{delta.1.215.dvsl} we compare the time-dependence 
of the quantum-classical differences 
in the case of the chaotic 
initial condition $\vec{\theta}(0) = (20^o,40^o,160^o,130^o)$ 
for quantum numbers $l=22$ (filled circles) and $l=220$ (open circles), 
using the same parameters as in Fig.\ \ref{delta.1.215.140.n200}. 
This demonstrates the remarkable fact that the exponential growth 
terminates when the difference measure reaches an essentially 
fixed magnitude ($\delta L_z \sim 1$ as for the case $l=154$), 
although the system dimension differs by an order of magnitude 
in the two cases.  

\begin{figure}[!t]
\begin{center}
\input{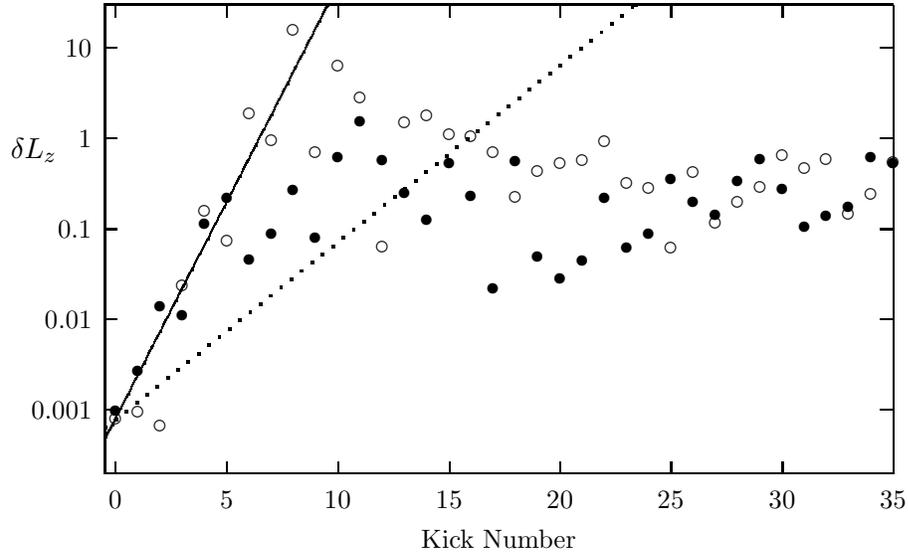}
\caption{
Growth of quantum-classical differences 
in the regime of global chaos, $\gamma=2.835$ and $r\simeq 1.1$,
with $l=154$, for the two initial conditions cited in text. 
The exponential growth rate 
(\ref{eqn:expansatz}) is plotted using the classical 
Lyapunov exponent, $\lambda_L =0.45$ (dotted line),  
and the exponent $\lambda_{qc}=1.1$ (solid line),   
which is obtained from a fit of (\ref{eqn:tb}) to the 
corresponding break-time data in Fig.\ \ref{break_time_p0.1}.
}
\label{delta.2.835.140}
\end{center}
\end{figure}

In Fig.\ \ref{delta.2.835.140} we consider the growth of the 
quantum-classical difference measure $\delta L_z(n)$ in 
a regime of global chaos, for $l=154$, 
and using the same set of parameters 
as those examined in Fig.\ \ref{vargrowth.2.835.140} 
($\gamma=2.835$ and $r\simeq1.1$).  
Again the upper envelope of the difference measure $\delta L_z(n)$ 
exhibits exponential growth at early times, though 
in this regime of global chaos the exponential growth 
persists only for a very 
short duration before saturation at $t^* \simeq 6$. 
The initial condition 
$\vec{\theta}(0) = (20^o,40^o,160^o,130^o)$ is a typical 
case (filled circles), 
where, as seen for the mixed regime parameters, 
the magnitude of the difference 
at the end of the exponential growth phase 
saturates at the value $\delta L_z(t^*) \simeq 1$, which does not scale 
with the system dimension (see Fig.\ \ref{deltamax.vs.l}).  
The initial condition 
$\vec{\theta}(0) = (45^o,70^o,135^o,70^o)$ (open circles) 
leads to an anomolously large deviation at the end  of the 
exponential growth phase, 
$\delta L_z(t^*) \simeq 10$, though still 
small relative to the system dimension $|{\bf L}| \simeq 154$. 
This deviation is transient however, and   
at later times the magnitude of quantum-classical differences fluctuates  
about the equilibrium value $\delta L_z \sim 1$. The quantum-classical 
differences are a factor 
of $1/l$ smaller than typical differences between the 
quantum expectation value and the single trajectory, 
which are of order system dimension 
(see Fig.\ \ref{qmlmnm}) as in the mixed regime case.

In all cases where the initial quantum and classical states are 
launched from a chaotic zone we find that the initial time-dependence 
of quantum-classical differences compares favorably with the 
exponential growth ansatz,
\begin{equation}
\label{eqn:expansatz}
	\delta L_z(n) 
\simeq { 1 \over 8 l } \exp ( \lambda_{qc} n )
 \; \;\;\;\;\;\;\; {\mathrm f}{\mathrm o}{\mathrm r}  \;\;\;  n < t^*,
\end{equation}
where the exponent $\lambda_{qc}$ is a new exponent subject 
to numerical measurement \cite{Ball98}.  
Since contributions from the initial differences in 
other mismatched moments  will generally mix under 
the dynamical flow, it is appropriate 
to consider an effective initial difference for 
the prefactor in (\ref{eqn:expansatz}).  
The prefactor $1/8l$ is obtained  
by accounting for the initial contributions from the 3 cartesian components,  
$ [ \delta^2 L_x(0) + \delta^2 L_y(0) + \delta^2 L_z(0) ]^{1/2} 
= 1/8l $. 

We are interested in whether the Lyapunov exponent 
$\lambda_L$ is a good approximation to $\lambda_{qc}$. 
In Fig.\ \ref{delta.1.215.dvsl} 
we plot (\ref{eqn:expansatz}) with $\lambda_{qc} = \lambda_L = 0.04$  
(dotted line) for $l=220$. Clearly 
the largest Lyapunov exponent severly underestimates  
the exponential growth rate of the quantum-classical differences, in this case 
by more than an order of magnitude.  
The growth rate of the state width, $\lambda_w = 0.13$ , is also 
several times smaller than the initial growth rate of the 
quantum-classical differences. 
In the case of Fig.\ \ref{delta.2.835.140}, corresponding to a 
regime of global chaos with a much larger Lyapunov exponent, 
we plot (\ref{eqn:expansatz}) with 
 $\lambda_{qc} = \lambda_L = 0.45$ (dotted line), 
demonstrating that, in this regime 
too the largest Lyapunov 
exponent underestimates the initial growth rate of the quantum-classical 
difference measure $\delta L_z(n)$. 


We also find, from inspection of our results, that the time $t^*$ 
at which the exponential growth (\ref{eqn:expansatz}) terminates
can be estimated from $t_{sat}$, the time-scale 
on which the distributions saturate at or near system size (\ref{eqn:nsat}). 
In the case of the chaotic initial condition of 
Fig.\ \ref{vargrowth.1.215.ric2}, for which $\gamma=1.215$, 
visual inspection of the figure suggests that 
$t_{sat} \simeq 18$. This should be compared with 
Fig.\ \ref{delta.1.215.140.n200}, where the exponential 
growth of $\delta L_z(n)$ ends rather abruptly at $t^* \simeq 15$. 
In Fig.\ \ref{vargrowth.2.835.140}, corresponding 
to a regime of global chaos ($\gamma=2.835$),  
the variance growth saturates much earlier, around   
$t_{sat} \simeq 6$ for both initial conditions. 
From Fig.\ \ref{delta.2.835.140} it is apparent that in 
this regime $t^* \simeq 6$. As we increase $\gamma$ further, 
we find that the exponential growth phase of quantum-classical 
differences $\delta L_z(n)$ is shortened, lasting only until 
the corresponding quantum and classical distributions 
saturate at system size. 
For $\gamma \simeq 12$, with $\lambda_L \simeq 1.65$, 
the chaos is sufficiently strong that the initial 
coherent state for $l=154$ spreads to cover ${\mathcal P}$ 
within a single time-step. Similarly the initial difference measure
$\delta L_z(0) \simeq 0.001$ grows to the magnitude $\delta L_z(1) \simeq 1$ 
within a single time-step and subsequently fluctuates about 
that equilibrium value. 
We have also inspected the variation of $t^*$ with the quantum numbers
and found it to be consistent with the logarithmic dependence 
of $t_{sat}$ in (\ref{eqn:nsat}).  



\section{Correspondence Scaling in the Macroscopic Limit}
\label{ch3:scaling}


We have assumed in (\ref{eqn:expansatz})  
that the exponent $\lambda_{qc}$ is independent of the quantum numbers. 
A convenient way of confirming this, and also estimating the numerical value 
of $\lambda_{qc}$, is by means of a break-time measure. The 
break-time is determined from the time $t_b(l,p) $ at 
which quantum-classical differences exceed some tolerance $p$,
which does not scale with the system size for each quantum model.  
The classical parameters and initial condition are held fixed.
Setting $\delta L_z ( t_b) = p $ in (\ref{eqn:expansatz}), we obtain  $t_b$ 
in terms of $p$, $l$ and $\lambda_{qc}$, 
\begin{equation}
\label{eqn:tb}
	t_b \simeq \lambda_{qc}^{-1} \ln ( 8 \; p \; l  )
\;\;\;\;\;\; provided  \;\; \;\;\  p < {\mathcal O}(1).
\end{equation}
The restriction $p < {\mathcal O}(1)$, which 
plays a crucial role in limiting the robustness 
of the break-time measure (\ref{eqn:tb}), is explained and 
motivated further below. 

The explicit form we have obtained for the argument of the logarithm 
in (\ref{eqn:tb}) is a direct result of our estimate that the initial 
quantum-classical differences arising from the Cartesian components 
of the spin provide the dominant contribution to the prefactor  
of the exponential growth ansatz (\ref{eqn:expansatz}). 
Differences in the mismatched higher order moments, 
as well as intrinsic differences between 
the quantum dynamics and classical dynamics, 
may also contribute to this effective prefactor.  
We have checked that the initial value 
$\delta L_z(0) \simeq 1 /8l$ is an adequate 
estimate by comparing the intercept of 
the quantum-classical data on a semilog plot 
with the prefactor of (\ref{eqn:expansatz}) 
for a variety of $l$ values (see {\it e.g}.\ Fig.\ \ref{delta.1.215.dvsl}).  

\begin{figure}[!t]
\begin{center}
\input{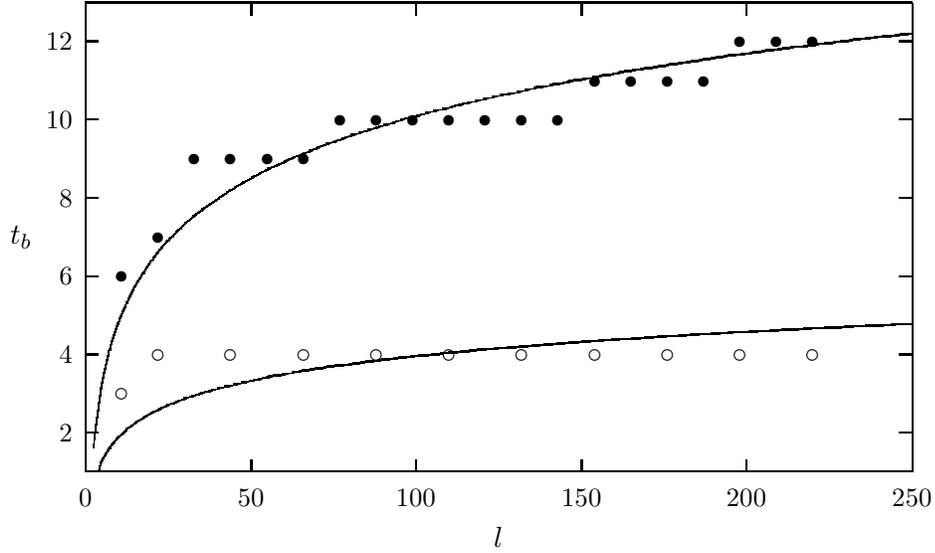}
\caption{
Scaling of the break-time using tolerance $p=0.1$ as a function   
of increasing quantum number for the mixed regime parameters 
$\gamma = 1.215$ and $r \simeq 1.1$ with 
$\vec{\theta}(0) =(20^o,40^o,160^o,130^o)$  (filled circles) 
and for the global chaos parameters $\gamma =2.835$ and $r\simeq 1.1$ with 
$\vec{\theta}(0) =(45^o,70^o,135^o,70^o)$
(open circles). 
We also plot the results of fits  
to the log rule (\ref{eqn:tb}), 
which produced exponents $\lambda_{qc} = 0.43$ 
for $\gamma=1.215$ and $\lambda_{qc}=1.1$ for $\gamma = 2.835$.
}
\label{break_time_p0.1}
\end{center}
\end{figure}

In Fig.\ \ref{break_time_p0.1} we examine the scaling of the break-time 
for $l$ values ranging from $11$ to $220$ and 
with fixed tolerance $p=0.1$. 
The break-time can assume only the integer values $t=n$ 
and thus the data exhibits a step-wise behaviour. 
For the mixed regime parameters, 
$\gamma=1.215$ and $r\simeq 1.1$ (filled circles), 
with initial condition $\vec{\theta}(0) = (20^o,40^o,160^o,130^o)$,  
a non-linear least squares fit to (\ref{eqn:tb}) gives 
$\lambda_{qc} = 0.43$. This fit result is 
plotted in the figure as a solid line. The close agreement between the data 
and the fit provides good evidence that 
the quantum-classical exponent $\lambda_{qc}$ is independent of 
the quantum numbers. 
To check this result against the time-dependent $\delta L_z(n)$ data, 
we have plotted the exponential curve (\ref{eqn:expansatz}) 
with $\lambda_{qc} = 0.43$ 
in Fig.\ \ref{delta.1.215.140.n200} using a solid line 
and  in Fig.\ \ref{delta.1.215.dvsl} using a solid line for $l=22$ and 
a dotted line for $l=220$. 
The exponent obtained from fitting (\ref{eqn:tb}) serves as an excellent 
approximation to the initial exponential growth (\ref{eqn:expansatz}) 
of the quantum-classical differences in each case. 

In Fig.\ \ref{break_time_p0.1} we also plot 
break-time results for the global chaos 
case $\gamma=2.835$ and $r\simeq 1.1$ (open circles) 
with initial condition $\vec{\theta}(0) = (45^o,70^o,135^o,70^o)$. 
In this regime the quantum-classical 
differences grow much more rapidly and, consequently, 
the break-time is very short and remains 
nearly constant over this range of computationally accessible quantum numbers. 
Due to this limited variation, in this regime we can not 
confirm (\ref{eqn:tb}), although the data are  
consistent with the predicted logarithmic dependence on $l$.
Moreover, the break-time 
results provide an effective method for estimating $\lambda_{qc}$ 
if we assume that (\ref{eqn:tb}) holds. 
The same fit procedure as detailed above yields the quantum-classical 
exponent $\lambda_{qc} = 1.1$. This fit result is 
plotted in  Fig.\ \ref{break_time_p0.1} as a solid line. More importantly, 
the exponential curve (\ref{eqn:expansatz}), plotted 
with fit result $\lambda_{qc} = 1.1$,  
can be seen to provide very good agreement with the initial growth rate 
of Fig.\ \ref{delta.2.835.140} for either initial condition, as expected. 

In the mixed regime ($\gamma=1.215$), 
the quantum-classical exponent $\lambda_{qc} = 0.43$ 
is an order of magnitude greater 
than the largest Lyapunov exponent $\lambda_L=0.04$ 
and about three times larger 
than the growth rate of the width $\lambda_w =0.13$. 
In the regime of global chaos ($\gamma=2.835$) 
the quantum-classical exponent $\lambda_{qc} = 1.1$ 
is a little more than twice as large as the 
largest Lyapunov exponent $\lambda_L=0.45$.

The condition $p< {\mathcal O}(1) $ is a very restrictive 
limitation on the domain of application of the log break-time 
(\ref{eqn:tb}) and it is worthwhile to explain its significance. 
In the mixed regime case of Fig.\ \ref{delta.1.215.140.n200},  
with $l=154$, we have plotted the tolerance values 
$p=0.1$ (dotted line) and $p=15.4$ (sparse dotted line). 
The tolerance $p=0.1$ is exceeded at $t=11$, while 
the quantum-classical differences are still growing exponentially, 
leading to a log break-time for this tolerance value. 
For the tolerance $p=15.4 \ll |{\bf L}|$, on the other hand,  
the break-time does not occur on a measurable time-scale, whereas 
according to the logarithmic rule  (\ref{eqn:tb}), with $l = 154$ 
and $\lambda_{qc} = 0.43$, we should expect a rather 
short break-time $t_b \simeq 23$. Consequently the break-time  
(\ref{eqn:tb}), applied 
to delimiting the end of the Liouville regime, 
is not a robust measure of quantum-classical 
correspondence. 

Our definition of the break-time (\ref{eqn:tb})  
requires holding the tolerance $p$ fixed as the system size increases 
(and not as a fraction of the system dimension as in \cite{Haake87}) 
when comparing systems with different quantum numbers. 
Had we chosen to compare systems using a fixed relative tolerance, $f$,  
then the break-time would be of the form 
$t_b \simeq \lambda_{qc}^{-1} \ln ( 8 \; f \; l^2  )$ 
and subject to the restriction $f < {\mathcal O}(1/l)$. 
Since $f \rightarrow 0$ in the classical limit,  
this form emphasizes 
that the log break-time applies only to differences that are 
vanishing fraction of the system dimension in that limit.

Although we have provided numerical evidence 
(in Fig.\ \ref{delta.1.215.dvsl}) of one mixed regime case 
in which the largest quantum-classical differences 
occuring at the end of the exponential growth period 
remain essentially constant for varying quantum numbers, 
$\delta L_z (t^*) \sim {\mathcal O}(1)$, we find that 
this behaviour represents the typical case for all parameters and initial
conditions which produce chaos classically.  
To demonstrate this behaviour we consider 
the scaling (with increasing quantum numbers) 
of the maximum values attained by $\delta L_z(n)$ 
over the first 200 kicks, $\delta L_z^{max}$. 
Since $t^* \ll 200$ over the range of $l$ values examined, 
the quantity $\delta L_z^{max}$ is a rigorous upper bound for 
$\delta L_z(t^*)$. 

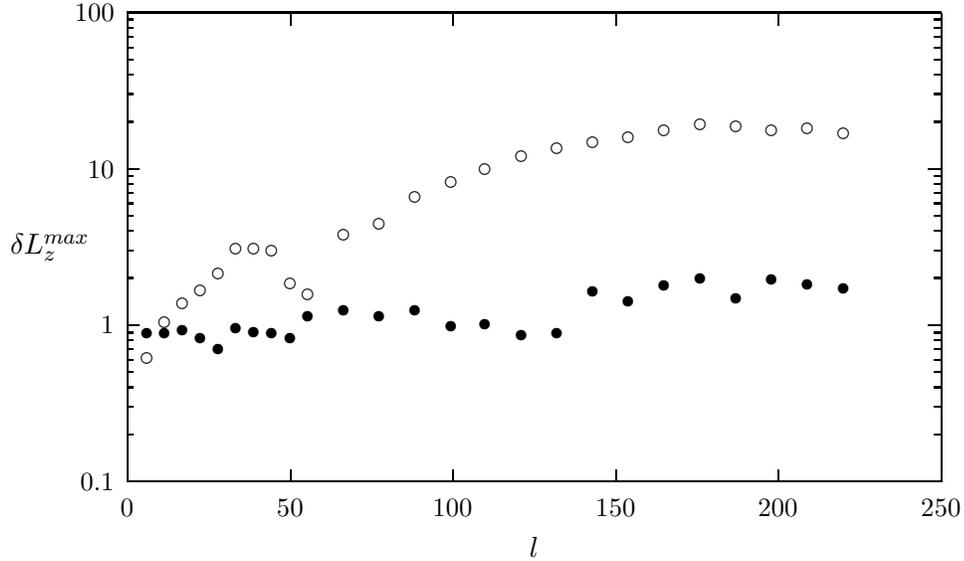
\begin{figure}[!t]
\begin{center}
\setlength{\unitlength}{0.240900pt}
\ifx\plotpoint\undefined\newsavebox{\plotpoint}\fi
\sbox{\plotpoint}{\rule[-0.200pt]{0.400pt}{0.400pt}}%
\begin{picture}(1500,900)(0,0)
\font\gnuplot=cmr10 at 10pt
\gnuplot
\sbox{\plotpoint}{\rule[-0.200pt]{0.400pt}{0.400pt}}%
\put(161.0,123.0){\rule[-0.200pt]{4.818pt}{0.400pt}}
\put(141,123){\makebox(0,0)[r]{0.1}}
\put(1419.0,123.0){\rule[-0.200pt]{4.818pt}{0.400pt}}
\put(161.0,197.0){\rule[-0.200pt]{2.409pt}{0.400pt}}
\put(1429.0,197.0){\rule[-0.200pt]{2.409pt}{0.400pt}}
\put(161.0,240.0){\rule[-0.200pt]{2.409pt}{0.400pt}}
\put(1429.0,240.0){\rule[-0.200pt]{2.409pt}{0.400pt}}
\put(161.0,271.0){\rule[-0.200pt]{2.409pt}{0.400pt}}
\put(1429.0,271.0){\rule[-0.200pt]{2.409pt}{0.400pt}}
\put(161.0,295.0){\rule[-0.200pt]{2.409pt}{0.400pt}}
\put(1429.0,295.0){\rule[-0.200pt]{2.409pt}{0.400pt}}
\put(161.0,314.0){\rule[-0.200pt]{2.409pt}{0.400pt}}
\put(1429.0,314.0){\rule[-0.200pt]{2.409pt}{0.400pt}}
\put(161.0,331.0){\rule[-0.200pt]{2.409pt}{0.400pt}}
\put(1429.0,331.0){\rule[-0.200pt]{2.409pt}{0.400pt}}
\put(161.0,345.0){\rule[-0.200pt]{2.409pt}{0.400pt}}
\put(1429.0,345.0){\rule[-0.200pt]{2.409pt}{0.400pt}}
\put(161.0,357.0){\rule[-0.200pt]{2.409pt}{0.400pt}}
\put(1429.0,357.0){\rule[-0.200pt]{2.409pt}{0.400pt}}
\put(161.0,369.0){\rule[-0.200pt]{4.818pt}{0.400pt}}
\put(141,369){\makebox(0,0)[r]{1}}
\put(1419.0,369.0){\rule[-0.200pt]{4.818pt}{0.400pt}}
\put(161.0,443.0){\rule[-0.200pt]{2.409pt}{0.400pt}}
\put(1429.0,443.0){\rule[-0.200pt]{2.409pt}{0.400pt}}
\put(161.0,486.0){\rule[-0.200pt]{2.409pt}{0.400pt}}
\put(1429.0,486.0){\rule[-0.200pt]{2.409pt}{0.400pt}}
\put(161.0,517.0){\rule[-0.200pt]{2.409pt}{0.400pt}}
\put(1429.0,517.0){\rule[-0.200pt]{2.409pt}{0.400pt}}
\put(161.0,540.0){\rule[-0.200pt]{2.409pt}{0.400pt}}
\put(1429.0,540.0){\rule[-0.200pt]{2.409pt}{0.400pt}}
\put(161.0,560.0){\rule[-0.200pt]{2.409pt}{0.400pt}}
\put(1429.0,560.0){\rule[-0.200pt]{2.409pt}{0.400pt}}
\put(161.0,576.0){\rule[-0.200pt]{2.409pt}{0.400pt}}
\put(1429.0,576.0){\rule[-0.200pt]{2.409pt}{0.400pt}}
\put(161.0,591.0){\rule[-0.200pt]{2.409pt}{0.400pt}}
\put(1429.0,591.0){\rule[-0.200pt]{2.409pt}{0.400pt}}
\put(161.0,603.0){\rule[-0.200pt]{2.409pt}{0.400pt}}
\put(1429.0,603.0){\rule[-0.200pt]{2.409pt}{0.400pt}}
\put(161.0,614.0){\rule[-0.200pt]{4.818pt}{0.400pt}}
\put(141,614){\makebox(0,0)[r]{10}}
\put(1419.0,614.0){\rule[-0.200pt]{4.818pt}{0.400pt}}
\put(161.0,688.0){\rule[-0.200pt]{2.409pt}{0.400pt}}
\put(1429.0,688.0){\rule[-0.200pt]{2.409pt}{0.400pt}}
\put(161.0,732.0){\rule[-0.200pt]{2.409pt}{0.400pt}}
\put(1429.0,732.0){\rule[-0.200pt]{2.409pt}{0.400pt}}
\put(161.0,762.0){\rule[-0.200pt]{2.409pt}{0.400pt}}
\put(1429.0,762.0){\rule[-0.200pt]{2.409pt}{0.400pt}}
\put(161.0,786.0){\rule[-0.200pt]{2.409pt}{0.400pt}}
\put(1429.0,786.0){\rule[-0.200pt]{2.409pt}{0.400pt}}
\put(161.0,805.0){\rule[-0.200pt]{2.409pt}{0.400pt}}
\put(1429.0,805.0){\rule[-0.200pt]{2.409pt}{0.400pt}}
\put(161.0,822.0){\rule[-0.200pt]{2.409pt}{0.400pt}}
\put(1429.0,822.0){\rule[-0.200pt]{2.409pt}{0.400pt}}
\put(161.0,836.0){\rule[-0.200pt]{2.409pt}{0.400pt}}
\put(1429.0,836.0){\rule[-0.200pt]{2.409pt}{0.400pt}}
\put(161.0,849.0){\rule[-0.200pt]{2.409pt}{0.400pt}}
\put(1429.0,849.0){\rule[-0.200pt]{2.409pt}{0.400pt}}
\put(161.0,860.0){\rule[-0.200pt]{4.818pt}{0.400pt}}
\put(141,860){\makebox(0,0)[r]{100}}
\put(1419.0,860.0){\rule[-0.200pt]{4.818pt}{0.400pt}}
\put(161.0,123.0){\rule[-0.200pt]{0.400pt}{4.818pt}}
\put(161,82){\makebox(0,0){0}}
\put(161.0,840.0){\rule[-0.200pt]{0.400pt}{4.818pt}}
\put(417.0,123.0){\rule[-0.200pt]{0.400pt}{4.818pt}}
\put(417,82){\makebox(0,0){50}}
\put(417.0,840.0){\rule[-0.200pt]{0.400pt}{4.818pt}}
\put(672.0,123.0){\rule[-0.200pt]{0.400pt}{4.818pt}}
\put(672,82){\makebox(0,0){100}}
\put(672.0,840.0){\rule[-0.200pt]{0.400pt}{4.818pt}}
\put(928.0,123.0){\rule[-0.200pt]{0.400pt}{4.818pt}}
\put(928,82){\makebox(0,0){150}}
\put(928.0,840.0){\rule[-0.200pt]{0.400pt}{4.818pt}}
\put(1183.0,123.0){\rule[-0.200pt]{0.400pt}{4.818pt}}
\put(1183,82){\makebox(0,0){200}}
\put(1183.0,840.0){\rule[-0.200pt]{0.400pt}{4.818pt}}
\put(1439.0,123.0){\rule[-0.200pt]{0.400pt}{4.818pt}}
\put(1439,82){\makebox(0,0){250}}
\put(1439.0,840.0){\rule[-0.200pt]{0.400pt}{4.818pt}}
\put(161.0,123.0){\rule[-0.200pt]{307.870pt}{0.400pt}}
\put(1439.0,123.0){\rule[-0.200pt]{0.400pt}{177.543pt}}
\put(161.0,860.0){\rule[-0.200pt]{307.870pt}{0.400pt}}
\put(40,491){\makebox(0,0){ $\delta L_z^{max}$ } }
\put(800,21){\makebox(0,0){ $l$ }}
\put(161.0,123.0){\rule[-0.200pt]{0.400pt}{177.543pt}}
\put(192,357){\circle*{18}}
\put(220,357){\circle*{18}}
\put(248,362){\circle*{18}}
\put(276,349){\circle*{18}}
\put(304,332){\circle*{18}}
\put(332,365){\circle*{18}}
\put(360,358){\circle*{18}}
\put(388,356){\circle*{18}}
\put(417,348){\circle*{18}}
\put(445,383){\circle*{18}}
\put(501,392){\circle*{18}}
\put(557,383){\circle*{18}}
\put(613,392){\circle*{18}}
\put(670,368){\circle*{18}}
\put(723,370){\circle*{18}}
\put(780,354){\circle*{18}}
\put(836,356){\circle*{18}}
\put(892,422){\circle*{18}}
\put(948,407){\circle*{18}}
\put(1004,432){\circle*{18}}
\put(1061,442){\circle*{18}}
\put(1117,411){\circle*{18}}
\put(1173,441){\circle*{18}}
\put(1229,433){\circle*{18}}
\put(1286,427){\circle*{18}}
\put(192,317){\circle{18}}
\put(220,373){\circle{18}}
\put(248,403){\circle{18}}
\put(276,423){\circle{18}}
\put(304,451){\circle{18}}
\put(332,490){\circle{18}}
\put(360,489){\circle{18}}
\put(388,487){\circle{18}}
\put(417,434){\circle{18}}
\put(445,418){\circle{18}}
\put(501,512){\circle{18}}
\put(557,528){\circle{18}}
\put(613,571){\circle{18}}
\put(670,594){\circle{18}}
\put(723,614){\circle{18}}
\put(780,634){\circle{18}}
\put(836,647){\circle{18}}
\put(892,656){\circle{18}}
\put(948,665){\circle{18}}
\put(1004,675){\circle{18}}
\put(1061,684){\circle{18}}
\put(1117,681){\circle{18}}
\put(1173,676){\circle{18}}
\put(1229,678){\circle{18}}
\put(1286,670){\circle{18}}
\end{picture}
\caption{
Maximum quantum-classical difference occuring over the first 200 kicks   
in the regime of global chaos ($\gamma=2.835$, $r\simeq 1.1$) 
plotted against increasing quantum number. These maximum values provide 
an upper bound on $\delta L_z(t^*)$ for each $l$. 
The data corresponding to the initial condition
$\vec{\theta}(0) =(20^o,40^o,160^o,130^o)$ (filled circles) 
represent a typical case in which 
the maximum quantum-classical differences 
do not vary significantly with $l$. 
The large deviations observed for the initial 
condition $\vec{\theta}(0) =(45^o,70^o,135^o,70^o)$ (open circles) 
are an exceptional case, with maximum differences growing rapidly for small 
quantum numbers but tending asymptotically toward independence of $l$. 
These curves provide an upper bound on the tolerance values 
$p$ for which the break-time measure scales logarithmically with $l$. 
}
\label{deltamax.vs.l}
\end{center}
\end{figure}

In Fig.\ \ref{deltamax.vs.l} we compare $\delta L_z^{max}$ 
for the two initial conditions of Fig.\ \ref{delta.2.835.140} 
and using the global chaos parameters ($\gamma =2.835$, $r \simeq 1.1$).  
The filled circles in Fig.\ \ref{deltamax.vs.l}  
correspond to the initial 
condition $\vec{\theta}(0) = (20^o,40^o,160^o,130^o)$.  
As in the mixed regime,  
the maximum deviations exhibit 
little or no scaling with increasing quantum number. This is the 
typical behaviour that we have observed  
for a variety of different initial conditions 
and parameter values. 
These results motivate the generic rule, 
\begin{equation}
\frac{\delta L_z(t^*)}{\sqrt{l(l+1)}} \le 
\frac{\delta L_z^{max}}{\sqrt{l(l+1)}} \sim {\mathcal O}(1/l). 
\end{equation}
Thus the magnitude of quantum-classical 
differences reached at the end of the exponential growth regime, 
expressed as a fraction of the system dimension, approaches  
zero in the classical limit. 

However, for a few combinations of parameters and initial conditions 
we do observe a `transient' discrepancy peak 
occuring at $ t \simeq t^* $ that exceeds ${\mathcal O}(1)$.  
This peak is quickly smoothed away 
by the subsequent relaxation of the quantum and classical distributions. 
This peak is apparent in 
Fig.\ \ref{delta.2.835.140} (open circles),  corresponding to 
the most conspicuous case that we have identified. This  
case is apparent as a small deviation in the normalized data of 
Fig.\ \ref{qmlmnm}. The scaling 
of the magnitude of this peak with increasing $l$ is 
plotted with open circles in Fig.\ \ref{deltamax.vs.l}. 
The magnitude of the peak initially increases rapidly but appears 
to become asymptotically independent of $l$. 
The other case that we have 
observed occurs for the classical parameters 
$\gamma =2.025$, with $r\simeq 1.1$ and $a=5$, and 
with initial condition $\vec{\theta}(0) = (20^o,40^o,160^o,130^o)$.  
We do not understand the 
mechanism leading to such transient peaks, although they are of 
considerable interest since they provide the most prominent 
examples of quantum-classical discrepancy that we have observed.

\section{Discussion}
\label{sec:discussion}

In this study of a non-integrable 
model of two interacting spins we have characterized the 
correspondence between quantum expectation values 
and classical ensemble averages for intially localised states.  
We have demonstrated that in chaotic states 
the quantum-classical differences initially grow 
exponentially with an exponent $\lambda_{qc}$ that is 
consistently larger than the largest Lyapunov exponent. 
In a study of the moments of the Henon-Heiles system, 
Ballentine and McRae \cite{Ball98,Ball00} have also shown 
that quantum-classical differences in chaotic states  
grow at an exponential rate with 
an exponent larger than the largest Lyapunov exponent. 
This exponential behaviour appears to be a generic feature of the 
short-time dynamics of quantum-classical differences in chaotic states. 


Since we have studied a spin system, we have been able to solve 
the quantum problem without truncation of the Hilbert space, 
subject only to numerical roundoff, and thus 
we are able to observe the dynamics of the quantum-classical differences 
well beyond the Ehrenfest regime. We have shown that the exponential growth 
phase of the quantum-classical differences 
terminates well before these differences
have reached system dimension. We find that the time-scale 
at which this occurs can be estimated  
from the time-scale at which the distribution widths 
approach the system dimension, 
$t_{sat} \simeq (2 \lambda_w)^{-1} \ln (l)$ for initial 
minimum uncertainty states. 
Due to the close correspondence in the growth rates of the 
quantum and classical distributions, this time-scale can 
be estimated from the classical physics alone. This is useful 
because the computational complexity of the problem does not grow  
with the system action in the classical case. Moreover, we find that 
the exponent $\lambda_w$ can be approximated by the largest 
Lyapunov exponent when the kinematic surface is predominantly chaotic. 

We have demonstrated that the exponent $\lambda_{qc}$ 
governing the initial growth rate of the quantum-classical differences 
is independent of the quantum numbers, and that the effective prefactor to 
this exponential growth decreases as $1/l$. These results 
imply that a log break-time rule (\ref{eqn:tb})
delimits the dynamical regime of Liouville correspondence.  
However, the exponential growth of quantum-classical differences 
persists only for short times and small differences, and thus this 
log break-time rule applies only in a similarly restricted domain. 
In particular, we have found that the magnitude of the differences 
occuring at the end of the initial exponential growth phase  
does not scale with the system dimension. 
A typical magnitude for these differences, 
relative to the system dimension, is ${\mathcal O}(1/l)$. 
Therefore, $\log(l)$ break-time rules characterizing 
the end of the Liouville 
regime are not robust, since they 
apply to quantum-classical differences only in a restricted domain, 
{\it i.e}.\ to relative differences that are smaller 
than ${\mathcal O}(1/l)$. 

This restricted domain effect does not arise for the better known 
log break-time rules describing the end of the Ehrenfest 
regime \cite{Ball94,BZ78,Haake87}.  The Ehrenfest log 
break-time remains robust 
for arbitrarily large tolerances since the corresponding differences 
grow roughly exponentially until saturation at the system 
dimension \cite{Fox94b,Fox94a}. Consequently,  
a $\log(l)$ break-time indeed implies a {\em breakdown} of 
Ehrenfest correspondence.
However, the logarithmic break-time rule characterizing  
the end of the Liouville regime does not  
imply a breakdown of Liouville correspondence 
because it does not apply to the observation 
of quantum-classical discrepancies larger than ${\mathcal O}(1/l)$. 
The appearance of residual ${\mathcal O}(1/l)$ quantum-classical 
discrepancies in the description of a macroscopic body 
is, of course, consistent with quantum mechanics having a 
proper classical limit. 

We have found, however, that for certain exceptional 
combinations of parameters and initial conditions 
there are relative quantum-classical differences 
occuring at the end of the exponential growth phase that  
can be larger than ${\mathcal O}(1/l)$, though still 
much smaller than the system dimension. In absolute terms, these  
transient peaks seem to grow with the system dimension 
for small quantum numbers but become asymptotically 
independent of the system dimension for larger quantum numbers. 
Therefore, even in these least favorable 
cases, the {\em fractional} differences between quantum and 
classical dynamics approach zero in the limit $l \rightarrow \infty$. 
This vanishing of fractional differences is sufficient to 
ensure a classical limit for our model.  

Finally, contrary to the results found in the present model, 
it has been suggested that a log break-time delimiting 
the Liouville regime implies 
that certain isolated macroscopic bodies in chaotic 
motion should exhibit non-classical behaviour on observable time scales. 
However, since such non-classical behaviour is not observed in the chaotic 
motion of macroscopic bodies,   
it is argued that the observed classical behaviour emerges 
from quantum mechanics only when the quantum description is expanded to 
include interactions with the many degrees-of-freedom of the 
ubiquitous environment \cite{ZP95a,Zurek98b}. (This effect, 
called decoherence, rapidly evolves a pure system state 
into a mixture that is essentially devoid of non-classical properties.)    
However, 
in our model the classical behaviour emerges in the macroscopic limit 
of an isolated few degree-of-freedom quantum system that is 
described by a pure state and subject only to unitary evolution.  
Quantum-classical correspondence at both early and 
late times arises in spite of the log break-time because this break-time 
rule applies only when the quantum-classical difference threshold 
is chosen smaller than ${\mathcal O}(\hbar)$.  
In this sense we find that the decoherence 
effects of the environment are not necessary for correspondence 
in the macroscopic limit. Of course the effect of decoherence may be 
experimentally significant in the quantum and mesoscopic domains, but 
it is not required {\em as a matter of principle} to ensure a classical 
limit.







\chapter{Correspondence for the Probability Distributions}

This chapter is taken from Emerson and Ballentine \cite{EB00b}.
\section{Introduction}

The study of chaos in quantum dynamics has led to differing views 
on the conditions required for demonstrating 
quantum-classical correspondence \cite{Ball94,Zurek98a}.   
Moreover, the criteria by which this correspondence should be 
measured have also been a subject of some 
controversy \cite{ZP94,CC95,ZP95a}.    
While much of the earlier work 
on this topic is concerned with characterizing 
the degree of correspondence between quantum expectation values and 
classical dynamical variables \cite{BZ78,Frahm85,Haake87}, the 
more recent approach is to focus on differences between 
the properties of quantum states and associated 
classical phase space densities 
evolved according to Liouville's equation 
\cite{Ball94,HB94,Fox94b,RBWG95,Ball98,EB00a}.  

Several authors have examined quantum-classical correspondence 
by considering the effects of interactions with a stochastic 
environment \cite{Dec96,HB96,Gong99}, a process sometimes 
called {\em decoherence}. While this process may improve the degree 
of quantum-classical correspondence for fixed quantum numbers, 
it has been further suggested that the  
limit of large quantum numbers is inadequate for correspondence, 
and that decoherence {\em must} be taken into account 
to generate classical appearances from quantum theory; 
this view has been argued to apply  
even in the case of macroscopic bodies that are described 
initially by well-localised states, 
provided their classical motion is chaotic \cite{ZP95a,ZP95b,Zurek98b}. 
In this chapter we examine how the degree of correspondence 
with Liouville dynamics scales specifically 
in the limit of {\em large} quantum numbers. 
This ``classical limit'' is distinct from a ``thermodynamic 
limit'', that is, a limit involving {\em many} quantum numbers. 

The degree of Liouville correspondence has been 
characterized previously by studying 
the differences between the means and variances of the 
dynamical variables \cite{Ball94,HB94,RBWG95,Ball98,EB00a,Ball00}. 
This involves a comparison of quantum expectation values and classical
ensemble averages. 
However, these low-order moments give only crude information about the 
differences between the quantum and classical states. 
Specifically, the quantum state may exhibit coarse structure which 
differs significantly 
from the classical state although the  
means and variances (for some simple observabes) 
are nearly the same for the quantum and classical states. 
Moreover, much of the previous work was concerned with correspondence 
at early times, or more precisely, in the Ehrenfest regime when the 
states are narrow compared to system dimensions \cite{Ball98,EB00a}.

Another approach is to identify quantum-classical differences with 
differences between the Wigner quasi-distribution and the 
classical phase space density \cite{Zurek98a}. 
This approach is objectionable 
because the Wigner quasi-distribution may take 
on negative values and therefore may not be interpreted as 
a ``classically observable'' phase space distribution. 
It is possible to consider instead smoothed quantum phase space 
distributions, but in this case the residual quantum-classical  
differences still do not have clear experimental  
significance.  

In this chapter, we characterize the degree of 
quantum-classical correspondence by comparing quantum 
probability distributions for dynamical variables 
with the corresponding classical marginal 
distributions for these dynamical variables. 
These are well-defined classical observables that describe the 
distribution of outcomes upon measurement of the 
given dynamical variable.  We are interested in the 
differences that arise on a {\em fine} scale and therefore 
characterize the typical quantum-classical deviations  
that arise in bins of width $\hbar$.  

The dynamics are generated by the model of interacting spins 
described in Chapter 2. The Hilbert space
is finite dimensional so no artificial truncation of the state is required. 
The quantum time-evolution is unitary and the classical 
motion is volume-preserving (symplectic).
In the case of classically chaotic motion, we follow 
initially localised states until they have evolved well beyond 
the relaxation time-scale of the classical density. 
Throughout the chapter we emphasize that the {\em quantum} signatures of 
chaos that appear in the quantum distributions 
are the same as those that appear in 
the marginal classical distributions. 
In particular, the quantum relaxation rates can be accurately estimated 
from the Liouville dynamics of an approximately 
matching initial phase space density. 
This purely classical approximation is surprisingly accurate  
even for small quantum numbers, 
but may be most useful for the theoretical description of 
mesoscopic systems since 
the purely classical calculations do not scale with the quantum numbers.   

The quantum and classical 
probability distributions remain close even after the states 
have spread to the system dimension. 
Specifically, in mixed regimes, 
the quantum distributions exhibit an 
equilibrium shape that reflects the details 
of the classical KAM surfaces.  
When the classical manifold is predominantly chaotic, the quantum  
and classical states relax close to the microcanonical state.  
However, in both of these chaotic regimes the equilibrium quantum 
distributions exhibit characteristic 
fluctuations away from the classical ones.  
We demonstrate that 
the standard deviation of these quantum-classical differences becomes 
vanishingly small in the classical limit, 
${\mathcal J}/\hbar \rightarrow \infty$, where 
${\mathcal J}$ is a characteristic system action. 


\section{Dynamical Behaviour of Probability Distributions}


In the case of a classical mixing system, initial densities 
with non-zero measure are expected to spread in an increasingly 
uniform manner throughout the accessible phase space. 
The term {\em uniform} is meant to apply specifically 
in a coarse-grained sense. For some simple maps, such as the baker's map, 
it is possible to show that this rate of relaxation to the equilibrium 
configuration occurs exponentially with time \cite{Dorfman}. 

The spin map we consider (\ref{eqn:map}) is not {\em mixing} on the 
accessible classical manifold ${\mathcal P}$, but has {\em mixed} dynamics:  
depending on the system parameters, 
the surface ${\mathcal P}$ can generally  
be decomposed into 
regions of regular dynamics and a connected region of chaotic dynamics 
\cite{EB00a}.  In parameter regimes 
that are predominantly chaotic, we expect 
behaviour on ${\mathcal P}$ that approximates that of a mixing system. 
In particular, initially localised Liouville densities 
should relax close towards the 
microcanonical measure at an exponential rate, on average. 
In this section we demonstrate that these signatures of chaos 
are exhibited also by the quantum dynamics. Most striking is the 
degree of similarity between the quantum and classical 
behaviours even in regimes with classically 
mixed dynamics.  

We are interested in the behaviour of quantum probability distributions 
that are associated with measurements of 
classical dynamical variables. The quantum probability distribution 
associated with the classical observable $L_z$ is given by, 
\begin{equation}
\label{eqn:plz}
P_{L_z}(m_l) =  
\langle \psi(n) | \; R_{l,m_l} \; | \psi(n) \rangle 
= {\mathrm T} {\mathrm r} \left[ \; | l,m_l \rangle \langle l,m_l 
| \rho^{(l)}(n)  \; \right], 
\end{equation}
where,
\begin{equation}
R_{l,m_l} =  1_s \otimes | l, m_l \rangle \langle l, m_l | 
\end{equation}
is a projection operator onto the eigenstates of $L_z$, and 
\begin{equation}
\label{eqn:redl}
\rho^{(l)}(n)= {\mathrm T} {\mathrm r}^{(s)} \left[ \; | \psi(n) \rangle 
\langle \psi(n) | \; \right],  
\end{equation}
is the reduced state operator for 
the spin ${\bf L}$ at time $n$ and ${\mathrm T}{\mathrm r}^{(s)}$ denotes 
a trace over the factor space ${\mathcal H}_s$. We have written 
out the explicit expression (\ref{eqn:plz}) to emphasize 
that the probability of obtaining each $m_l$ value is associated with 
a projector onto a subspace of the {\em factor} space ${\mathcal H}_l$. 

For reasons related to this fact (which we will make clear in later sections), 
we are also interested in examining the 
probability distributions associated with components of 
the {\it total} angular momentum ${\bf J} = {\bf S} + {\bf L}$.  
The probability of obtaining a given $m_j$ value upon measurement of 
$J_z$ is given by, 
\begin{equation}
\label{eqn:pjz}
P_{J_z}(m_j) = \sum_{m_s} \! 
| \langle \psi(n) |  s, l,m_s,m_j - m_s \rangle |^2,   
\end{equation}
where $|  s, l,m_s,m_j - m_s \rangle$ is an element of the orthonormal basis 
(\ref{eqn:basis}).
The probability $P_{J_z}(m_j)$ 
is associated with a projector onto a   
subspace of the {\em full} Hilbert space ${\mathcal H}$. 
The dimension of each subspace  
is given by the number of pairs $(m_s,m_l)$ that yield 
a given value of $m_j = m_s + m_l$.

The classical probability distributions associated with dynamical variables 
are obtained by partial integration over the accessible phase space. 
In the case of $L_z$, the continuous marginal distribution is given by, 
\begin{equation}
\label{eqn:pcclz}
	P(L_z) = \int \! \int \! \int dS_z d\phi_s d\phi_l 
\; \rho_c(S_z,\phi_s,L_z,\phi_l), \nonumber
\end{equation}
where for notational convenience 
we have suppressed reference to the time-dependence. 
The marginal probability distribution for the total spin component $J_z$ 
is obtained by integration subject to the constraint $S_z + L_z = J_z$,   
\begin{equation}
\label{eqn:pccjz}
	P(J_z) = \int \! \int \! \int \! \int dS_z d\phi_s d L_z d\phi_l 
\; \rho_c(S_z,\phi_s,L_z,\phi_l) \; \delta(S_z + L_z - J_z). \nonumber
\end{equation}
These classical distributions are continuous, though their quantum 
counter-parts are intrinsically discrete. 
To construct a meaningful quantum-classical comparison 
it is useful to discretize the classical distrbutions 
by integrating the continuous probabilities 
over intervals of width $\hbar =1$ centered on 
the quantum eigenvalues. 
In the case of the component $L_z$, 
the quantum probability $P_{L_z}(m_l)$ is then associated with 
the classical probability of finding $L_z$ in the 
interval $[m_l-1/2,m_l+1/2]$. This is given by  
\begin{equation}
\label{eqn:pclz}
	P_{L_z}^c(m_l)  = \int_{m_l - 1/2}^{m_l + 1/2} P(L_z). 
\end{equation}
Similarly, in the case of $J_z$, we compare each quantum $P_{J_z}(m_j)$ with 
the {\em discrete classical} probability, 
\begin{equation}
\label{eqn:pcjz}
P_{J_z}^c(m_j) = \int_{m_j - 1/2}^{m_j + 1/2} P(J_z).
\end{equation}

In the following discussion of the numerical results we will 
emphasize that, for chaotic states, 
the steady-state shape of the quantum and classical 
distributions should be compared 
with the marginal distributions derived from the microcanonical state. 
Our model  
is non-autonomous, but the spin magnitudes are conserved. The 
appropriate {\em classical} microcanonical measure is  
a constant on the accessible manifold ${\mathcal P} = S^2 \times S^2$.  
This follows 
from the usual equilibrium hypothesis that all accessible microstates 
are equiprobable, where equiprobability is defined 
with respect to the invariant measure (\ref{eqn:measure}). 
This microcanonical density projected onto the $L_z$-axis 
produces the discrete, flat distribution, 
\begin{equation}
\label{eqn:pmclz}
P_{L_z}^{mc}(m_l)=(2l+1)^{-1}. 
\end{equation}
However, projected along $J_z$, 
the microcanonical distribution is not flat, but has a tent-shape, 
\begin{eqnarray}
\label{eqn:pmcjz}
P_{J_z}^{mc}(m_j) & =  &  \frac{l+s+1-|m_j|}{(2s+1)(2l+1)} 
\;\;\; {\mathrm f} {\mathrm o}{\mathrm r} \;\;\;  
|m_j| \geq l-s  \nonumber \\ 
& = &  \frac{1}{2l+1}  \;\;\; {\mathrm f}{\mathrm o}{\mathrm r} 
\;\;\; |m_j| \leq l-s. 
\end{eqnarray}
In quantum mechanics, the equiprobability hypothesis 
implies that the appropriate microcanonical state 
is an equal-weight mixture. This microcanonical state, sometimes 
called a random state, is  
proportional to the identity in the full Hilbert space 
${\mathcal H} = {\mathcal H}_s \otimes {\mathcal H}_l$. 
It produces the same projected 
microcanonical distributions, 
{\it i.e}.\ (\ref{eqn:pmclz}) for $L_z$ 
and (\ref{eqn:pmcjz}) for $J_z$, as the classical microcanonical state. 

\subsection{Mixed Regime Chaos}

We consider first a classical parameter regime ($\gamma = 1.215$, 
$r=1.1$, and $a=5$) for which the kinematically 
accessible phase space ${\mathcal P}$ is highly mixed. 
The chaotic region appears to be connected (all chaotic 
initial conditions have the same 
largest Lyapunov exponent $\lambda_L = 0.04$) and covers 
about half of the kinematic surface. A projection of only the 
chaotic initial conditions onto the plane spanned by $S_z$ and $L_z$ 
reveals large regular islands surrounding the stable parallel fixed points 
$(\pm S_z, \pm L_z)$, with chaotic regions spreading out from 
the unstable anti-parallel fixed points $(\pm S_z,\mp L_z)$.  
A similar projection of the regular initial conditions 
shows points not only clustered about the parallel fixed points but 
also spread along the axis $\tilde{S}_z = \tilde{L}_z$. 

\begin{figure}[!t]
\vspace{0.5in} 
\hspace{0.25in}
\input{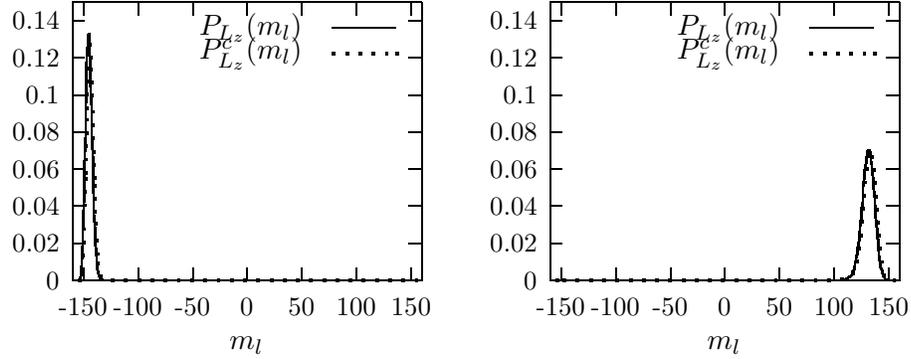}
\vspace{-0.3in}
\caption{
Quantum and classical probability distributions for $L_z$ with $l=154$  
in chaotic zone of mixed regime ($\gamma=1.215,r=1.1,a=5$). 
The dots are visible because they are shifted to 
the right by half of their width. 
The figure on the left is the initial state ( $n=0$) 
and that on the right is at time-step $n=6$.
}
\label{g1.215.plz.140.00.06}
\end{figure}

\begin{figure}[!t]
\vspace{0.5in} 
\hspace{0.25in}
\input{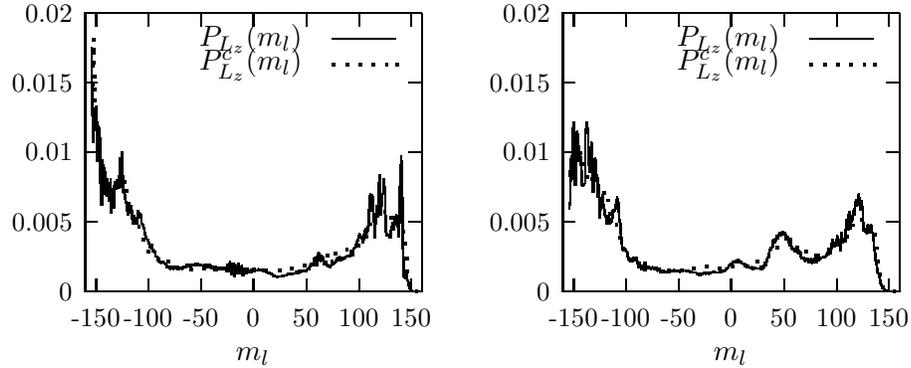}
\vspace{-0.3in}
\caption{Same as Fig.\ \ref{g1.215.plz.140.00.06} but for time-steps 
$n=99$ on the left and $n=100$ on the right.  Both quantum and 
classical distributions have reached the system dimension 
and are relaxing towards equilibrium.}
\label{g1.215.plz.140.100}
\end{figure}

\begin{figure}[!t]
\vspace{0.5in}
\hspace{0.25in}
\input{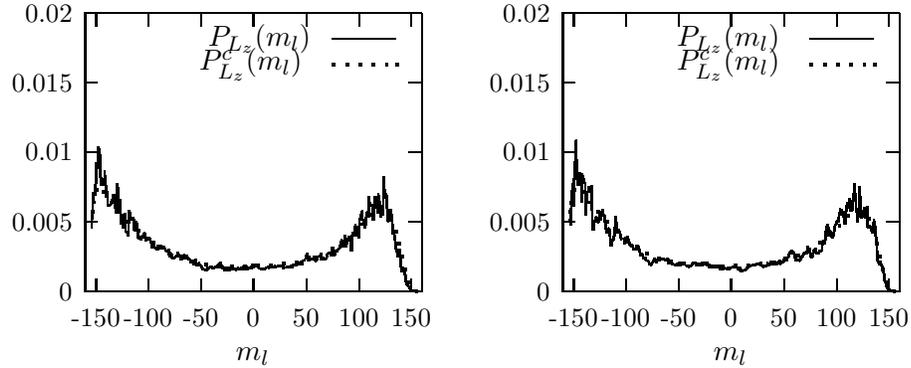}
\vspace{-0.3in}
\caption{Same as Fig.\ \ref{g1.215.plz.140.00.06} but for $n=199$ on the 
left and $n=200$ on the right.
The quantum distribution is fluctuating about a classical steady-state.  
}
\label{g1.215.plz.140.200}
\end{figure}

We now consider the time-evolution of quantum and classical states 
concentrated in the chaotic zone near one of the unstable anti-parallel 
fixed points,  with initial centroids directed along 
$\theta(0)=(20^o,40^o,160^o,130^o)$. 
The quantum dynamics are calculated using quantum numbers $s=140$ 
and $l=154$.  
As shown in Fig. \ref{g1.215.plz.140.00.06}, at early times 
both the quantum distribution $P_{L_z}(m_l)$ (solid line) 
and the corresponding classical distribution $P_{L_z}^c(m_l)$ (dots)
remain well-localised. Their initial differences are not 
distinguishable on the scale of the figures. 
(The dots are shifted to the right by half of their width.) 
By time-step $n=20$ both quantum and classical 
distributions have broadened to the system dimension and 
begin to exhibit noticeable differences. 
As shown in Fig.\ \ref{g1.215.plz.140.100}, 
around $n=100$ the distributions have begun to settle close 
to an equilibrium shape. 
In Fig.\ \ref{g1.215.plz.140.200} the successive 
time steps $n=199$ and $n=200$ show that,  although both the 
quantum and classical distributions have relaxed very close 
to the same equilibrium distribution, the quantum distribution 
exhibits rapidly oscillating fluctuations about the classical 
steady-state. 

\begin{figure}[!t]
\vspace{0.5in}
\hspace{0.25in}
\input{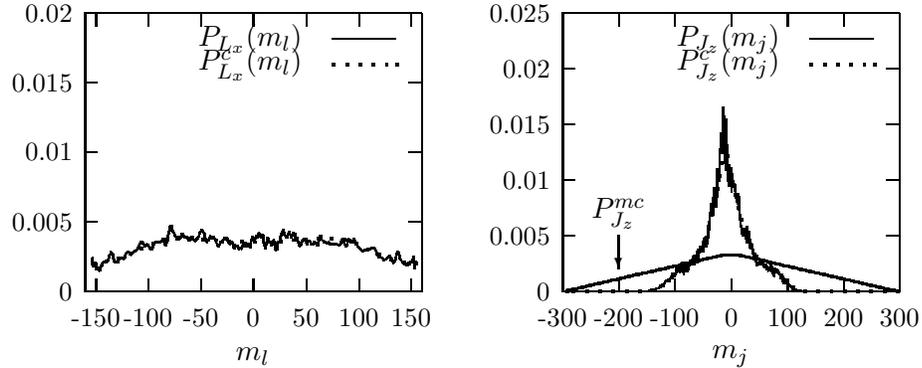}
\vspace{-0.3in}
\caption{
Same as in the previous figure, but for 
$P_{L_x}(m_l)$ on the left and 
$P_{J_z}(m_j)$ on the right, at time-step $n=200$.  
Both $P_{J_z}(m_j)$ and $P_{J_z}^c(m_j)$ 
are localised relative to the 
projected microcanonical distribution $P_{J_z}^{mc}(m_j)$.   
}
\label{g1.215.plx.pjz.140}
\end{figure}

Both the quantum and classical 
equilibrium distributions (projected along $L_z$) 
show significant deviation from the 
microcanonical distribution  (\ref{eqn:pmclz}).
This is also true of the distribution projected 
along $L_x$, which has a different non-uniform 
equilibrium distribution than 
that observed when projecting onto $L_z$ (see the 
left box of Fig.\ \ref{g1.215.plx.pjz.140}). 
Uniform marginal distributions would be expected 
if the classical mapping was mixing, in which case arbitrary 
initial densities (with non-vanishing measure) would relax to  
the microcanonical distribution. Since the accessible 
kinematic surface has large KAM surfaces in this parameter 
regime, the coarse-grained {\em classical} equilibrium 
distributions are not expected to be flat.  
An unexpected feature of the results is the observation 
that the shape of the equilibrium 
{\em quantum} distributions so accurately reflects the 
details of the KAM structure  in the classical phase space. 
This feature is most striking in the case of the distributions 
projected along $J_z$ (see the right box 
in Fig.\ \ref{g1.215.plx.pjz.140}).
The steady-state quantum and classical 
probability distributions $P_{J_z}(m_j)$ and $P_{J_z}^c(m_j)$ 
are both sharply peaked about $m_j = 0$. 
This equilibrium shape is much more sharply peaked 
than the tent-shape of the 
projected {\em microcanonical} distribution, 
$P_{J_z}^{mc}(m_j)$, given by (\ref{eqn:pmcjz}) and 
also plotted in the right box in Fig.\ \ref{g1.215.plx.pjz.140}. 
The important point is that the 
additional localization of the {\em quantum} distribution 
can be understood from a standard 
fixed-point analysis of the {\em classical} map \cite{EB00a}:    
the presence of KAM surfaces arising due to the stability 
of the parallel fixed points 
prevents the chaotic classical spins 
from aligning in parallel along the $z$-axis. 
Most remarkably, we find that 
the steady-state quantum distributions accurately reproduce this  
parameter-dependent structure of the mixed classical phase space 
even for much smaller quantum numbers. We examine 
how the accuracy of this correspondence 
scales with the quantum numbers in section \ref{ch4:scaling}.

\subsection{Regime of Global Chaos}

If we hold $a=5$ and $r=1.1$ fixed and increase the coupling strength 
to the value $\gamma=2.835$,  
then all four of the fixed points mentioned above 
become unstable \cite{EB00a}. 
Under these conditions less than $0.1\%$ of the surface ${\mathcal P}$
is covered with regular islands; the remainder of the surface 
produces a connected chaotic zone with 
largest Lyapunov exponent $ \lambda_L = 0.45 $.  We will sometimes 
refer to this parameter regime as one of {\em global chaos} since the 
kinematically accessible phase space is predominantly chaotic.  

\begin{figure}[!t]
\vspace{0.5in}
\hspace{0.25in}
\input{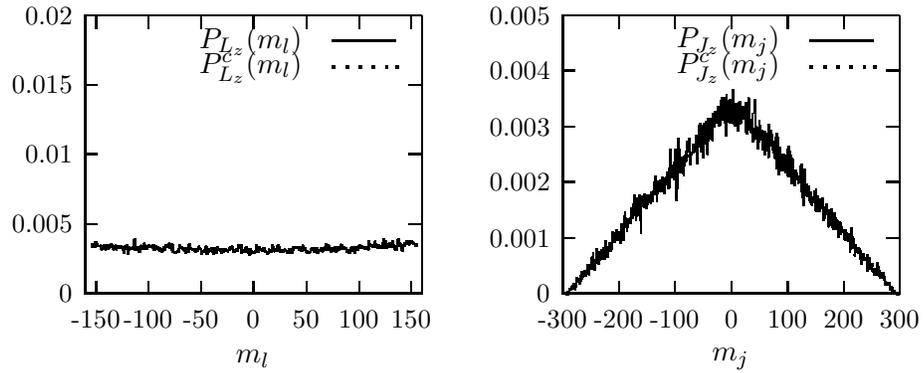}
\vspace{-0.3in}
\caption{
The equilibrium shapes of $P_{L_z}(m_l)$ 
and $P_{J_z}(m_j)$ at time-step $n=50$ with $l=154$ 
for a state launched in the global chaos regime 
($\gamma=2.835,r=1.1,a=5$). 
The quantum distributions exhibit small rapidly 
oscillating fluctuations about the 
projected microcanonical distributions. 
The classical distributions
are not visible since the points 
lie within the 
fluctuating quantum data.  
}
\label{g2.835.plz.pjz.140.50}
\end{figure}

The dynamics of the classical and quantum distributions are much simpler 
in this regime. We find that initially localised 
distributions, launched from 
arbitrary initial conditions, relax to the microcanonical 
distribution on a very short time-scale. To demonstrate this, 
we consider the dynamics of 
an initial quantum state with $s=140$ and $l=154$, and a 
corresponding classical 
density, launched from $\theta(0) = (20^o, 40^o, 160^o, 130^o)$. 
Though the initial distributions are the same as in the mixed regime,  
by time-step $n \simeq 6$ 
the quantum and classical distributions 
have already spread to the system dimension and begin 
to exhibit noticeable 
differences. By time-step $n \simeq 12$
both distributions have relaxed 
very close to the microcanonical distributions.  
We plot the equilibrium quantum and classical 
projected distributions $P_{L_z}(m_j)$ and $P_{J_z}(m_j)$ 
in Fig.\ \ref{g2.835.plz.pjz.140.50} for time-step $n=50$.
The projected {\em classical} distributions are nearly indistinguishable 
from the microcanonical forms, $P_{L_z}^{mc}(m_j)$ and $P_{L_z}^{mc}(m_j)$,  
and  the {\em quantum}
distributions again exhibit small fluctuations 
about the classical distributions. 
We have found that these equilibrium 
quantum-classical differences asymptote to  
a non-vanishing minimum when the measure of KAM surfaces becomes negligible. 
These minimum quantum fluctuations 
reflect characteristic deviations from the microcanonical state that arise 
because the equilibrium quantum state is a sequence of pure states, 
whereas the microcanonical state corresponds to a random mixture. 


\clearpage
\section{Rates of Relaxation to Equilibrium}

In order to characterize the time-scale of relaxation 
to equilibrium it is convenient to study the time-dependence 
of a scalar 
measure that is sensitive to deviations from the equilibrium 
state.  A conventional indicator of this rate of 
approach to equilibrium is the coarse-grained entropy,  
\begin{equation}
\label{eqn:hb}
	H = - \sum_i P_i \ln P_i. 
\end{equation} 
Here the $\{P_i\}$ stand for the quantum probabilities associated 
with projectors onto some basis of microstates ({\it e.g}.\ the projected 
distributions discussed in the previous section). 
The sum (\ref{eqn:hb}) is a standard measure of the 
information contained in a probability distribution 
and is sometimes called the Shannon entropy.  

The Shannon entropy has a number of useful properties. 
First, unlike the von Neumann entropy ${\mathrm Tr} [ \rho \ln \rho]$, 
the Shannon entropy is basis-dependent. It reduces to the von Neumann 
entropy if the `chosen' basis diagonalizes the state operator. 
However, this basis, or, more precisely, the set of projectors onto the 
(time-dependent) spectral decomposition of the state operator, 
does not necessarily correspond to a 
set of classically meaningful observables.   
Our main interest is to examine correspondence 
at the level of classical dynamical variables, so we consider 
probabilty distributions associated with projectors onto the 
eigenstates of classically well-defined operators.  
The classical counterparts to these probability distributions 
are associated  with some fixed partioning of the phase space  
into cells of width $\hbar$ along the axes of the associated dynamical 
variable. 

Second, whereas the von Neuman entropy 
of the total system is constant in time 
(${\mathrm Tr} [ \rho \ln \rho]=0$ since 
$\rho = |\psi \rangle \langle \psi |$), 
the basis-dependent Shannon entropy 
may have time-dependence even if the quantum state is pure. 
Thus (\ref{eqn:hb})  may be applied to examine the rate of relaxation 
of either {\em pure} or {\em mixed} quantum states. 
It is in this sense that we use the term {\em relaxation},  
although the time-evolution is unitary in the quantum model 
(and volume-preserving in the classical model). 

Given some fixed partioning of the phase space, if a 
classical state remains evenly spread through the phase space cells 
it occupies, and  
spreads through the phase-space exponentially with time, then 
an entropy like (\ref{eqn:hb}) 
should grow linearly with time. 
In this section we show that this argument 
holds approximately also for quantum states  
launched from a classically chaotic region of phase space. 
The actual rate of relaxation of the quantum states  
is accurately predicted by the classical entropy even for small 
quantum numbers. 

\begin{figure}[!t]
\begin{center}
\input{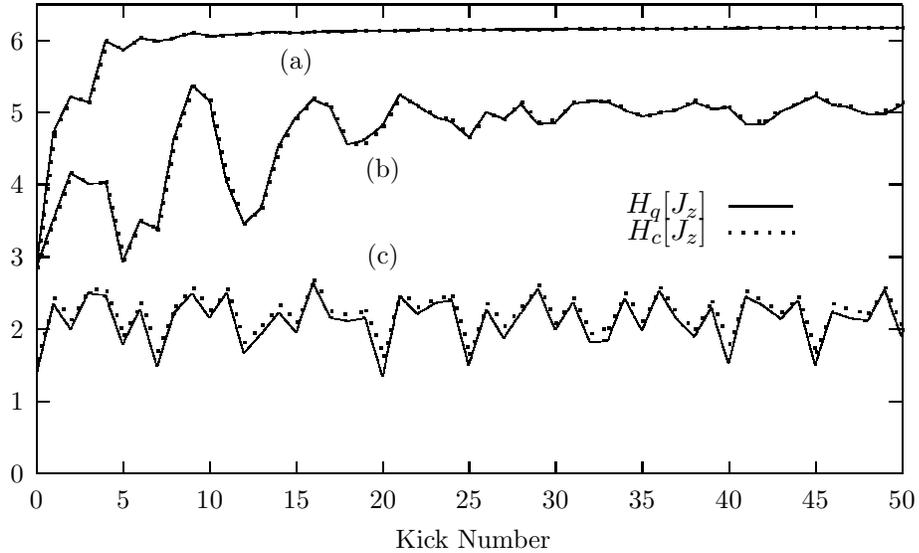}
\caption
[Comparison of the quantum and classical entropies 
for $s=140$ and $l=154$ in 
(a) regime of global chaos ($\gamma=2.835$); (b) chaotic zone of 
the mixed regime ($\gamma=1.215$); 
(c) regular zone of the mixed regime ($\gamma=1.215$).]
{
Comparison of the quantum and classical entropies 
$H[J_z] = - \sum_{m_j} P_{J_z}(m_j) \log  P_{J_z}(m_j)$ 
for $s=140$ and $l=154$ in 
(a) regime of global chaos ($\gamma=2.835$); (b) chaotic zone of 
the mixed regime ($\gamma=1.215$); 
(c) regular zone of the mixed regime ($\gamma=1.215$).
}
\label{HJz.294}
\end{center}
\end{figure}

We demonstrate this behaviour by first considering the 
quantum entropy $H_q[J_z]$ of 
the probabilities associated with the eigenvalues $m_j$ 
of $J_z$, {\it i.e}.\ the probabilities defined in (\ref{eqn:pjz}).
The corresponding classical entropy, $H_c[J_z]$, is calculated using 
the discrete classical probabilities (\ref{eqn:pcjz}).  
In Fig.\ \ref{HJz.294} we compare the 
time-development of the quantum and classical entropies 
using quantum numbers $s=140$ and $l=154$. 
For these quantum numbers,  
the microcanonical ({\it i.e}.\ maximum) value of 
the entropy is $H^{mc}[J_z]=6.2$. 
In case (c), corresponding to 
a regular zone of the mixed regime 
($\theta(0) =(5^o,5^o,5^o,5^o)$,$\gamma=1.215$), we actually see 
the greatest amount of difference between the quantum 
($H_q[J_z]$) and classical ($H_c[J_z]$) entropies. 
$H_q$ exhibits a quasi-periodic 
oscillation about its initial value whereas for $H_c$ these oscillations 
eventually dampen. 
For smaller quantum numbers, and thus broader initial states,  
$H_c$ dampens much more rapidly although $H_q$ continues to exhibit 
a pronounced quasi-periodic behaviour.  
In case (b), 
with initial centroid $\theta(0)= (20^o,40^o,160^o,130^o)$ 
set in a chaotic region of the equally mixed regime, 
both $H_q$ and $H_c$ oscillate about an initially increasing average
before relaxing towards a constant value that lies well below 
the microcanonical maximum $H^{mc}[J_z] = 6.2$. 
This saturation away from the maximum is expected  
in the classical model since a large fraction of the kinematic surface 
is covered with regular islands and remains inaccessible.  
In case (a), corresponding  
to the regime of global chaos ($\gamma =2.835$) 
and with the same initial state as (b), 
the entropies are nearly identical. Both grow much more quickly 
than in the mixed regime case, roughly linearly, 
until saturating very near the maximum value. 

\begin{figure}[!t]
\begin{center}
\input{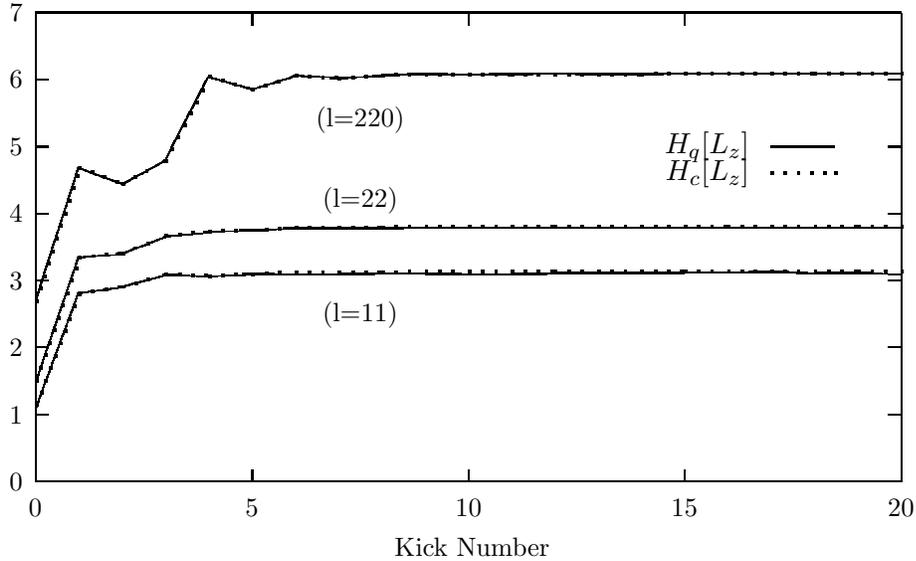}
\caption[Comparison of the quantum and classical subsystem entropies 
for increasing system sizes in the global chaos regime of 
Fig.\ \ref{HJz.294}.]{
Comparison of the quantum and classical subsystem entropies 
$H[L_z] = - \sum_{m_j} P_{L_z}(m_l) \ln  P_{L_z}(m_l)$ 
for increasing system sizes in the global chaos regime of 
Fig.\ \ref{HJz.294}.
}
\label{Hent.Lz.11.22.220}
\end{center}
\end{figure}

The quantum entropy is very well approximated by its  
classical counterpart also for smaller quantum numbers. 
In Fig.\ \ref{Hent.Lz.11.22.220} we display the growth rates 
of the quantum and classical entropies of the probabilty 
distributions associated with the observable $L_z$ for 
three sizes of quantum system ($l=11$, $l=22$, $l=220$) using 
the same parameters and initial condition as for 
data-set (a) in Fig.\ \ref{HJz.294}. 
In each case the quantum entropy is essentially identical 
to the corresponding classical entropy. The initial rate 
of growth is similar in each case, roughly linear, 
and of order the Lyapunov exponent, $\lambda_L =0.45$.  

These results extend previous work demonstrating 
that the widths of quantum states grow exponentially with 
time, on average, until saturation at the system dimension 
\cite{Fox94b,Ball98,EB00a}. Modulo the small quantum 
fluctuations, for both quantum and classical 
models we find that the subsequent relaxation to an 
equilibrium  configuration occurs on the time-scale,  
\begin{equation}
t_{rel} \sim t_{sat} + {\mathcal O}(\lambda_L^{-1}),  
\end{equation}
where $t_{sat} \simeq \lambda_{w}^{-1} \ln l$  estimates 
the time it takes the initial coherent state 
to reach the system dimension.  
The exponent $\lambda_{w}$ is the exponent governing  
the growth rate of the state width \cite{EB00a}.   
The last term ${\mathcal O}(\lambda_L^{-1})$ approximates  
the additional time-required for the state 
to become more or less uniformly spread 
over the accessible phase space. 
In predominantly chaotic regimes we have found that  
$\lambda_w \simeq \lambda_L$, 
though in mixed regimes $\lambda_w$ is generally a few times 
larger than the largest Lyapunov exponent. 

\clearpage
\section{Time-dependence of Quantum-Classical Differences} 

Before examing the scaling of quantum-classical differences 
with increasing quantum numbers, it is useful to determine 
first their time-domain characteristics 
under the different types of classical behviour. 
Previous work has shown that quantum-classical differences 
for low-order moments, though initially 
small, grow exponentially with time when the classical motion is chaotic  
\cite{Ball98,Ball00} until the states approach the system size 
\cite{EB00a}. On this saturation 
time-scale those quantum-classical differences reach   
their maximum magnitude, 
but surprisingly this maximum was small, ${\cal O}(\hbar)$. 
More specifically, it did not scale with the quantum numbers. 
Of course, two distributions can be altogether different even when   
the differences between their means and variances are quite small, 
and therefore it is useful to examine the differences between 
the quantum and classical states in a more sensitive way. 

In this section we examine the time-dependence of 
bin-wise deviations 
between the quantum and classical probability distributions.  
For the observable $L_z$ 
this indicator takes the form,  
\begin{equation}
\label{eqn:siglz}
\sigma[L_z] = \sqrt{ {1 \over (2l+1)} \sum_{m_l = -l}^l  
[ P_{L_z}(m_l) - P_{L_z}^c(m_l) ]^2 }.
\end{equation}
This standard deviation estimates typical 
quantum-classical differences on the scale of $\hbar$ along 
the $L_z$-axis. Each interval is centered on a quantum eigenvalue.
The $ P_{L_z}^c(m_l)$ correspond to 
a measurement, or coarse-graining, of the classical density 
on an extremely fine-scale. 

\begin{figure}[!t]
\begin{center}
\input{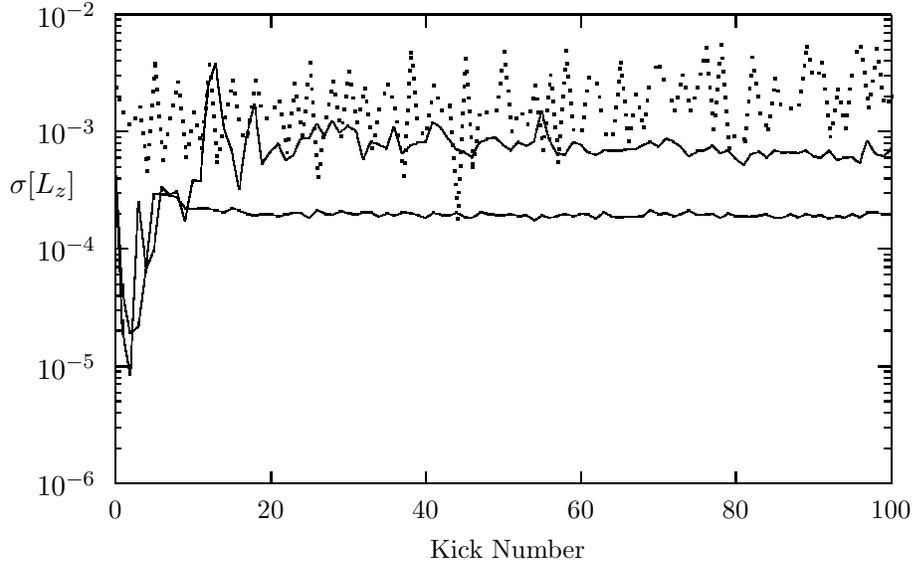}
\caption{
Time dependence of the standard deviation $\sigma[L_z]$ of 
quantum-classical differences (\ref{eqn:siglz})
for states launched from a regular zone (dotted line) 
of the mixed regime ($\gamma=1.215$), from a chaotic zone of the same 
mixed regime (middle solid line), 
and from the regime of global chaos (lower solid line,$\gamma=2.835$).  
The initial discrepancy 
is relatively large, but quickly decreases, and then increases until 
reaching an asymptotic equilibrium value. This occurs more slowly for 
the mixed regime case, for which the asymptotic value is also larger. 
In all cases $s=140$ and $l=154$. 
}
\label{D2.Lz.154}
\end{center}
\end{figure}

In Fig.\ \ref{D2.Lz.154} we examine the time dependence of $\sigma[L_z]$ for 
the same three classical sets of parameters and initial conditions 
displayed in Fig.\ \ref{HJz.294}. 
The initial value of $\sigma[L_z]$ is generally 
not zero since it is not possible to match all the marginal distributions 
exactly in the case of the SU(2) coherent states \cite{EB00a}. 
The actual magnitude of the initial discrepancy depends 
on the angle between the axis of measurement, {\it e.g}.\ $L_z$, and 
the direction of polarization of the initial state.  
For both chaotic states the differences initially decrease from their 
angle-dependent value  
and then increase until saturation at a steady-state value. 
This steady-state value is  reached much later 
in the mixed regime (upper solid line), than in 
the global chaos regime (lower solid line).  
It occurs on the time-scale, $t_{rel}$, on which    
the underlying distributions have reached their steady-state 
configurations (modulo the quantum fluctuations). 

As shown in Fig.\ \ref{D2.Lz.154}, the quantum-classical differences are 
actually largest for the regular state (dotted line) of the mixed regime 
($\gamma=1.215$) at both early and late times (relative to the 
relaxation time-scale). 
The steady-state magnitude of the differences for the global chaos 
regime ($\gamma=2.835$) 
is significantly smaller than the typical magnitude 
for the mixed regime. 
However, for larger values of the classical 
perturbation strength $\gamma$,
this average steady-state magnitude does not decrease further   
(with the quantum numbers held fixed) but has reached 
a non-vanishing minimum. 
The magnitude of the minimum steady-state fluctuations,  
$\sigma_{L_z} \simeq 2 \times 10^{-4}$, should be 
compared with a typical magnitude 
of the quantum and classical distributions, 
$ P_{L_z}(m_l) \simeq 3 \times 10^{-3}$.  
In the following section 
we examine how these fluctuations scale with increasing 
quantum numbers.

\begin{figure}[!t]
\begin{center}
\input{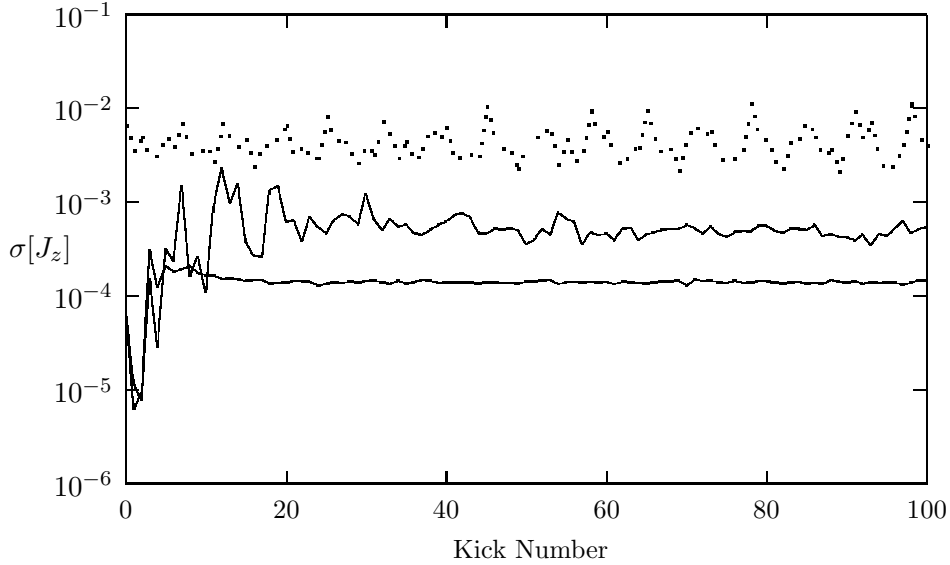}
\caption{
Same as Fig.\ \ref{D2.Lz.154} but for the standard deviation of 
quantum-classical differences of the total angular momentum, 
$\sigma[J_z]$, given by (\ref{eqn:sigjz}).
}
\label{D2.Jz.154}
\end{center}
\end{figure}

Above we have considered quantum-classical differences 
for observables (projectors onto subspaces) associated with the  
factor space ${\mathcal H}_l$. 
In this factor space the state is initially pure but 
becomes mixed as a result of dynamical interacions with the 
other subsystem. 
It is interesting to check if the dynamical 
behaviours of the differences are an artefact of this dynamical mixing.    
Therefore we consider also bin-wise quantum-classical differences 
for an observable ($J_z = S_z + L_z$) that acts non-trivially 
on the full Hilbert space ${\mathcal H} = {\mathcal H}_s 
\otimes {\mathcal H}_l$. The quantum 
state in the full Hilbert space remains pure throughout 
the time-evolution.  
We construct the same standard deviation of the bin-wise 
differences between the quantum and 
classical probability distributions as above, 
\begin{equation}
\label{eqn:sigjz}
\sigma[J_z] = \sqrt{ {1 \over [2(s+l)+1)]} \sum_{m_j}   
[ P_{J_z}(m_j) - P_{J_z}^c(m_j) ]^2 }, 
\end{equation}
where $m_j \in \{l+s, l+s-1, \dots, -(l+s) \}$. 
In Fig.\ \ref{D2.Jz.154} we compare $\sigma[J_z]$ in 
the same three classical regimes examined in Fig.\ \ref{D2.Lz.154}. 
Once again the regular state (dotted line) exhibits the 
largest quantum-classical differences, and the differences for 
both chaotic states (middle and lower solid line) 
grow to a steady-state value on the time-scale at which 
the underlying distributions relax to their equilibrium configurations.
As above, the average value of the differences for 
$\gamma=2.835$ (lower solid line) correspond to a 
non-vanishing minimum, that is,  
the average value does not 
noticeably decrease for larger values of $\gamma$. 
The minimum quantum fluctuations are again small 
when compared with the average height of the probability 
distribution, $[2(s+l) +1]^{-1} \simeq 2 \times 10^{-3}$. 

For $\gamma \simeq 2.835$, the measure of regular islands 
is already very close to zero and the classical system 
is nearly ergodic on ${\mathcal P}$.  Similarly, the quantum 
state is no longer constrained by any invariant classical 
structures but spreads {\em almost} evenly about the accessible 
Hilbert space. We find that the standard deviations of the 
quantum fluctuations 
that account for the equilibrium quantum-classical differences 
approach a non-vanishing minimum as the classical dynamics 
approach ergodicity on ${\mathcal P}$.
These equilibrium differences can not vanish (for fixed quantum numbers) 
because the total quantum state remains pure under the unitary dynamics,  
whereas the microcanonical equilibrium corresponds to an 
equal-weight mixture.  


\clearpage
\section{Correspondence in the Classical Limit}
\label{ch4:scaling}

We now turn to an examination of the classical limit, 
${\mathcal J}/\hbar \rightarrow \infty$, where ${\mathcal J}$ 
is characteristic system action. Since the quantum-classical 
differences grow to their largest values once the states 
have spread to the system size and subsequently fluctuate about this 
magnitude, we will examine the scaling of the differences 
in this late time-domain, that is,  
when the states 
have relaxed close to their equilibrum configurations. 
Moreover, these scaling results will then 
complement previous work 
that has focussed on correspondence at early times \cite{EB00a}, 
in the Ehrenfest regime when the states 
are narrow relative to the system dimensions. 

We wish to determine if the standard deviation of the 
quantum-classical differences (defined in the previous 
section) decreases in magnitude 
with increasing quantum numbers. 
When comparing models with increasing quantum 
numbers, we hold the width of each probability 
bin fixed (at $\hbar=1$). Since the number of bins 
will increase with the quantum number, it follows that  
the height of the probability distribution in a 
given bin will also decrease. Consequently, 
we construct a scale-independent, or relative, measure of 
the bin-wise quantum fluctuations by taking the ratio of the standard 
deviation to the average value of the probability distribution. 
For the observable $L_z$ this takes the form,   
\begin{equation}
\label{eqn:rlz}
R [L_z(n)] =  \frac{\sigma [L_z(n)]}{{\overline P}_{L_z}}  = N_l \; \sigma[L_z(n)]. 
\end{equation}
where the average value $  {\overline P}_{L_z} = 1/(2l+1) = 1/N_l$. 
If this relative measure approaches zero in the classical limit then 
the quantum probabilty distribution converges  
to the corresponding classical one in that limit. 

\begin{figure}[!t]
\begin{center}
\input{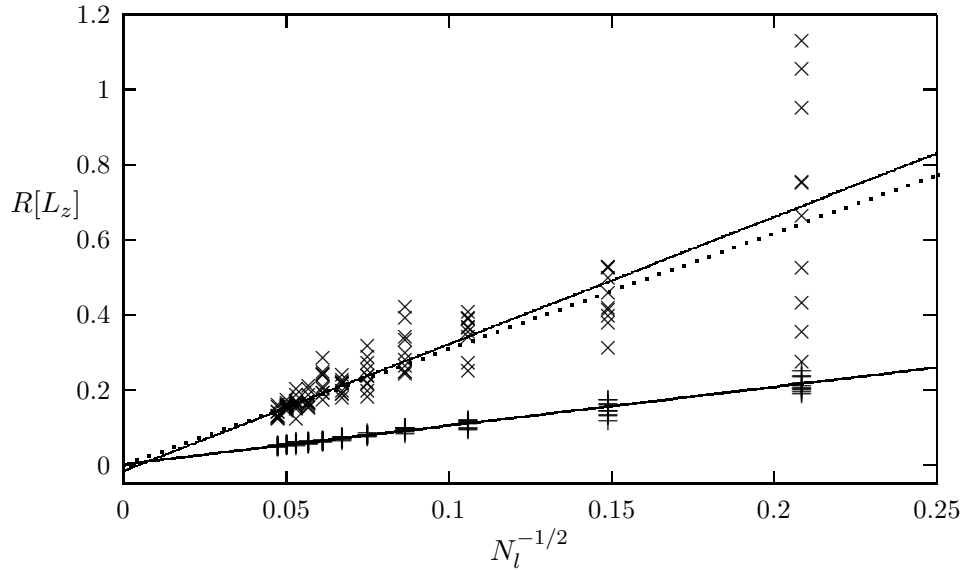}
\caption{
Scaling of relative quantum-classical differences (\ref{eqn:rlz}  
in the equilibrum time-domain versus increasing system size. 
Scatter of crosses corresponds to time-steps  $191 \leq n \leq 200$,  
for a state launched in 
the chaotic zone of the mixed regime ($\gamma=1.215$). 
Scatter of plus signs corresponds to time-steps  $41 \leq n \leq 50$,  
for a state launched in the global chaos regime. 
Data sets in both of these regimes are consistent with the scaling law 
$R \simeq {N_l}^{-1/2}$, where $N_l =2l+1$.  
}
\label{Rvssqrtnl}
\end{center}
\end{figure}

In Fig.\ \ref{Rvssqrtnl} we consider typical equilibrium 
values of $R[L_z(n)]$ plotted against $1/\sqrt{N_l}$.  
We study the scaling using $N_l$ because it 
is equal to the dimension of the factor space ${\mathcal H}_l$ and 
it is also proportional to the subsystem size $N_l \simeq 2 |{\bf L}|$. 
We first consider a state launched in the global chaos regime 
($\gamma=2.835$, $r=1.1$), with initial condition 
$\theta(0) = (45^o,70^o,135^o,70^o)$. The scatter of plus 
signs for each $N_l = 2l+1$ value 
corresponds to time-steps $n$ such that 
$41 \leq n \leq 50$.  These time-step values are 
chosen because they occur well after the relaxation time 
$t_{rel} \simeq 6$. In this regime the data exhibits very 
little scatter. A least-squares 
fit to the curve $ R = A / \sqrt{N_l} + B $ yields a value for the 
intercept $B$ that is 
consistent with zero ($ B =0.001  \pm 0.001$) and a slope of order 
unity ($A = 1.032 \pm 0.02$). 
An intercept consistent with zero implies   
that quantum-classical differences vanish in the classical limit, {\it i.e}.\ 
$P_{L_z}(m_l) \rightarrow P_{L_z}^c(m_l)$ as $l \rightarrow \infty$. 
This result is especially remarkable since we have considered 
the differences that arise given classical measurements  
which resolve the observable $L_z$ with the rather extraordinary 
precision of $\hbar=1$.  

We next consider a state launched from the chaotic zone of 
the mixed regime ($\gamma=1.215$, $r=1.1$) 
with $\theta(0) = (20^o,40^o,160^o,130^o)$.
The scatter of crosses in  Fig.\ \ref{Rvssqrtnl} corresponds to 
time-steps 
$191 \leq n \leq 200$ $n$, again chosen well after the relaxation time 
$ t_{rel}$ for the range of quantum numbers considered. 
The scatter of quantum-classical differences at each $N_l$ value  
is much more significant in this regime in which the 
equilibrium distributions reflect a much more complex 
phase space structure. 
However, the relative differences exhibit, on average, a similar dependence 
on the quantum numbers as in the predominantly chaotic regime. 
In this regime a least-squares fit to the curve  
$ R = A / {N_l}^{1/2} + B $ yields a slope of order 
unity ($A = 3.39 \pm 0.15$) but a negative 
value for the intercept ($B=- 0.017 \pm 0.009$) within 
two-standard deviations of zero. 
A negative intercept is not physically meaningful 
(since $R$ is a positive definite 
quantity) and we assume it arises as a consequence 
of the statistical scatter in the data.  
Also plotted is the curve $R =  C/N_l^{1/2}$, 
with slope $C =3.09 \pm 0.04$ also determined from a least-squares fit. 
Both fits are good, with reduced $\chi^2$ values of order unity. 

\begin{figure}[!t]
\begin{center}
\input{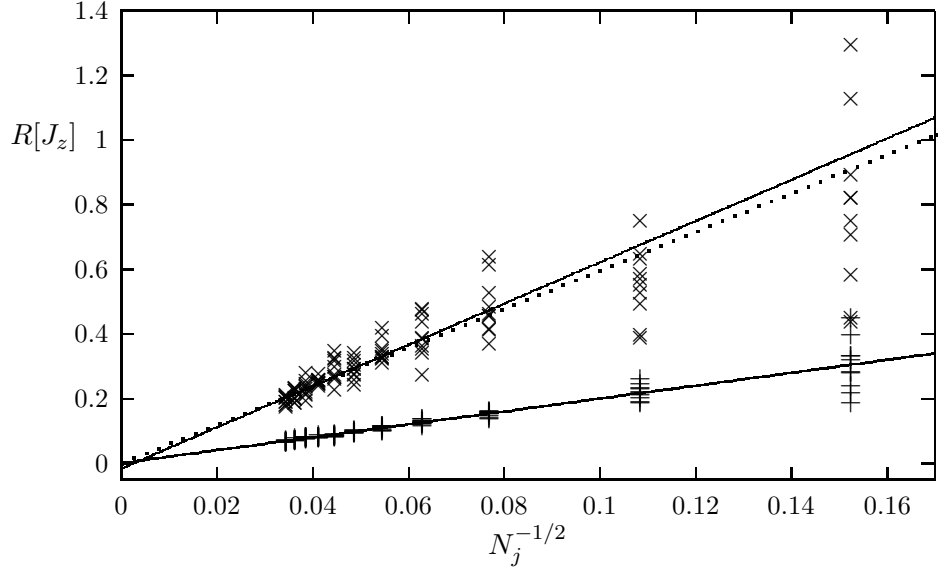}
\caption{
Same as Fig.\ \ref{Rvssqrtnl} but using $R[J_z]$, as given by 
(\ref{eqn:rjz}). Data sets in both types of chaotic regime  
are consistent with the scaling law 
$R \simeq {N_j}^{-1/2}$, where $N_j = 2(l+s)+1$.  
}
\label{Rvssqrtnj}
\end{center}
\end{figure}

As we noted in the last section, the subsystem 
states do not remain pure, because of dynamically induced  
entanglement between the subsystems. 
Since the subsystem state (\ref{eqn:redl}) 
in the factor space ${\mathcal H}_l$ 
is not pure, but highly mixed in the equilibrium time-domain, 
it is possible that the scaling with $N_l$ 
that we observe is related to the purity-loss from 
this entanglement. 
Consequently, 
it is useful to examine the scaling of the quantum-classical 
differences for the total spin $J_z$. 
The operator $J_z$ acts non-trivially in 
the full Hilbert space ${\mathcal H}$.
In this Hilbert space the system is described 
by a pure state vector at all times. 
In Fig.\ \ref{Rvssqrtnj} we consider the scaling of the ratio,   
\begin{equation}
\label{eqn:rjz}
R [J_z(n)] =  \frac{\sigma [J_z(n)]}{{\overline P}_{J_z}}  
= N_j \; \sigma[J_z(n)],  
\end{equation}
where ${\overline P}_{J_z}  = [2(s+l)+1]^{-1}= N_j^{-1}$ is the 
average value of either distribution,    
versus the dimension $N_j$.  
Here $N_j$ is the number of subspaces 
associated with distinct eigenvalues ($m_j$) of the quantum 
operator ($J_z$). In contrast with $N_l$, $N_j$ is not equal to 
the dimension of the corresponding Hilbert space, though it 
is a measure of the system size since $N_j \simeq 2 |{\bf J}|$. 
The parameters and initial conditions shown  in  Fig.\ \ref{Rvssqrtnj} 
are the same as in Fig.\ \ref{Rvssqrtnl}. 
The same fit procedure as above, but applied to the 
function $R =  A/N_j^{1/2} + B$,  yields a  value for $B$
that is again consistent with zero ($B= 0.00038 \pm 0.0016$) and a 
positive slope of order unity ($A =  2.00 \pm 0.04$)
in the predominantly chaotic regime (scatter of plus signs). 
Thus the relative standard deviation for $J_z$ 
also decreases as the square of the quantum numbers 
and fits to an intercept that is consistent with zero. This implies 
that the fluctuating quantum distributions 
approach the classical equilibrium, even for a few degree-of-freedom  
system, which is described at all times by a pure state.
In a chaotic state of the mixed regime (scatter of crosses), the 
fluctuations are larger, and the same fit 
procedure as above gives ($B = -0.016 \pm 0.012$, $A = 6.4 \pm 0.3$), 
where the negative value for $B$ lies within two standard deviations of 
zero and is presumed to result from the statistical 
scatter of the data.    
Also plotted is the equation $R =  C/N_j^{1/2}$, with $C =5.97 \pm 0.06$ 
determined from a least-squares fit. The fits to both equations 
are good, with reduced $\chi^2$ values of order unity.


\clearpage
\section{Discussion}

We have shown that, in classically chaotic regimes, initially 
localised quantum states relax 
to an equilibrium configuration that reflects the details of 
the classical phase-space structure. 
We find a remarkable degree of correspondence between the quantum and 
classical relaxation rates, even for small quantum numbers. 
Moreover, contrary 
to results obtained for the low-order moments \cite{Ball98,EB00a}, 
the degree of difference between  
the probability distributions is actually smaller for the chaotic 
states than the regular states. 

The equilibrium quantum distributions exhibit small rapidly oscillating  
fluctuations about the coarse-grained classical equilibrium. 
As the measure of regular islands on the classical manifold 
approaches zero, the quantum and classical equilibrium configurations  
approach their microcanonical forms, and the quantum fluctuations 
about the classical equilibrum approach a non-vanishing minimum. 
This minimum arises because we consider total quantum states 
that are pure, whereas the microcanonical configuration 
is produced by an equal-weight mixture. 

For the distributions associated with the 
subsystem observable ${\bf L}$, 
the standard deviation of these 
differences, relative to the average value, decreases as $N_l^{-1/2}$,  
where $N_l=2l+1 \simeq 2 |{\bf L}|$ 
is the dimension of the factor space, 
and becomes vanishingly small in the limit 
of large quantum numbers ({\it i.e}.\ large spins). 
These results suggest that 
correspondence with classical Liouville mechanics emerges 
in the classical limit for 
time-scales much longer than the Ehrenfest time.  

A great deal of recent work has emphasized that  
the loss of purity resulting from interactions 
with a quantum environment removes characteristic 
quantum effects and improves the degree of quantum-classical 
correspondence \cite{HB96,Zurek98a,Gong99}. 
While this is certainly the case 
for small quantum systems, it has been further argued 
that these {\em decoherence} effects 
must be taken into consideration to see the emergence of classical 
properties from quantum mechanics,  
even in the limit of large quantum numbers, 
if the classical motion is chaotic \cite{ZP95a,Zurek98b}. 

Since our model is comprised of interacting subsystems,  
initially separable pure states become entangled dynamically;  
the subsystem states (in each factor space) 
do not remain pure but become mixed. 
This entanglement process has 
an effect that is analogous to the process of decoherence. 
Hence one might suspect that 
the emergent classical behaviour that we have observed 
for the properties of the subsystem ${\bf L}$ 
may be strictly the result of a ``decoherence'' effect arising 
from entanglement with the other subsystem.
To address this possibility, we have 
considered also the quantum-classical differences that arise 
in the probability distributions for a total system observable, $J_z$. 
In this case the quantum observables are 
projectors onto subspaces of the full Hilbert space, rather 
than merely a factor space.  The quantum state 
in this full Hilbert space is not subject to any entanglement 
or decoherence and remains pure throughout the unitary time-evolution.  
We have found that the scale-independent standard deviations for 
these quantum-classical 
differences decrease as $1/\sqrt{N_j}$, where 
$N_j =2 (s+l) +1 \simeq 2 |{\bf J}|$ is a measure of the system size 
and $N = (2s+1)(2l+1)$ is the dimension of the Hilbert space.  
The bin-wise quantum-classical differences 
become increasingly difficult 
to observe, in the limit of large quantum numbers, even for 
system observables that are isolated from the effects of decoherence.
In this sense the process of decoherence 
is not {\em necessary} to produce quantum-classical  
correspondence in the classical limit. 







\chapter{Breakdown of Ehrenfest Correspondence}

\section{Introduction}

The quantum-classical correspondence principle expresses the 
view that the predictions of classical mechanics must emerge 
from quantum mechanics in the macroscopic limit.  
A typical statement of this principle is given by Messiah \cite{Messiah}: 
``In the limit $\hbar \to 0$, the laws of Quantum Mechanics 
must reduce to those of Classical Mechanics.'' 
Here the expression ``$\hbar \to 0$'' is shorthand notation for 
the limit where the characteristic 
system action is much larger than Planck's constant. 
Quantum-classical correspondence is {\it required} in the   
macroscopic limit if classical mechanics provides a 
valid approximation to the observed phenomena in this limit,  
whereas quantum mechanics provides correct  
predictions on all action scales. 
In order to test whether this principle holds 
in relevant physical situations, it is necessary to make 
more precise statements about which classical theory of mechanics, 
or, put differently,  
which classical properties, should be expected to emerge 
from the underlying quantum description.  
In this context, it is useful to introduce and distinguish 
two different 
interpretations of the quantum-classical correspondence principle. 

The first of these, which I shall call {\it Ehrenfest correspondence}, 
is the claim that the phase space {\it trajectories} described  
by Hamilton's equations of motion 
are required to 
emerge from quantum mechanics in the macroscopic limit.  
This approach to characterizing correspondence 
was originally devised by Ehrenfest \cite{Ehrenfest} and 
forms the basis for most textbook discussions of the correspondence 
principle \cite{Liboff, Sakurai, Merzbacher,Schiff,Messiah}. 
For a generic Hamiltonian 
system of the form $H= p^2/2m + V(q)$, this theorem states that  
the time-dependence of the expectation values 
$\langle q(t) \rangle = \langle \psi(t) | q | \psi(t) \rangle$ 
and $\langle p(t) \rangle= \langle \psi(t) | p | \psi(t) \rangle$, 
in an arbitrary quantum state $\psi(t)$, is prescribed 
by the differential equations, 
\begin{eqnarray}
\label{ch5:qdiff}
\frac{d \langle q(t) \rangle }{d t} & = &
\frac{\langle p(t) \rangle}{m} \nonumber \\
\frac{d \langle p(t) \rangle }{d t} & = & 
\langle F(q) \rangle. 
\end{eqnarray}
If the quantum state is sufficiently well-localised, then 
the approximation, 
\begin{equation}
\langle F(q) \rangle \simeq F(\langle q \rangle) 
\end{equation}
holds, and the differential equations 
for the quantum expectation values are then 
well approximated by Hamilton's equations of motion \cite{Goldstein}. 
A clear statement of this interpretation of the correspondence 
principle is articulated by Merzbacher \cite{Merzbacher}: 
``We require that 
the classical motion of a particle be approximated by the 
average behaviour of a wave packet with a fairly sharp 
peak and as precise a momentum as the uncertainty principle 
permits and that the expectation values of the dynamical variables,  
calculated for such a wave packet, satisfy the laws of classical 
mechanics.'' 

The second interpretation of the correspondence principle, 
which I will refer to as {\it Liouville correspondence}, 
is the weaker proposition that only the 
statistical properties of Liouville 
mechanics are required to 
emerge from the quantum description in the macroscopic limit. 
This view may be characterized by the condition that 
the quantum expectation values, $\langle A(q,p) \rangle$, for 
classically well-defined observables, 
should approach classical ensemble averages for the associated dynamical 
variables, $\langle A(q,p) \rangle_c$. In this correspondence 
picture, the classical ensemble averages are 
calculated from the prescription, 
\begin{equation}
	\langle A \rangle_c = \int dq \int dp A(q,p) \rho_c(q,p,t)
\end{equation}
where the classical density, $\rho_c(q,p)$,  
describes the possible configurations for a 
system with incompletely specifed phase space coordinates. 
This density evolves with time according to the 
Liouville equation, 
\begin{equation}
	\frac{\partial \rho(q,p,t)}{\partial t} = \{ \rho(q,p,t), H \}
\end{equation}
where $\{\cdot,\cdot \}$ is the Poission bracket. 
In the previous chapters I have examined the characteristics of 
Liouville correspondence in considerable detail. 
I have demonstrated, for the model system described in Chapter 2, 
that the statistical properties of Liouville mechanics 
emerge from quantum mechanics with a degree of approximation that 
improves as the system action increases. 
In particular, I examined this correspondence 
for quantum expectation values and classical 
ensemble averages associated with 
the system dynamical variables, their variances, and 
characteristic functions defined over extremely small intervals 
along the phase space axes corresponding to these variables. 
Even for the chaotic states of the model, the differences between 
quantum and Liouville mechanics for these observables 
have been shown to become vanishingly small 
in the limit ``$\hbar \to 0$'' over physically relevant time-scales. 
Most significantly, the correspondence with Liouville mechanics 
remained valid not just at early times (for well-localised quantum states), 
but also at late time 
(well after the width of the quantum state had grown to the system size). 
The conditions of Liouville correspondence are therefore satisfied 
in the macroscopic limit, 
in the sense that the observables of quantum mechanics are 
well approximated by the statistical predictions of classical 
Liouville mechanics, over physically accessible time-scales, even 
in the case of chaotic motion. 

In this chapter I will demonstrate that the conditions of 
Ehrenfest correspondence may not be satisfied in the macroscopic limit 
when the classical motion is chaotic.  
In particular, I will argue 
that Ehrenfest correspondence may fail on physically relevant 
time-scales for some chaotic 
macroscopic systems, and, therefore, that quantum mechanics 
may be unable to describe the observable, deterministic 
motion of some macroscopic bodies. The implications of this 
result for the interpretation of the quantum state will 
be drawn out in the discussion at the end of this chapter. 

%


This chaper is organized as follows. 
First I will present a theoretical argument indicating that  
the presence of chaos in general Hamiltonian models 
leads to macroscopic differences 
on a time-scale that grows only logarithmically with 
increasing system action.  Specifically, I will show 
the quantum state centroids deviate   
from the predicted Newtonian trajectory 
in an experimentally observable manner.  
This time-scale may be experimentally accessible for 
some real macroscopic systems, leading to  
an observable breakdown of Ehrenfest correspondence.  
I will then illustrate the features of this general argument 
in the specific case of the coupled spin system described in Chapter 2.
This example will also clarify how, contrary to a recent 
claim in the literature \cite{Zurek98b}, the effects of decoherence 
are unable to restore this anticipated breakdown of 
Ehrenfest correspondence. 
In the discusssion I will explain how, 
on the assumption that individual macroscopic systems conserve 
energy and 
remain well-localised over experimentally verifiable time-scales, 
this argument leads to the conclusion that quantum mechanics is unable 
to provide a valid description of macroscopic dynamics, and is 
therefore not a complete theory.  


\section{Ehrenfest Correspondence Conditions}

The conditions for Ehrenfest correspondence 
are summarised succinctly by Messiah \cite{Messiah}: 
``In order that this picture 
be satisfactory, it is necessary that: (a) the mean values follow 
the classical laws of motion to a good approximation; (b) the dimension 
of the wave packet be small with respect to the characteristic dimensions 
of the problem, and that they remain small in the course of time.'' 
It is useful to formulate these requirements as mathematical conditions. 

Let ${\bf x}_c =(q_c^1, \dots, q_c^N, p_c^1, \dots, p_c^N)$, 
denote the $2N$ phase space coordinates of an $N$ degree of freedom system, 
let $\langle {\bf x} \rangle = 
\langle \psi(t) | {\bf x}| \psi(t) \rangle $  denote 
the corresponding quantum expectation values, and let 
$\xi^i = ( \langle x^i \rangle - x_c^i )$ stand for the   
Ehrenfest differences.  
Further, let $\chi^i$ denote some prescribed set of thresholds 
that characterize 
adequate agreement between the quantum and classical predictions 
(e.g., determined by the resolution of the experiment measurements), and 
let $t_{obs}$ stand for a characteristic time-scale over which 
the system may be subject to experimental observation. 
I will use ${\cal J}$ to denote a characteristic action 
of a physical system, 
with the {\it macroscopic} limit characterized by a very large 
value of the system action relative to Planck's constant, 
${\cal J}/\hbar >> 1$. 
In order for Newtonian trajectories to emerge from 
the quantum description, the Ehrenfest differences must remain 
small relative to the prescribed threshold,  
\begin{equation}
\label{ch5:Ehrconda}
  | \xi^i(t)|  = | \langle x^i \rangle - x_c^i | < \chi^i
	\;\;\;\; {\mathrm f} {\mathrm o} {\mathrm r}  
\;\;\;\ t < t_{obs} \;\;\;  {\mathrm a} {\mathrm n} {\mathrm d} 
\;\;\; {\cal J}/\hbar >> 1.   
\end{equation}
For a generic Hamiltonian system, this is only possible if  
the quantum state remains sufficiently well-localised, 
\begin{equation}
\label{ch5:Ehrcondb}
   \Delta x^i < \chi^i \;\;\;\; {\mathrm f} {\mathrm o} {\mathrm r}  
\;\;\;\ t < t_{obs} \;\;\;  {\mathrm a} {\mathrm n} {\mathrm d} 
\;\;\; {\cal J}/\hbar >> 1,   
\end{equation}
where for notational convenience I have used the same 
tolerance thresholds $\chi^i$ for the quantum state widths  
$ \Delta x^i = [\langle (x^i)^2 \rangle - \langle x^i \rangle^2 ]^{1/2}$ 
as for the Ehrenfest differences.  
These Ehrenfest conditions may be taken to define the proposition   
that quantum mechanics can describe the deterministic motion 
of a single macroscopic system, to within some prescribed 
accuracy and over physically observable time-scales, 
through the centroids of a well-localised state. 
I will now show that these Ehrenfest 
conditions may {\em break down} on physically relevant time-scales 
for macroscopic bodies that are subject to 
classically chaotic dynamics. 

\section{The Ehrenfest Break-Time}
\label{sect:tehr}

Let the Ehrenfest break-time, $t_{Ehr}$, 
be defined as the shortest 
time upon which one of the Ehrenfest differences $\xi^i$ 
grows larger than one of the thresholds in Eq.\ (\ref{ch5:Ehrconda}).  
The question under consideration is whether $t_{Ehr} < t_{obs}$ 
holds in relevant macroscopic situations.

A simple argument suffices to estimate how the  
time-scale $t_{Ehr}$ grows with 
with increasing system action ${\cal J}$. 
In the general case of 
an autonomous classical flow, the time-dependence of 
the $2N$ phase space variables are prescibed by the 
following set of first-order differential equations, 
\begin{equation}
\frac{d {\bf x}_c} {d t}= \{ {\bf x}_c, H \} = {\bf F}({\bf x}_c). 
\end{equation}
where $\{\cdot,\cdot \}$ is the 
Poisson bracket and $H$ stands for the Hamiltonian. 
The time-evolution of the corresponding Heisenberg 
operators, ${\bf x}_q$,  is given by the same set of 
differential equations, 
\begin{equation}
\frac{d {\bf x}_q} {d t}=  \frac{i}{\hbar} [ x_q, H] = {\bf F}({\bf x}_q),  
\end{equation}
where $[ \cdot , \cdot]$ denotes the commutator for two operators.  
Expanding the function ${\bf F}(x_q)$ 
about the {\it classical} trajectory, ${\bf x}_c$, gives, 
\begin{equation}
\label{ch5:taylor}
\frac{d {\bf x}_q} {d t}= {\bf F}({\bf x}_c) + ( x_q^i - x_c^i ) 
\frac{\partial {\bf F}} {\partial x_q^i} + {\it O}( [x_q^i - x_c^i ]^2). 
\end{equation}
Taking the expectation value of both sides it follows that, 
\begin{equation}
\frac{d {\bf \xi}} {d t}=  {\bf \xi}
\frac{\partial {\bf F}} {\partial x^i}|_{x^i=x_c^i} 
+ {\it O}( \xi^2). 
\end{equation}
While the quantum-classical differences remain sufficiently small,  
\begin{equation}
\label{ch5:small}
\xi^j \xi^k \frac{\partial^2 F^i } {\partial x_j \partial x_k} <<  
\frac{\partial F^i} {\partial x^l} \xi^l,  
\end{equation}
the higher-order terms in the expansion (\ref{ch5:taylor}) 
remain negligible, and 
the growth of the Ehrenfest differences is governed by the matrix,
\begin{equation}
 M_{ij} = \frac{\partial F_i} {\partial x^j}, 
\end{equation}
which is just the classical tangent map. The time-dependent 
eigenvalues of this matrix, evaluated along a fiducial 
trajectory, determine the stability of the classical flow. 
Consequently, in the case of classically {\it regular} motion, 
the quantum-classical differences will grow 
at most as a small power of time, 
\begin{equation}
\label{ch5:power}
{\bf \xi}(t) \simeq \xi_o t^\alpha,  
\end{equation}
where $\alpha \simeq 1$.
On the other hand, in classically chaotic regions, 
the local flow will exhibit expansion along some directions 
and contraction along others, 
and the very definition of the classical Lyapunov exponents \cite{LL}
entails that the quantum-classical 
differences should grow, on average, exponentially with time,  
\begin{equation}
\label{ch5:exp}
{\bf \xi}(t) \simeq \xi_o \exp ( \lambda_L t ),  
\end{equation}
where $\lambda_L$ is the largest Lyapunov exponent. 
I have used the symbol $\xi_o$ to denote 
the non-vanishing difference 
that will immediately arise 
between the quantum expectation values and classical 
dynamical variables due to the presence of a non-zero  
quantum-state width in either position or momentum \cite{Ball94}.
Since $\chi^i$ denotes  
the threshold that defines a break between the quantum and 
classical predictions, it follows that 
this break arises on the time-scale, 
\begin{equation}
\label{ch5:tlog}
 t_{Ehr} \sim \lambda_L^{-1} \ln( \chi^i / \xi_o^i). 
\end{equation}

The result (\ref{ch5:tlog}) can be expressed in a 
slightly more useful form. The ``classical limit'' of 
quantum mechanics corresponds to a sequence of quantum models 
that describe physical systems with 
increasing size (e.g., increasing quantum numbers), but with 
other parameters adjusted so that each model of this sequence 
is associated with the same classical system. 
The duration of the correspondence between the quantum and classical 
expectation values is {\it optimized} if the initial states are 
always chosen to be minimum uncertainty states (i.e., the initial 
variances are not increased in proportion to the system size but held 
fixed), whereas the break thresholds $\chi^i$ 
are always chosen as {\it fixed fraction} of the 
characteristic system size for each model in this sequence. 
Under these conditions, 
the ratio $(\chi^i/\xi_o^i)$ will scale as a power of the ratio 
$({\cal J}/\hbar)$, 
where ${\cal J}$ is a characteristic system action, and the break 
between the quantum and classical theories arises on the time-scale,  
\begin{equation}
\label{ch5:tEhr}
 t_{Ehr} \sim \lambda_L^{-1} \ln( {\cal J} / \hbar),  
\end{equation}
which estimates the optimal duration 
of {\it Ehrenfest} correspondence for chaotic systems.  
The time-scale (\ref{ch5:tEhr}) has been derived previously for a number 
of specific model systems \cite{BZ78,Berry79a,Berry79b,
Zas81,CC81,Frahm85,Haake87} 
through a variety of different techniques.  
Although some of these authors have interpreted the 
break-time (\ref{ch5:tEhr}) as a time-scale during which ``classical'' 
behaviour 
emerges from the quantum description \cite{Zas81,Chi88,Frahm85,Haake87,CC}, 
this interpretation is imprecise \cite{Ball94,EB00a,EB00b}, and  
I will refer to (\ref{ch5:tEhr}) specifically as the {\it 
Ehrenfest} break-time since it is defined as the  
time-scale during which the {\it Ehrenfest} correspondence conditions 
(\ref{ch5:Ehrconda}) remain valid. 

\section{The Breakdown of Ehrenfest Correspondence}
\label{sect:breakdown}

Some authors have maintained that the break-time scaling 
in Eq.\ (\ref{ch5:tEhr}) is compatible with the requirement of 
quantum-classical correspondence 
since $t_{Ehr} \to \infty$  as ``$\hbar \to 0$'' \cite{Zas81,Chi88,CC}. 
Of course, for real physical systems 
$\hbar$ is a constant, and the relevant question 
is whether (\ref{ch5:tEhr}) is sufficiently long to accomodate the 
Ehrenfest correspondence conditions for all chaotic macroscopic systems. 
Assuming the thresholds $\chi^i$ designate  
macroscopically observable differences, then the 
ratio $(\chi^i / \xi_o^i)$ is, of course, enormous. 
However, Zurek and Paz \cite{ZP95a,Zurek98b} have noted that  
the factor $\log({\cal J}/\hbar)$, appearing in the break-time 
expression (\ref{ch5:tEhr}), 
may be quite small for some macroscopic systems,  
even ones with astronomically large values  
of the ratio $({\cal J} / \hbar)$, since this ratio 
appears as an argument inside the logarithm. 
Consequently, the Ehrenfest break-time may be short compared to 
actual time-scales of observation. As an explicit example, these authors 
consider Hyperion, a moon of Saturn which is believed to exhibit a 
chaotic tumble \cite{Hyperion84,Hyperion94}.  They estimate the upper 
bound $t \simeq 20$ years for the onset of a massive 
discrepency between the quantum and classical predictions. 

Zurek \cite{Zurek98b} has suggested that such a short break-time 
raises the possibility of an observable  
breakdown of the correspondence principle for classically chaotic 
systems. 
Although I will demonstrate 
below that macroscopic chaotic motion certainly raises the possibility 
of an observable 
breakdown of the {\it Ehrenfest correspondence conditions}, it 
is important to emphasize that these situations 
do not suggest a breakdown of the correspondence principle, since 
the possibility of observing macroscopic 
differences between quantum mechanics and classical Liouville 
mechanics appears to be extremely remote \cite{EB00a,EB00b}. 
In the following section I will 
draw out some of the implicit assumptions 
of this breakdown argument that will help clarify 
this distinction. 

\section{Macroscopic Deviations and Kinematic Constraints}
\label{sect:add}


The argument leading to (\ref{ch5:exp}) 
leaves room for significant deviations from the 
exponential growth rate over short times, since the Lyapunov 
exponent is strictly defined as an average taken in 
the limit $t \to \infty$.
However, the exponential growth rate (\ref{ch5:exp}) and 
the $\log({\cal J}/\hbar)$ break-time should hold with a degree 
of approximation that improves as the initial difference 
$\xi_o$ decreases relative to the characteristic system size.   
A more critical assumption of the breakdown argument 
is that Eq.\ (\ref{ch5:tEhr}) describes 
the onset of {\it macroscopic} Ehrenfest differences. 
This follows only if the exponential growth rule (\ref{ch5:exp}) 
remains valid until the Ehrenfest differences grow to a significant 
fraction of the system size. 
But it is clear that the exponential growth of the Ehrenfest 
differences must  
eventually become invalidated. For example, 
in the case of bounded systems, Eq.\ (\ref{ch5:exp}) can remain valid 
at most until the differences grow to the size 
of the accessible phase space, at which time the growth will completely 
saturate.  Although from the condition (\ref{ch5:small}) 
it appears that any characteristic threshold at which 
this growth rate terminates may be expected to scale with the system size,   
in the absence of a rigorous argument, it is necessary  
to confirm this expectation through explicit calculation in 
specific model systems.   
Moreover, for any given system, it is necessary to determine 
if the Ehrenfest differences grow to a sufficiently large 
fraction of the system size that these differences 
become larger than the resolution of the macroscopic measurements. 


It may be objected that, even if the Ehrenfest differences 
grow to a macroscopic size on a physically relevant time-scale, 
these differences may not be experimentally 
significant as a result of the extreme sensitivity to initial 
conditions exhibited by the chaotic classical dynamics. 
Let $\Delta q(0) \Delta p (0) \neq 0$ denote the non-vanishing 
area of phase space from which the classical trajectory is known 
to have originated. 
From a purely classical point of view, at some later time 
$t$, it will be impossible to predict the location of the 
classical trajectory  
to within an area smaller than $\Delta q(t) \Delta p(t) 
\simeq \Delta q(0) \Delta p(0) \exp(2 \lambda_L t)$ as a result 
of the exponential growth of the imprecision in the 
initial phase space coordinates. 
It is therefore impossible to experimentally confirm the Newtonian 
prediction with a precision better than 
$\Delta q(0) \Delta p(0) \exp(2 \lambda_L t)$. 
However, according to (\ref{ch5:exp}), 
the Ehrenfest differences are expected 
to grow at the exponential rate, 
$\xi_q \xi_p \simeq \xi_q(0) \xi_p(0) \exp(2 \lambda_L t)$,  
which is the same rate of growth governing 
the imprecision of the classical prediction. 
Therefore the correctness of the classical (Newtonian) theory may not 
be experimentally 
confirmed with a precision that is smaller than the Ehrenfest differences. 
According to this objection, even if the predictions of the 
quantum and classical theories are macroscopically 
different, these differences are experimentally indistiguishable, and 
the theoretical prediction of a macroscopic discrepancy does not 
indicate an observable breakdown of the Ehrenfest 
correspondence principle. 

However, 
this practical objection may be overcome by considering certain kinematic 
constraints satisfied by the classical dynamical variables. 
For an autonomous system with energy $E({\bf q},{\bf p})$, 
although a prediction of the classical dynamical variables 
${\bf q}(t)$ and ${\bf p}(t)$ remains  
subject to an exponentially growing imprecision, 
these variables must satisfy the time-independent 
constraint $E({\bf q},{\bf p})=E_o$, since 
the energy is a constant of the motion. 
Consequently, at any given time,  the 
experimentally measured values of the coordinates,  
${\bf q}_m(t) \pm \delta_q $ and ${\bf p}_m(t) \pm 
\delta_p$, are predicted to  
satisfy the constraint $E({\bf q}_m(t),{\bf p}_m(t))=E_o $ on the basis 
of the classical theory. 

On the basis of the quantum theory, 
the expectation value of the energy 
$\langle E({\bf q},{\bf p}) \rangle = E_o$ is also a constant 
of the motion since the energy operator commutes with the Hamiltonian. 
However, the quantum mechanical prediction for the 
time-dependence of the function  
$E(\langle {\bf q}(t) \rangle,\langle {\bf p}(t) \rangle)$ 
is not subject to the same constraint as the classical function, 
and may exhibit macroscopic deviations from that constraint. 
Consequently, if one can show that the quantum 
prediction $E(\langle {\bf q}(t) \rangle,\langle {\bf p}(t) \rangle)$ 
differs from the classical prediction $E({\bf q}(t),{\bf p}(t)) = E_o$ by a 
macroscopically large magnitude, then 
experimental observation of the phase space 
coordinates (with adequate precision) 
will distinguish the predictions  of quantum theory 
from those of the classical (Newtonian) theory. 
If the function $E({\bf q},{\bf p})$ is reasonably well-behaved, then 
a set of macroscopically large Ehrenfest differences, arising 
on the time-scale $t_{Ehr}$, should lead 
to a macroscopically large deviation between the quantum prediction,  
$E(\langle {\bf q}(t_{Ehr}) \rangle,\langle {\bf p}(t_{Ehr}) \rangle)$,  
and the classical one, $E({\bf q}(t_{Ehr}),{\bf p}(t_{Ehr}))= E_o$. 
This argument may be generalized to the case of a non-autonomous system 
provided the system possesses any constant of the motion, 
$I({\bf q},{\bf p})$, i.e., a function that satisfies, 
\begin{equation}
 \frac{d I({\bf q},{\bf p})}{d t } =  
\{ I({\bf q},{\bf p}),H({\bf q},{\bf p},t) \} =0.  
\end{equation} 

\section{Case Study: Coupled Spins}

I will examine these features of Ehrenfest correspondence 
by explicit calculation of the quantum and classical dynamics for the 
model of nonlinearly coupled spins described in Chapter 2,  
\begin{equation}
\label{ch5:ham}
	H  = a (S_z + L_z) +  c S_x L_x \sum_n \delta(t-\tau n).  
\end{equation}
The key feature of the theoretical argument of section \ref{sect:breakdown} 
that must be checked by explicit calculation 
is whether the exponential growth of the Ehrenfest differences 
persists until reaching a magnitude 
that scales with the system size.

\begin{figure}[!t]
\begin{center}
\setlength{\unitlength}{0.240900pt}
\ifx\plotpoint\undefined\newsavebox{\plotpoint}\fi
\begin{picture}(1500,900)(0,0)
\font\gnuplot=cmr10 at 10pt
\gnuplot
\sbox{\plotpoint}{\rule[-0.200pt]{0.400pt}{0.400pt}}%
\put(181.0,123.0){\rule[-0.200pt]{4.818pt}{0.400pt}}
\put(161,123){\makebox(0,0)[r]{0.1}}
\put(1419.0,123.0){\rule[-0.200pt]{4.818pt}{0.400pt}}
\put(181.0,178.0){\rule[-0.200pt]{2.409pt}{0.400pt}}
\put(1429.0,178.0){\rule[-0.200pt]{2.409pt}{0.400pt}}
\put(181.0,211.0){\rule[-0.200pt]{2.409pt}{0.400pt}}
\put(1429.0,211.0){\rule[-0.200pt]{2.409pt}{0.400pt}}
\put(181.0,234.0){\rule[-0.200pt]{2.409pt}{0.400pt}}
\put(1429.0,234.0){\rule[-0.200pt]{2.409pt}{0.400pt}}
\put(181.0,252.0){\rule[-0.200pt]{2.409pt}{0.400pt}}
\put(1429.0,252.0){\rule[-0.200pt]{2.409pt}{0.400pt}}
\put(181.0,266.0){\rule[-0.200pt]{2.409pt}{0.400pt}}
\put(1429.0,266.0){\rule[-0.200pt]{2.409pt}{0.400pt}}
\put(181.0,279.0){\rule[-0.200pt]{2.409pt}{0.400pt}}
\put(1429.0,279.0){\rule[-0.200pt]{2.409pt}{0.400pt}}
\put(181.0,289.0){\rule[-0.200pt]{2.409pt}{0.400pt}}
\put(1429.0,289.0){\rule[-0.200pt]{2.409pt}{0.400pt}}
\put(181.0,299.0){\rule[-0.200pt]{2.409pt}{0.400pt}}
\put(1429.0,299.0){\rule[-0.200pt]{2.409pt}{0.400pt}}
\put(181.0,307.0){\rule[-0.200pt]{4.818pt}{0.400pt}}
\put(161,307){\makebox(0,0)[r]{1}}
\put(1419.0,307.0){\rule[-0.200pt]{4.818pt}{0.400pt}}
\put(181.0,363.0){\rule[-0.200pt]{2.409pt}{0.400pt}}
\put(1429.0,363.0){\rule[-0.200pt]{2.409pt}{0.400pt}}
\put(181.0,395.0){\rule[-0.200pt]{2.409pt}{0.400pt}}
\put(1429.0,395.0){\rule[-0.200pt]{2.409pt}{0.400pt}}
\put(181.0,418.0){\rule[-0.200pt]{2.409pt}{0.400pt}}
\put(1429.0,418.0){\rule[-0.200pt]{2.409pt}{0.400pt}}
\put(181.0,436.0){\rule[-0.200pt]{2.409pt}{0.400pt}}
\put(1429.0,436.0){\rule[-0.200pt]{2.409pt}{0.400pt}}
\put(181.0,451.0){\rule[-0.200pt]{2.409pt}{0.400pt}}
\put(1429.0,451.0){\rule[-0.200pt]{2.409pt}{0.400pt}}
\put(181.0,463.0){\rule[-0.200pt]{2.409pt}{0.400pt}}
\put(1429.0,463.0){\rule[-0.200pt]{2.409pt}{0.400pt}}
\put(181.0,474.0){\rule[-0.200pt]{2.409pt}{0.400pt}}
\put(1429.0,474.0){\rule[-0.200pt]{2.409pt}{0.400pt}}
\put(181.0,483.0){\rule[-0.200pt]{2.409pt}{0.400pt}}
\put(1429.0,483.0){\rule[-0.200pt]{2.409pt}{0.400pt}}
\put(181.0,492.0){\rule[-0.200pt]{4.818pt}{0.400pt}}
\put(161,492){\makebox(0,0)[r]{10}}
\put(1419.0,492.0){\rule[-0.200pt]{4.818pt}{0.400pt}}
\put(181.0,547.0){\rule[-0.200pt]{2.409pt}{0.400pt}}
\put(1429.0,547.0){\rule[-0.200pt]{2.409pt}{0.400pt}}
\put(181.0,579.0){\rule[-0.200pt]{2.409pt}{0.400pt}}
\put(1429.0,579.0){\rule[-0.200pt]{2.409pt}{0.400pt}}
\put(181.0,602.0){\rule[-0.200pt]{2.409pt}{0.400pt}}
\put(1429.0,602.0){\rule[-0.200pt]{2.409pt}{0.400pt}}
\put(181.0,620.0){\rule[-0.200pt]{2.409pt}{0.400pt}}
\put(1429.0,620.0){\rule[-0.200pt]{2.409pt}{0.400pt}}
\put(181.0,635.0){\rule[-0.200pt]{2.409pt}{0.400pt}}
\put(1429.0,635.0){\rule[-0.200pt]{2.409pt}{0.400pt}}
\put(181.0,647.0){\rule[-0.200pt]{2.409pt}{0.400pt}}
\put(1429.0,647.0){\rule[-0.200pt]{2.409pt}{0.400pt}}
\put(181.0,658.0){\rule[-0.200pt]{2.409pt}{0.400pt}}
\put(1429.0,658.0){\rule[-0.200pt]{2.409pt}{0.400pt}}
\put(181.0,667.0){\rule[-0.200pt]{2.409pt}{0.400pt}}
\put(1429.0,667.0){\rule[-0.200pt]{2.409pt}{0.400pt}}
\put(181.0,676.0){\rule[-0.200pt]{4.818pt}{0.400pt}}
\put(161,676){\makebox(0,0)[r]{$10^2$}}
\put(1419.0,676.0){\rule[-0.200pt]{4.818pt}{0.400pt}}
\put(181.0,731.0){\rule[-0.200pt]{2.409pt}{0.400pt}}
\put(1429.0,731.0){\rule[-0.200pt]{2.409pt}{0.400pt}}
\put(181.0,764.0){\rule[-0.200pt]{2.409pt}{0.400pt}}
\put(1429.0,764.0){\rule[-0.200pt]{2.409pt}{0.400pt}}
\put(181.0,787.0){\rule[-0.200pt]{2.409pt}{0.400pt}}
\put(1429.0,787.0){\rule[-0.200pt]{2.409pt}{0.400pt}}
\put(181.0,805.0){\rule[-0.200pt]{2.409pt}{0.400pt}}
\put(1429.0,805.0){\rule[-0.200pt]{2.409pt}{0.400pt}}
\put(181.0,819.0){\rule[-0.200pt]{2.409pt}{0.400pt}}
\put(1429.0,819.0){\rule[-0.200pt]{2.409pt}{0.400pt}}
\put(181.0,831.0){\rule[-0.200pt]{2.409pt}{0.400pt}}
\put(1429.0,831.0){\rule[-0.200pt]{2.409pt}{0.400pt}}
\put(181.0,842.0){\rule[-0.200pt]{2.409pt}{0.400pt}}
\put(1429.0,842.0){\rule[-0.200pt]{2.409pt}{0.400pt}}
\put(181.0,852.0){\rule[-0.200pt]{2.409pt}{0.400pt}}
\put(1429.0,852.0){\rule[-0.200pt]{2.409pt}{0.400pt}}
\put(181.0,860.0){\rule[-0.200pt]{4.818pt}{0.400pt}}
\put(161,860){\makebox(0,0)[r]{$10^3$}}
\put(1419.0,860.0){\rule[-0.200pt]{4.818pt}{0.400pt}}
\put(222.0,123.0){\rule[-0.200pt]{0.400pt}{4.818pt}}
\put(222,82){\makebox(0,0){0}}
\put(222.0,840.0){\rule[-0.200pt]{0.400pt}{4.818pt}}
\put(384.0,123.0){\rule[-0.200pt]{0.400pt}{4.818pt}}
\put(384,82){\makebox(0,0){2}}
\put(384.0,840.0){\rule[-0.200pt]{0.400pt}{4.818pt}}
\put(546.0,123.0){\rule[-0.200pt]{0.400pt}{4.818pt}}
\put(546,82){\makebox(0,0){4}}
\put(546.0,840.0){\rule[-0.200pt]{0.400pt}{4.818pt}}
\put(709.0,123.0){\rule[-0.200pt]{0.400pt}{4.818pt}}
\put(709,82){\makebox(0,0){6}}
\put(709.0,840.0){\rule[-0.200pt]{0.400pt}{4.818pt}}
\put(871.0,123.0){\rule[-0.200pt]{0.400pt}{4.818pt}}
\put(871,82){\makebox(0,0){8}}
\put(871.0,840.0){\rule[-0.200pt]{0.400pt}{4.818pt}}
\put(1033.0,123.0){\rule[-0.200pt]{0.400pt}{4.818pt}}
\put(1033,82){\makebox(0,0){10}}
\put(1033.0,840.0){\rule[-0.200pt]{0.400pt}{4.818pt}}
\put(1196.0,123.0){\rule[-0.200pt]{0.400pt}{4.818pt}}
\put(1196,82){\makebox(0,0){12}}
\put(1196.0,840.0){\rule[-0.200pt]{0.400pt}{4.818pt}}
\put(1358.0,123.0){\rule[-0.200pt]{0.400pt}{4.818pt}}
\put(1358,82){\makebox(0,0){14}}
\put(1358.0,840.0){\rule[-0.200pt]{0.400pt}{4.818pt}}
\put(181.0,123.0){\rule[-0.200pt]{303.052pt}{0.400pt}}
\put(1439.0,123.0){\rule[-0.200pt]{0.400pt}{177.543pt}}
\put(181.0,860.0){\rule[-0.200pt]{303.052pt}{0.400pt}}
\put(50,591){\makebox(0,0){ $ | L_z - \langle L_z \rangle| $}}
\put(810,21){\makebox(0,0){ {Kick Number }}}
\put(181.0,123.0){\rule[-0.200pt]{0.400pt}{177.543pt}}
\put(1176,307){\makebox(0,0)[r]{s=20, l=22}}
\put(303,275){\circle{18}}
\put(384,434){\circle{18}}
\put(465,515){\circle{18}}
\put(546,560){\circle{18}}
\put(627,397){\circle{18}}
\put(709,328){\circle{18}}
\put(790,547){\circle{18}}
\put(871,521){\circle{18}}
\put(952,547){\circle{18}}
\put(1033,554){\circle{18}}
\put(1114,526){\circle{18}}
\put(1196,426){\circle{18}}
\put(1277,485){\circle{18}}
\put(1358,526){\circle{18}}
\put(1439,365){\circle{18}}
\put(1246,307){\circle{18}}
\sbox{\plotpoint}{\rule[-0.600pt]{1.200pt}{1.200pt}}%
\put(1176,266){\makebox(0,0)[r]{s=200, l=220}}
\put(303,281){\circle*{18}}
\put(384,469){\circle*{18}}
\put(465,581){\circle*{18}}
\put(546,718){\circle*{18}}
\put(627,660){\circle*{18}}
\put(709,293){\circle*{18}}
\put(790,716){\circle*{18}}
\put(871,701){\circle*{18}}
\put(952,714){\circle*{18}}
\put(1033,706){\circle*{18}}
\put(1114,710){\circle*{18}}
\put(1196,717){\circle*{18}}
\put(1277,657){\circle*{18}}
\put(1358,680){\circle*{18}}
\put(1439,578){\circle*{18}}
\put(1246,266){\circle*{18}}
\sbox{\plotpoint}{\rule[-0.200pt]{0.400pt}{0.400pt}}%
\put(181,556){\usebox{\plotpoint}}
\put(181.0,556.0){\rule[-0.200pt]{303.052pt}{0.400pt}}
\put(181,739){\usebox{\plotpoint}}
\put(181.0,739.0){\rule[-0.200pt]{303.052pt}{0.400pt}}
\end{picture}
\caption{
The time-development of the Ehrefest differences for $L_z$ 
for classical parameters $\gamma=2.835$, 
$r=1.1$ and $a=5$ for which 
the phase space is predominantly chaotic. 
Open circles correspond to quantum numbers 
$(s,l) = (20,22)$ and filled circle correspond  
to $(s,l) = (200,220)$.  
In both cases the approximately exponential growth of the 
Ehrenfest differences persists until saturation at the 
system size $|{\bf L}|$, indicated by horizontal lines. 
}
\label{g2.835.cic2.s20.s200.lz}
\end{center}
\end{figure}
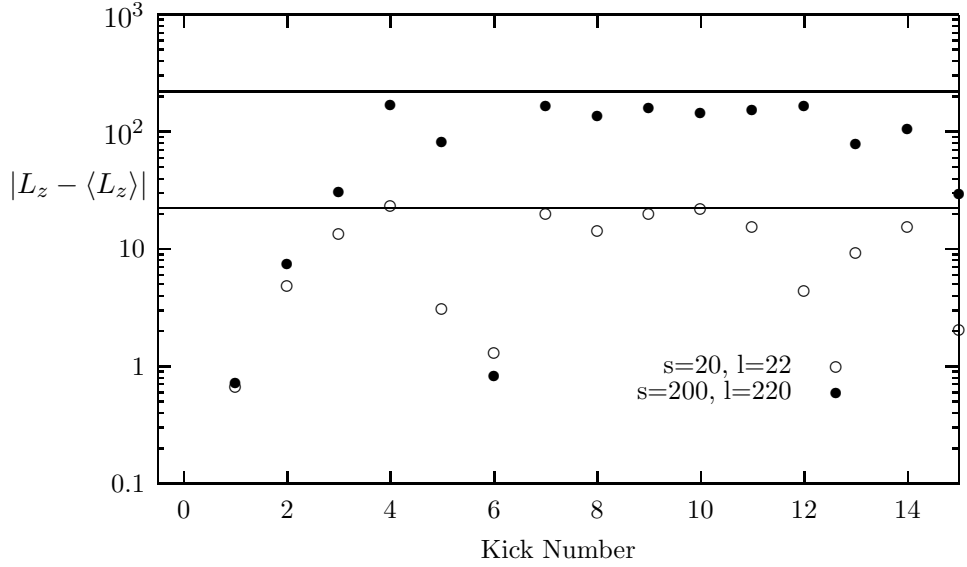

I will consider this correspondence in the predominantly chaotic 
regime associated with the 
classical parameters $\gamma=2.835, r=1.1$, and $a=5$, and 
with largest classical Lyapunov exponent $\lambda_L = 0.45$. 
The quantum dynamics are calculated using an initial coherent 
state centered at  $\vec{\theta}(0) = (20^o, 40^o, 160^o, 130^o)$.
The initial Ehrenfest differences may be set to zero if 
the classical initial conditions are set equal to 
the initial quantum state centroids,
\begin{equation}
 |{\bf L}(0)| = |\langle {\bf L}(0) \rangle | = l. 
\end{equation}
In the case of the coordinate $L_z$, the time-dependence 
of the Ehrenfest difference, 
\begin{equation}
\label{ch5:Ehrdiff}
\xi_{L_i}(t) = \langle L_i(t) \rangle - L_i(t), 
\end{equation}
is plotted in Fig.\ \ref{g2.835.cic2.s20.s200.lz} for quantum numbers 
$l=22$ and $l=220$ on a semi-log scale. 
The upper envelope of the 
Ehrenfest differences indeed corresponds to approximately  
exponential growth, as expected from Eq.\ (\ref{ch5:exp}).  
More importantly, 
this exponential growth persists until saturation at the 
system size that is associated with each quantum model. The system sizes 
corresponding to $l=22$ and $l=220$ are indicated 
by the horizontal solid lines in the figure. 
These results are consistent
with the general picture of the chaotic quantum dynamics developed 
in Chapters 3 and 4, where it was shown that the expectation values 
of dynamical variables approach their microcanonical equilibrium 
values, e.g., $\langle L_z \rangle \to 0$, on the saturation time-scale.     
 
As explained in section \ref{sect:add}, 
even macroscopic values of Ehrenfest differences 
in the coordinates,  such as (\ref{ch5:Ehrdiff}), 
arising on a physically accessible time-scale, 
do not automatically imply the possibility of 
an observable (experimentally significant) breakdown of correspondence,  
due to the exponential loss of precision that characterizes 
the chaotic classical dynamics.  
This objection may be overcome by considering the constraints 
imposed on the dynamics by the classical 
constant of the motion.  For the model (\ref{ch5:ham}), 
the magnitudes of the subsystem spins provide 
the required constants of the motion,  
\begin{eqnarray}
	\frac{d {\bf L}^2}{dt} & = & \{ {\bf L}^2,H \} = 0, \nonumber \\
	\frac{d {\bf S}^2}{dt} & = & \{ {\bf S}^2,H \} = 0. 
\end{eqnarray}
As a result of these constant functions, the time-dependent 
spin components are subject to the time-independent 
contraints, 
\begin{eqnarray}
\label{ch5:LSconstraints}
	L_x^2(t) + L_y^2(t) + L_z^2(t) & = & |{\bf L}|^2 \nonumber \\
	S_x^2(t) + S_y^2(t) + S_z^2(t) & = & |{\bf S}|^2
\end{eqnarray}
In addition, the magnitude of the 
total system angular momentum, though time-dependent, 
is subject to the constraint, 
\begin{equation}
\label{ch5:Jconstraint}
	{\bf L}^2 + {\bf S}^2 - 2 |{\bf L}| |{\bf S}| < {\bf J}^2  
< {\bf L}^2 + {\bf S}^2 + 2|{\bf L}||{\bf S}|.
\end{equation}

In the case of the quantum model, the expectation values of 
the operators $ {\bf L}^2$ and ${\bf S}^2$ are 
conserved, but the associated quadratic functions,
\begin{eqnarray}
\label{ch5:LSqmag}
	\langle {\bf L} \rangle^2 & = & 
	\langle L_x(t) \rangle^2 + \langle L_y(t)\rangle^2 + 
\langle L_z(t)\rangle^2, \nonumber \\
	\langle {\bf S} \rangle^2 & = & 
\langle S_x(t)\rangle^2 + \langle S_y(t)\rangle^2 + 
\langle S_z(t)\rangle^2,
\end{eqnarray}
are not conserved quantities.  
Similary, the magnitude of the total angular momentum, 
\begin{equation}
\label{ch5:Jqmag} 
 \langle {\bf J} \rangle^2 = \langle J_x \rangle^2 + \langle J_y \rangle^2 
+ \langle J_z \rangle^2,  
\end{equation}
is not bounded by the classical constraint (\ref{ch5:Jconstraint}).

\begin{figure}[!t]
\begin{center}
\setlength{\unitlength}{0.240900pt}
\ifx\plotpoint\undefined\newsavebox{\plotpoint}\fi
\sbox{\plotpoint}{\rule[-0.200pt]{0.400pt}{0.400pt}}%
\begin{picture}(1500,900)(0,0)
\font\gnuplot=cmr10 at 10pt
\gnuplot
\sbox{\plotpoint}{\rule[-0.200pt]{0.400pt}{0.400pt}}%
\put(221.0,123.0){\rule[-0.200pt]{4.818pt}{0.400pt}}
\put(201,123){\makebox(0,0)[r]{$10$}}
\put(1419.0,123.0){\rule[-0.200pt]{4.818pt}{0.400pt}}
\put(221.0,178.0){\rule[-0.200pt]{2.409pt}{0.400pt}}
\put(1429.0,178.0){\rule[-0.200pt]{2.409pt}{0.400pt}}
\put(221.0,252.0){\rule[-0.200pt]{2.409pt}{0.400pt}}
\put(1429.0,252.0){\rule[-0.200pt]{2.409pt}{0.400pt}}
\put(221.0,289.0){\rule[-0.200pt]{2.409pt}{0.400pt}}
\put(1429.0,289.0){\rule[-0.200pt]{2.409pt}{0.400pt}}
\put(221.0,307.0){\rule[-0.200pt]{4.818pt}{0.400pt}}
\put(201,307){\makebox(0,0)[r]{$10^2$}}
\put(1419.0,307.0){\rule[-0.200pt]{4.818pt}{0.400pt}}
\put(221.0,363.0){\rule[-0.200pt]{2.409pt}{0.400pt}}
\put(1429.0,363.0){\rule[-0.200pt]{2.409pt}{0.400pt}}
\put(221.0,436.0){\rule[-0.200pt]{2.409pt}{0.400pt}}
\put(1429.0,436.0){\rule[-0.200pt]{2.409pt}{0.400pt}}
\put(221.0,474.0){\rule[-0.200pt]{2.409pt}{0.400pt}}
\put(1429.0,474.0){\rule[-0.200pt]{2.409pt}{0.400pt}}
\put(221.0,492.0){\rule[-0.200pt]{4.818pt}{0.400pt}}
\put(201,492){\makebox(0,0)[r]{$10^3$}}
\put(1419.0,492.0){\rule[-0.200pt]{4.818pt}{0.400pt}}
\put(221.0,547.0){\rule[-0.200pt]{2.409pt}{0.400pt}}
\put(1429.0,547.0){\rule[-0.200pt]{2.409pt}{0.400pt}}
\put(221.0,620.0){\rule[-0.200pt]{2.409pt}{0.400pt}}
\put(1429.0,620.0){\rule[-0.200pt]{2.409pt}{0.400pt}}
\put(221.0,658.0){\rule[-0.200pt]{2.409pt}{0.400pt}}
\put(1429.0,658.0){\rule[-0.200pt]{2.409pt}{0.400pt}}
\put(221.0,676.0){\rule[-0.200pt]{4.818pt}{0.400pt}}
\put(201,676){\makebox(0,0)[r]{$10^4$}}
\put(1419.0,676.0){\rule[-0.200pt]{4.818pt}{0.400pt}}
\put(221.0,731.0){\rule[-0.200pt]{2.409pt}{0.400pt}}
\put(1429.0,731.0){\rule[-0.200pt]{2.409pt}{0.400pt}}
\put(221.0,805.0){\rule[-0.200pt]{2.409pt}{0.400pt}}
\put(1429.0,805.0){\rule[-0.200pt]{2.409pt}{0.400pt}}
\put(221.0,842.0){\rule[-0.200pt]{2.409pt}{0.400pt}}
\put(1429.0,842.0){\rule[-0.200pt]{2.409pt}{0.400pt}}
\put(221.0,860.0){\rule[-0.200pt]{4.818pt}{0.400pt}}
\put(201,860){\makebox(0,0)[r]{$10^5$}}
\put(1419.0,860.0){\rule[-0.200pt]{4.818pt}{0.400pt}}
\put(260.0,123.0){\rule[-0.200pt]{0.400pt}{4.818pt}}
\put(260,82){\makebox(0,0){0}}
\put(260.0,840.0){\rule[-0.200pt]{0.400pt}{4.818pt}}
\put(417.0,123.0){\rule[-0.200pt]{0.400pt}{4.818pt}}
\put(417,82){\makebox(0,0){2}}
\put(417.0,840.0){\rule[-0.200pt]{0.400pt}{4.818pt}}
\put(575.0,123.0){\rule[-0.200pt]{0.400pt}{4.818pt}}
\put(575,82){\makebox(0,0){4}}
\put(575.0,840.0){\rule[-0.200pt]{0.400pt}{4.818pt}}
\put(732.0,123.0){\rule[-0.200pt]{0.400pt}{4.818pt}}
\put(732,82){\makebox(0,0){6}}
\put(732.0,840.0){\rule[-0.200pt]{0.400pt}{4.818pt}}
\put(889.0,123.0){\rule[-0.200pt]{0.400pt}{4.818pt}}
\put(889,82){\makebox(0,0){8}}
\put(889.0,840.0){\rule[-0.200pt]{0.400pt}{4.818pt}}
\put(1046.0,123.0){\rule[-0.200pt]{0.400pt}{4.818pt}}
\put(1046,82){\makebox(0,0){10}}
\put(1046.0,840.0){\rule[-0.200pt]{0.400pt}{4.818pt}}
\put(1203.0,123.0){\rule[-0.200pt]{0.400pt}{4.818pt}}
\put(1203,82){\makebox(0,0){12}}
\put(1203.0,840.0){\rule[-0.200pt]{0.400pt}{4.818pt}}
\put(1360.0,123.0){\rule[-0.200pt]{0.400pt}{4.818pt}}
\put(1360,82){\makebox(0,0){14}}
\put(1360.0,840.0){\rule[-0.200pt]{0.400pt}{4.818pt}}
\put(221.0,123.0){\rule[-0.200pt]{293.416pt}{0.400pt}}
\put(1439.0,123.0){\rule[-0.200pt]{0.400pt}{177.543pt}}
\put(221.0,860.0){\rule[-0.200pt]{293.416pt}{0.400pt}}
\put(80,571){\makebox(0,0){ $ {\bf L}^2 - \langle {\bf L} \rangle^2 $}}
\put(830,21){\makebox(0,0){ {Kick Number }}}
\put(221.0,123.0){\rule[-0.200pt]{0.400pt}{177.543pt}}
\put(1183,676){\makebox(0,0)[r]{s=20, l=22}}
\put(339,285){\circle{18}}
\put(417,352){\circle{18}}
\put(496,393){\circle{18}}
\put(575,418){\circle{18}}
\put(653,425){\circle{18}}
\put(732,427){\circle{18}}
\put(810,432){\circle{18}}
\put(889,432){\circle{18}}
\put(968,433){\circle{18}}
\put(1046,433){\circle{18}}
\put(1125,433){\circle{18}}
\put(1203,433){\circle{18}}
\put(1282,433){\circle{18}}
\put(1360,433){\circle{18}}
\put(1439,433){\circle{18}}
\put(1253,676){\circle{18}}
\sbox{\plotpoint}{\rule[-0.600pt]{1.200pt}{1.200pt}}%
\put(1183,635){\makebox(0,0)[r]{s=200, l=220}}
\put(339,478){\circle*{18}}
\put(417,571){\circle*{18}}
\put(496,678){\circle*{18}}
\put(575,795){\circle*{18}}
\put(653,790){\circle*{18}}
\put(732,801){\circle*{18}}
\put(810,799){\circle*{18}}
\put(889,801){\circle*{18}}
\put(968,801){\circle*{18}}
\put(1046,801){\circle*{18}}
\put(1125,801){\circle*{18}}
\put(1203,802){\circle*{18}}
\put(1282,802){\circle*{18}}
\put(1360,802){\circle*{18}}
\put(1439,802){\circle*{18}}
\put(1253,635){\circle*{18}}
\sbox{\plotpoint}{\rule[-0.500pt]{1.000pt}{1.000pt}}%
\put(221,150){\usebox{\plotpoint}}
\put(221.00,150.00){\usebox{\plotpoint}}
\put(236.29,164.04){\usebox{\plotpoint}}
\put(251.56,178.10){\usebox{\plotpoint}}
\put(266.86,192.12){\usebox{\plotpoint}}
\put(282.59,205.65){\usebox{\plotpoint}}
\put(297.40,220.20){\usebox{\plotpoint}}
\put(312.70,234.22){\usebox{\plotpoint}}
\put(327.97,248.28){\usebox{\plotpoint}}
\put(343.25,262.32){\usebox{\plotpoint}}
\put(358.64,276.24){\usebox{\plotpoint}}
\put(374.08,290.08){\usebox{\plotpoint}}
\put(389.09,304.42){\usebox{\plotpoint}}
\put(404.80,317.98){\usebox{\plotpoint}}
\put(420.05,332.05){\usebox{\plotpoint}}
\put(434.93,346.52){\usebox{\plotpoint}}
\put(450.52,360.21){\usebox{\plotpoint}}
\put(465.53,374.53){\usebox{\plotpoint}}
\put(480.83,388.55){\usebox{\plotpoint}}
\put(496.51,402.13){\usebox{\plotpoint}}
\put(511.49,416.49){\usebox{\plotpoint}}
\put(526.98,430.29){\usebox{\plotpoint}}
\put(542.29,444.29){\usebox{\plotpoint}}
\put(557.14,458.79){\usebox{\plotpoint}}
\put(572.70,472.52){\usebox{\plotpoint}}
\put(588.18,486.34){\usebox{\plotpoint}}
\put(603.01,500.85){\usebox{\plotpoint}}
\put(618.72,514.41){\usebox{\plotpoint}}
\put(634.02,528.44){\usebox{\plotpoint}}
\put(649.29,542.50){\usebox{\plotpoint}}
\put(664.58,556.53){\usebox{\plotpoint}}
\put(679.87,570.57){\usebox{\plotpoint}}
\put(695.14,584.63){\usebox{\plotpoint}}
\put(710.44,598.65){\usebox{\plotpoint}}
\put(725.74,612.68){\usebox{\plotpoint}}
\put(741.00,626.75){\usebox{\plotpoint}}
\put(756.30,640.77){\usebox{\plotpoint}}
\put(771.94,654.41){\usebox{\plotpoint}}
\put(786.84,668.84){\usebox{\plotpoint}}
\put(802.24,682.75){\usebox{\plotpoint}}
\put(817.64,696.64){\usebox{\plotpoint}}
\put(832.67,710.95){\usebox{\plotpoint}}
\put(847.97,724.97){\usebox{\plotpoint}}
\put(863.61,738.61){\usebox{\plotpoint}}
\put(878.51,753.05){\usebox{\plotpoint}}
\put(894.12,766.72){\usebox{\plotpoint}}
\put(909.08,781.08){\usebox{\plotpoint}}
\put(924.43,795.05){\usebox{\plotpoint}}
\put(940.09,808.67){\usebox{\plotpoint}}
\put(955.05,823.05){\usebox{\plotpoint}}
\put(970.18,837.25){\usebox{\plotpoint}}
\put(985.93,850.77){\usebox{\plotpoint}}
\put(996,860){\usebox{\plotpoint}}
\put(221,335){\usebox{\plotpoint}}
\put(221.00,335.00){\usebox{\plotpoint}}
\put(236.42,348.89){\usebox{\plotpoint}}
\put(252.05,362.54){\usebox{\plotpoint}}
\put(266.97,376.97){\usebox{\plotpoint}}
\put(282.57,390.64){\usebox{\plotpoint}}
\put(297.88,404.64){\usebox{\plotpoint}}
\put(312.93,418.93){\usebox{\plotpoint}}
\put(328.29,432.86){\usebox{\plotpoint}}
\put(343.72,446.74){\usebox{\plotpoint}}
\put(358.60,461.20){\usebox{\plotpoint}}
\put(374.26,474.82){\usebox{\plotpoint}}
\put(389.56,488.84){\usebox{\plotpoint}}
\put(404.82,502.91){\usebox{\plotpoint}}
\put(420.12,516.94){\usebox{\plotpoint}}
\put(435.42,530.96){\usebox{\plotpoint}}
\put(450.69,545.02){\usebox{\plotpoint}}
\put(465.97,559.06){\usebox{\plotpoint}}
\put(481.27,573.09){\usebox{\plotpoint}}
\put(496.53,587.15){\usebox{\plotpoint}}
\put(511.83,601.18){\usebox{\plotpoint}}
\put(527.53,614.75){\usebox{\plotpoint}}
\put(542.37,629.26){\usebox{\plotpoint}}
\put(557.67,643.28){\usebox{\plotpoint}}
\put(572.94,657.33){\usebox{\plotpoint}}
\put(588.23,671.38){\usebox{\plotpoint}}
\put(603.58,685.34){\usebox{\plotpoint}}
\put(619.10,699.10){\usebox{\plotpoint}}
\put(634.06,713.48){\usebox{\plotpoint}}
\put(649.73,727.08){\usebox{\plotpoint}}
\put(665.07,741.07){\usebox{\plotpoint}}
\put(680.04,755.42){\usebox{\plotpoint}}
\put(695.65,769.09){\usebox{\plotpoint}}
\put(710.54,783.54){\usebox{\plotpoint}}
\put(725.76,797.65){\usebox{\plotpoint}}
\put(741.48,811.19){\usebox{\plotpoint}}
\put(756.51,825.51){\usebox{\plotpoint}}
\put(771.92,839.39){\usebox{\plotpoint}}
\put(787.31,853.31){\usebox{\plotpoint}}
\put(794,860){\usebox{\plotpoint}}
\sbox{\plotpoint}{\rule[-0.200pt]{0.400pt}{0.400pt}}%
\put(221,437){\usebox{\plotpoint}}
\put(221.0,437.0){\rule[-0.200pt]{293.416pt}{0.400pt}}
\put(221,802){\usebox{\plotpoint}}
\put(221.0,802.0){\rule[-0.200pt]{293.416pt}{0.400pt}}
\end{picture}
\caption{
The time-development of the differences between the classical 
constant of the motion ${\bf L}^2 = l^2$ and the quantum  
prediction for $\langle L \rangle^2$ for the same 
classical parameters as in Fig.\ \ref{g2.835.cic2.s20.s200.lz}. 
The data are compared 
to an exponential growth rate given  by twice 
the classical Lyapunov exponent, $\lambda_L =0.45$. 
Open circles correspond to quantum numbers
$(s,l) = (20,22)$ and filled circle correspond 
to $(s,l) = (200,220)$.  
In both cases the exponential growth of the 
Ehrenfest differences persists until saturation at the 
maximum possible value ${\bf L}^2$, indicated by horizontal lines. 
}
\label{g2.835.cic2.s20.s200.ll}
\end{center}
\end{figure}
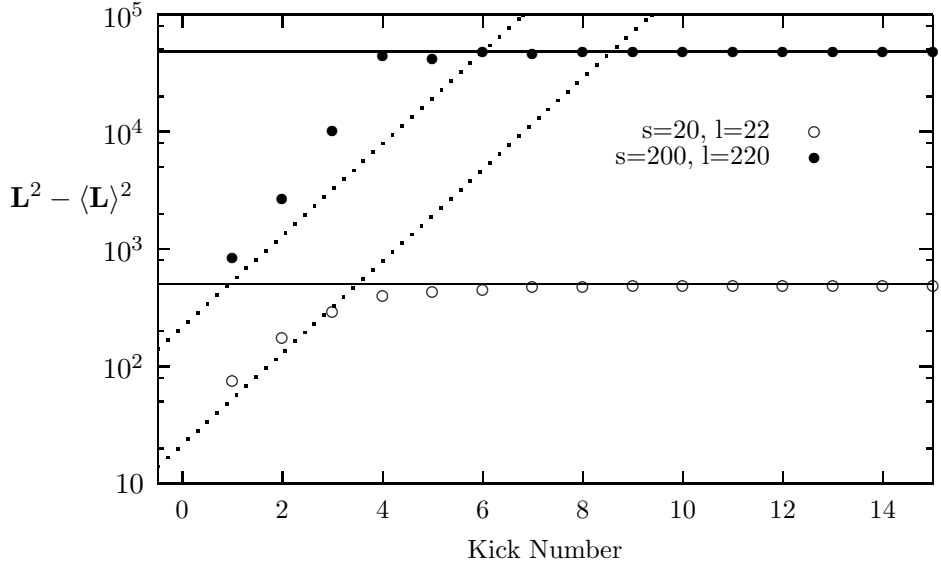

In Fig.\ \ref{g2.835.cic2.s20.s200.ll} the time-dependent differences 
${\bf L}^2 - \langle {\bf L} \rangle^2$  
are plotted for $l=22$ and $l=220$.  
The growth of these 
differences is approximately exponential, and governed by a rate 
that appears to be independent of the system size. This growth may be 
estimated on the basis of the analytical argument predicting 
exponential growth (\ref{ch5:exp})  
for the Ehrenfest differences for the coordinates, 
$\xi = {\bf L} - \langle {\bf L} \rangle$.
Writing ${\bf L}(n) = \langle {\bf L}(n) \rangle + \xi(n)$ and squaring 
both sides gives, 
\begin{equation}
\label{ch5:test}
	{\bf L}^2 - \langle {\bf L} \rangle^2 = \xi^2 - 2 
\xi \cdot \langle {\bf L} \rangle.
\end{equation}
Since $\xi^2(n) \simeq \xi_o^2  \exp( 2 \lambda_L n)$ grows 
very rapidly, and assuming 
the second term in (\ref{ch5:test}) 
may therefore be neglected for large $n$, it follows that  
\begin{equation}
\label{ch5:test2}
	{\bf L}^2 - \langle {\bf L} \rangle^2 \sim \xi_o^2 
\exp( 2 \lambda_L n). 
\end{equation}

An alternative approach to characterizing 
the growth rate for the Ehrenfest difference 
${\bf L}^2 - \langle {\bf L} \rangle^2$ follows from comparison with 
the numerically measured exponential rate of growth 
of the quantum variance (\ref{eqn:expvar}). 
The magnitude of quantum variance $(\Delta {\bf L})^2$ 
provides a good approximation to the time-dependent Ehrenfest difference, 
${\bf L}^2 - \langle {\bf L} \rangle^2$, for large $l$, that is, 
\begin{equation}
\label{ch5:var2diff}
(\Delta {\bf L})^2 \simeq {\bf L}^2 - \langle {\bf L} \rangle^2,     
\end{equation}
since, 
\begin{equation}
{\bf L^2} =l^2 \simeq \langle {\bf L}^2 \rangle = l(l+1),  
\end{equation}
where quantities on the left correspond to the classical magnitude 
of the spin and those on the right to the quantum magnitude. 
Therefore, from (\ref{ch5:var2diff}) and (\ref{eqn:expvar}), 
it may be expected that 
the Ehrenfest differences for the quadratic functions grow according to,  
\begin{equation}
\label{ch5:expvar} 
	{\bf L}^2 - \langle {\bf L} \rangle^2  \simeq l \exp( 2 \lambda_w n), 
\end{equation}
where $\lambda_w \simeq \lambda_L$ in regimes 
of predominantly chaotic dynamics  (as shown in Chapter 3). 
This estimate of the growth rate is the same as (\ref{ch5:test2}) obtained  
on the basis of the  
analytical argument leading to (\ref{ch5:exp}), but also 
provides an estimate of the prefactor $\xi_o^2 \simeq l$. 
The exponential rate (\ref{ch5:expvar}) is plotted 
in Fig.\ \ref{g2.835.cic2.s20.s200.ll}, 
for $l=22$ and $l=220$, with $\lambda_w = \lambda_L =0.45$, 
which slightly underestimates the growth rate of the data.  

The most important feature of the data 
in Fig.\ \ref{g2.835.cic2.s20.s200.ll} is that 
this exponential growth of these observable Ehrenfest differences 
does not saturate until the differences reach the magnitude 
of the corresponding classical constant of the motion ${\bf L}^2$. 
The classical magnitudes 
${\bf L}^2 = 484$ and ${\bf L}^2 \simeq 48400$, 
for ${\bf L} = 22 $ and ${\bf L} = 220$, respectively, 
are plotted as solid lines in Fig.\ \ref{g2.835.cic2.s20.s200.ll}.  
These are the maximum 
values that may be expected on purely kinematic grounds, since,  
\begin{equation}
	0 \leq | \langle {\bf L} \rangle^2 - {\bf L}^2 | \leq {\bf L}^2. 
\end{equation}

\begin{figure}[!t]
\begin{center}
\input{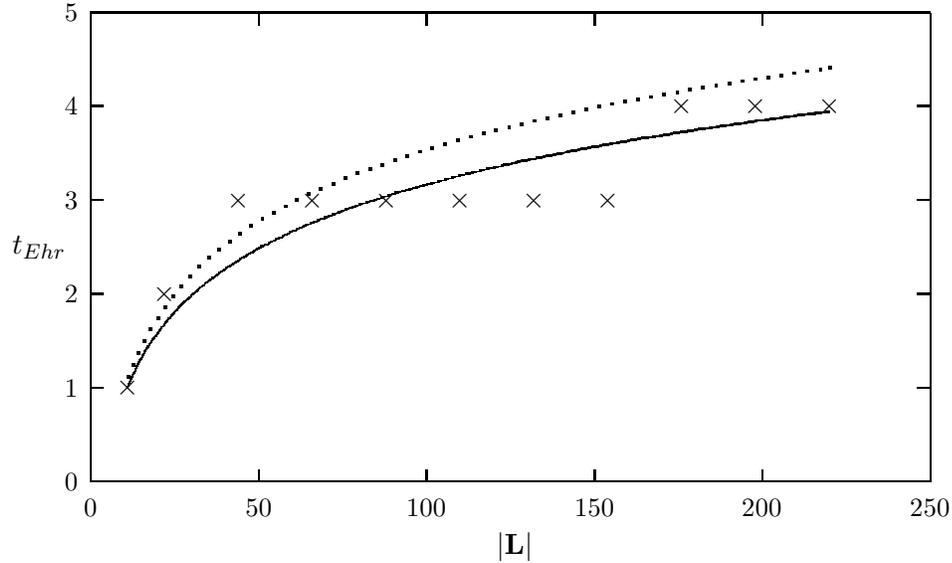}
\caption{
The break time obtained by the condition (\ref{ch5:break}), using 
a threshold $ f {\bf L}^2= 0.25 {\bf L}^2$, 
plotted against quantum number $l$ ranging from 11 to 220. 
The data are obtained using the same classical 
parameters as in Fig.\ \ref{g2.835.cic2.s20.s200.ll} 
and initial condition $\vec{\theta} = 20^o,40^o,160^o,130^o)$.
The solid line corresponds to Eq.\ \ref{ch5:log} 
with exponent $\lambda = 0.51$ obtained by a least-squares fit. 
The dotted line corresponds to Eq.\ \ref{ch5:log} 
using the largest Lyapunov exponent $\lambda_L = 0.45$. 
}
\label{tehr.2.835.chi.25.ll}
\end{center}
\end{figure}

The break-time for the  experimentally observable 
differences ${\bf L}^2 - \langle {\bf L} \rangle^2$
is determined numerically from the earliest time at which these differences 
exceed a fraction $f$ of the classical magnitude, 
i.e., the condition, 
\begin{equation}
\label{ch5:break}
|{\bf L}^2 - \langle {\bf L}\rangle^2| > f {\bf L}^2.  
\end{equation}
This calculated break-time is plotted against increasing 
system size, using $f = 0.25$, in Fig.\ (\ref{tehr.2.835.chi.25.ll}). 
The numerical data exhibit a step-wise behaviour since 
the time, $n$, is confined to integer values. 
This scaling behaviour of the break time may be estimated by setting 
the LHS of the analytical prediction (\ref{ch5:expvar}) equal to the 
threshold $f {\bf L}^2$, and solving for the time, to give, 
\begin{equation}
\label{ch5:log}
	t_{Ehr} \simeq 1 / (2 \lambda_w) \log ( f {\bf L} ).  
\end{equation}
This curve is plotted in Fig.\ \ref{tehr.2.835.chi.25.ll} 
using $\lambda_w = \lambda_L = 0.45$ (dotted line),  
which slightly over-estimates the break-time results.  
A least squares fit to (\ref{ch5:log}) gives $\lambda_w = 0.51$, which 
provides an excellent fit to the data (solid line).  
The scaling demonstrated by the break-time data in this figure 
corresponds to that of a typical initial coherent state 
evolved under parameters that produce a predominantly chaotic 
classical phase space. 
These results explicitly confirm that the $\log({\cal J}/\hbar)$  
dependence of the Ehrenfest break-time holds even for 
{\it experimentally observable} differences that have grown to a 
significant fraction of the relevant classical magnitude. 

\section{Discussion}


The argument outlined above should apply to a wide 
variety of chaotic systems. In particular, 
the quantum expectation values 
may be experimentally distinguished from the classical coordinates 
on the short time-scale (\ref{ch5:tEhr}) 
provided the classical model admits a constant of the motion 
which is some ``reasonable'' function of the coordinates.
For example, the differences between the quantum 
and classical predictions for any 
polynomial function of the coordinates should grow 
exponentially as long as the differences between the coordinates are 
themselves subject to exponential growth. 

Though I have considered this problem in a specific model 
system, there is also evidence in other model systems 
that the break-time (\ref{ch5:tEhr}) should apply to Ehrenfest differences 
of order the system size when the phase space 
is predominantly chaotic. 
Fox and coworkers \cite{Fox94a,Fox94b} have shown that the 
quantum variances 
in the kicked top and the kicked rotor grow at an 
approximately exponential rate which subsides only 
when the variances have reached the 
system size. By virtue of the connection between 
the quantum variance 
and the time-development of the expectation 
values expressed in (\ref{eqn:qdiff}), it follows that 
these systems will also exhibit Ehrenfest differences 
of order the system size on the time-scale 
(\ref{ch5:tEhr}). Unfortunately, these authors have not 
explicitly reported how their results scale with increasing 
system size. 

\subsection{Liouville Correspondence Does Not Breakdown}

In section \ref{sect:breakdown} I emphasized that the possibility 
of a breakdown of the correspondence principle was based on demonstrating 
that {\it macroscopic} quantum-classical differences arose 
on an experimentally accessible time-scale. While I have provided 
evidence above that this is indeed the case for the Ehrenfest 
differences, the magnitude of which grow to the system size 
on the short time-scale (\ref{ch5:tEhr}), I have also demonstrated, 
in Chapters 3 and 4,  
that the quantum-Liouville differences that arise on this 
short time-scale do not scale with the system size. 
In particular, in Chapter 3  
it was shown that the Liouville break-time grows 
logarithmically with the system size only when applied to differences 
that remain a factor $(\hbar/{\cal J})$ smaller than the system size. 
The presence of such small quantum-classical   
differences in the description of a macroscopic body 
is not an indication of a {\em breakdown} of 
the correspondence principle, since this principle 
requires only approximate agreement between the quantum 
and classical theories in the macroscopic limit.


\subsection{Can Decoherence Restore Ehrenfest Correspondence?}



In the model that I have considered, each  
spin is a subsystem of a larger, isolated system. The time-evolution 
of each subsystem  
may be described, in general, by some non-unitary  
map as a result of the interactions with 
the other subsystem. The initial coherent state for each 
subsystem rapidly evolves into a nearly  
random mixture. These properties may be considered to 
arise as a result of the ``decoherence'' provided 
from interactions with the other subsystem. 
The important point is that 
the quantum environment provided by the other system 
only serves to increase the state width of each subsystem, 
and therefore actually leads to a faster breakdown of the 
Ehrenfest conjecture (\ref{ch5:Ehrconda}), 
but certainly does not help to preserve that conjecture.
Since I have 
formulated a statement of the correspondence problem  
specifically in terms of the  
conjecture (\ref{ch5:Ehrconda}), as a constraint on the system coordinates,  
these results demonstrate explicitly that decoherence considerations 
do not circumvent the breakdown of Ehrenfest correspondence. 

\subsection{Consequences for Interpretation of the Quantum State}

As explained in the Introduction, the inapplicability 
of the conjecture (\ref{ch5:Ehrconda}) to the class of chaotic 
macroscopic bodies, does not 
entail a failure of the correspondence principle, 
but a failure of a particular interpretation of the 
correspondence principle.  This view is reinforced 
by noting that the conjecture (\ref{ch5:Ehrconda}) is 
not implied by Born's postulate, $P(q) = |\psi(q)|^2$, but 
consists of an independent assumption. (However, 
Born's postulate, combined with the assumption (\ref{ch5:Ehrcondb}), 
does entail (\ref{ch5:Ehrconda}).) 

The Ehrenfest correspondence conditions, (\ref{ch5:Ehrconda}) and 
(\ref{ch5:Ehrcondb}), are motivated 
by a particular interpretation of the quantum formalism, namely, 
the view that quantum mechanics can provide a complete description 
of an individual physical system. 
Here it is useful to distinguish between two 
different interpretations of the quantum state \cite{BallQM}:

\vspace{0.2in}
\noindent 
(i) a pure state provides a complete and exhaustive description 
of an individual system; and,  

\vspace{0.2in}
\noindent 
(ii) a pure state provides a description of certain statistical 
properties of an ensemble of similarly prepared systems.
\vspace{0.2in}

The results of Chapters 3 and 4 provide strong evidence in support 
of interpretation (ii), whereas the argument of the present 
chapter suggests the possibility of experimentally refuting 
interpretation (i).

\section{Is Quantum Mechanics Complete?}

According to the general argument provided in the Introduction, 
the quantum expectation values may be expected to deviate from 
the Newtonian coordinates on 
a time-scale that may be short compared to physically relevant 
time-scales for a wide class of chaotic macroscopic systems. 
If one is tempted to maintain that 
the quantum expectation values actually describe 
the dynamics of a single chaotic macroscopic system, as suggested 
by the Ehrenfest Correspondence conjecture (\ref{ch5:Ehrconda}), then 
one is lead to the implausible view that 
significant deviations from classical energy conservation 
should be expected to arise, very rapidly, in the case of chaotic 
macroscopic systems. 
Since something as fundamental as energy conservation 
for an isolated system is likely to be confirmed under 
a wide variety of experimental conditions, 
it is natural to assume 
that it is the ``expectation value trajectory'' predicted from 
the conjecture (\ref{ch5:Ehrconda}), rather than the 
Newtonian trajectory, that will be experimentally invalidated.

On the assumption that individual macroscopic systems do not violate 
energy conservation on the time-scale (\ref{ch5:tEhr}), 
the argument outlined in this chapter leads to the conclusion 
that quantum mechanics is unable to provide a complete description 
of the motion of individual macroscopic systems. 
Therefore, quantum theory does not provide 
a complete description of reality. This is the conclusion derived 
by Einstein, Podolsky, and Rosen (EPR) a long time ago in a very different 
way \cite{EPR}. In order to derive their conclusion 
EPR adopted the following principle \cite{EPR}:  

\vspace{0.2in}
{\it A necessary condition for a 
complete theory is that ``every element of physical reality must have 
a counterpart in the physical theory.''} 
\vspace{0.2in}

In this chapter I have argued that the coordinates of macroscopic 
chaotic systems are not correctly described by 
the quantum expectation values since the latter deviate from   
these coordinates in a physically implausible and 
experimentally significant manner.
But since this interpretative framework for 
quantum theory fails to provide a complete description of 
macroscopic physical reality, 
is there then any physical 
interpretation of quantum theory that can meet the challenge 
of providing a complete 
and correct description of the chaotic dynamics of 
individual macroscopic bodies? 



\appendix
\chapter{Nonclassical Moments for SU(2) Coherent States}

Ideally we would like to construct an initial classical 
density that reproduces all of the moments of the 
initial quantum coherent states. This is possible 
in a Euclidean phase space, in which case all Weyl-ordered moments of 
the coherent state can be matched exactly by the moments of a 
Gaussian classical distribution. 
However, below we prove that no classical density $\rho_c(\theta,\phi)$
that describes an ensemble of spins of fixed length $|{\bf J}|$ 
can be constructed with marginal distributions that match   
those of the SU(2) coherent states (\ref{eqn:cs}). 
Specifically, we consider the set of  
distributions on ${\mathcal S}^2$ with continuous 
independent variables $\theta \in [ 0 , \pi ]$ 
and $ \phi \in [ 0,2\pi) $, 
measure  $ d \mu = \sin  \theta d \theta d \phi$, and subject to 
the usual normalization, 
\begin{equation}  
	\int_{{\mathcal S}^2} d \mu \; \rho_c(\theta,\phi) = 1.  
\end{equation}

For convenience we choose the coherent state 
to be polarized along the positive $z$-axis, $\rho = 
| j,j \rangle \langle j,j |$. This state is axially symmetric:  
rotations about the $z$-axis by an arbitrary 
angle $\phi$ leave the state operator invariant. 
Consequently we require axially symmetry of the  
corresponding classical distribution, 
\begin{equation}
\label{eqn:cazisymm}
\rho_c(\theta,\phi)= \rho_c(\theta).
\end{equation}

We use the expectation of the quadratic operator,  
  $ \langle {\bf J}^2 \rangle = j(j+1) $,
to fix the length of the classical spins,  
\begin{equation}
\label{eqn:length}
	|{\bf J}| = \sqrt{\langle J^2 \rangle_c} = \sqrt{j(j+1)}.  
\end{equation}
Furthermore, 
the coherent state $|j,j\rangle$ is an eigenstate of $J_z$ with 
moments along the $z$-axis given by
	$\langle J_z^n \rangle = j^n$
for integer $n$. 
Therefore we require that the classical distribution 
produces the moments, 
\begin{equation}
\label{eqn:cjz}
	\langle J_z^n \rangle_c = j^n. 
\end{equation}

These requirements are satisfied by 
the $\delta$-function distribution, 
\begin{equation}
\label{eqn:vm}
	 \rho_v(\theta) =  { \delta(\theta - 
	\theta_o) \over 2 \pi \sin \theta_o },   
\end{equation}
where $\cos \theta_o = j/|{\bf J}|$ defines $\theta_o$. 
This distribution is the 
familiar vector model of the old quantum theory 
corresponding to the intersection of a cone 
with the surface of the sphere. 

However, in order to derive an inconsistency between the quantum 
and classical moments we do not need to assume 
that the classical distribution is given explicitly 
by (\ref{eqn:vm}); we only need to make use of the  
the azimuthal invariance condition (\ref{eqn:cazisymm}), 
the length condition (\ref{eqn:length}), 
and the first two even moments of (\ref{eqn:cjz}).  

First we calculate some of the quantum coherent state moments 
along the $x$-axis (or any axis orthogonal to $z$),  
\begin{eqnarray}
\label{eqn:xmoments}
	\langle J_x^m \rangle & = & 0   \; \; \; 
{\mathrm f}{\mathrm o}{\mathrm r} 
\; {\mathrm o}{\mathrm d}{\mathrm d} \;\; m  \nonumber \\
	\langle J_x^2 \rangle & = & j/2 \nonumber \\ 
	\langle J_x^4 \rangle & = & 3j^2/4 -j/4 \nonumber.   
\end{eqnarray}
In the classical case, these moments are of the form,  
\begin{equation}
\label{eqn:jxm} 
\langle J_x^m \rangle_c  =  
\int d J_z \int d \phi \rho_c(\theta)  
|{\bf J}|^m \cos^m(\phi) \sin^m(\theta).  
\end{equation}
For $m$ odd the integral over $\phi$ vanishes, as required 
for correspondence with the odd quantum moments. For $m$ even 
we can evaluate (\ref{eqn:xmoments})
by expressing the r.h.s.\ as a linear combination 
of the $z$-axis moments (\ref{eqn:cjz}) of equal and lower order. 
For $m=2$ this requires substituting $\sin^2(\theta) = 1 - \cos^2(\theta) $ 
into (\ref{eqn:jxm}) and then integrating over $\phi$ to obtain 
\begin{eqnarray}
\langle J_x^2 \rangle_c  & = &   
\pi \int d J_z \rho_c(\theta) |{\bf J}|^2 
- \pi \int d J_z \rho_c(\theta) |{\bf J}|^2 \cos^2(\theta) \nonumber \\
& = & |{\bf J}| /2 - \langle J_z^2 \rangle /2. \nonumber 
\end{eqnarray}
Since $\langle J_z^2 \rangle$  is determined by 
(\ref{eqn:cjz}) and the length is fixed from (\ref{eqn:length})
we can deduce the classical value without knowing $\rho(\theta)$,  
\begin{equation}   
	\langle J_x^2 \rangle_c  =  j/2.  
\end{equation}
This agrees with the value of corresponding quantum moment. 
For $m=4$, however, by a similar procedure we deduce 
\begin{equation}   
	\langle J_x^4 \rangle_c =  3 j^2/8, 
\end{equation}
that differs from the quantum moment 
$ \langle J_x^4 \rangle $ 
by the factor, 
\begin{equation}
	\delta J_x^4 = 
| \langle J_x^4 \rangle - \langle J_x^4 \rangle_c | =  | 3 j^2/8 - j/4 |, 
\end{equation}
concluding our proof that 
no classical distribution on ${\mathcal S}^2$ can reproduce 
the quantum moments.

\chapter{Wigner-Weyl Representation of Quantum Mechanics}


The Wigner function is constructed from a quantum state 
operator $\rho$ through the transform \cite{Tak89,Wigner}, 
\begin{equation}
	\rho_w(q,p,t) = \frac{1}{2\pi\hbar} \int dy 
\langle q- y/2|  \rho(t) | q+y/2\rangle e^{iyp/\hbar}. 
\end{equation}
The Wigner function, $\rho_w(q,p,t)$, 
which is a real ``c-number'' function, 
plays the role of the quantum state operator in the 
Wigner-Weyl representation of quantum mechanics. 
It generally takes on negative values in non-vanishing regions 
of its ``phase space'' domain, $\{q,p\} \in R$, and 
therefore may not be interpreted as a probability 
distribution.  

Each Hermitian operator, $A(q,p)$, 
in the standard representation, is also 
associated with a ``c-number'' function, 
\begin{equation}
	A_w(q,p) = \int dy 
\langle q- y/2|  A(q,p) | q+y/2\rangle e^{iyp/\hbar}.
\end{equation}
Observable quantities are obtained from the prescription,  
\begin{equation}
\label{eqn:wexp}
	\langle A(q,p) \rangle  = \int dq  \int dp
			\; \rho_w(q,p) \; A_w(q,p).  
\end{equation}
The Wigner-Weyl representation is an exact representation of 
quantum mechanics in the sense that,
\begin{equation}
	\int dq  \int dp
			\; \rho_w(q,p) \; A_w(q,p) 
= {\mathrm T}{\mathrm r}[ A(q,p) \rho ],  
\end{equation}
for any Weyl-ordered operator $A(q,p)$. 


The time-dependence of the Wigner function can be expressed in 
the following form \cite{Moyal}, 
\begin{equation}
\label{eqn:moyal}
	\frac{d \rho_W(q,p) } {d t} 
= \{ H , \rho_W(q,p) \} + \sum_{n=1}^{\infty} 
\left(\frac{\hbar}{2}\right)^{2n} \frac{(-1)^n}{(2n+1)!}
\frac {\partial^{2n+1} \rho_W(q,p)} {\partial p^{2n+1}} 
\frac {\partial^{2n+1} H} {\partial q^{2n+1}},
\end{equation}
with an obvious connection to the time-dependence of the Liouville 
ensemble. 
Here $\{\cdot,\cdot \}$ denotes the classical Poisson bracket. 
As noted by Takahashi \cite{Tak89} and others, 
the quantum corrections to the Liouville 
flow (sometimes called the ``Moyal 
terms'') do not trivially vanish as ``$\hbar \to 0$'', 
in spite of their explicit proportionality to powers of $\hbar$. 
The Wigner distribution 
has an implicit dependence on $\hbar$, and factors of 
$(q/\hbar)^n$ may be expected from the $\partial^n/\partial_{p^n}$ 
derivatives in Eq.\ (\ref{eqn:moyal}). 

A clearer picture of the origin of quantum corrections to 
the Liouville picture is obtained by considering the differences 
that arise in the 
time-development of ``observable'' quantities. 
The time-development of the quantum expectation value  
$\langle A(q,p) \rangle$, corresponding to a time-independent 
Weyl-ordered operator, $A(q,p)$, follows from 
Eqs.\ (\ref{eqn:moyal}) and  (\ref{eqn:wexp}): after 
integrating by parts, and assuming the surface terms vanish 
(for physically realistic states), 
\begin{equation}
	\left[\frac{\partial^{k} \rho_W(q,p)}{\partial p^{k}} 
\frac {\partial^{j} A} {\partial p^{j}} \right]_{\infty}^{-\infty} = 0,  
\end{equation}
it follows that, 
\begin{equation}
	\frac{d \langle A_w(q,p) \rangle } {d t} 
= \{ H_w , A_w \} + \sum_{n=1}^{\infty} 
\left(\frac{\hbar}{2}\right)^{2n} \frac{(-1)^{n+1}}{(2n+1)!}
\frac {\partial^{2n+1} A} {\partial p^{2n+1}} 
\frac {\partial^{2n+1} H} {\partial q^{2n+1}}. 
\end{equation}
This expression demonstrates that 
quantum {\it dynamical} corrections to the classical Liouville flow 
do not arise unless the Hamiltonian 
contains at least cubic powers of $q$, and, furthermore,  
{\em explicit} corrections do not arise unless the observable $A$ 
contains at least cubic powers of $p$.  These conclusions were  
determined previously by Ballentine and McRae in the specific 
case of the Henon-Heiles model \cite{Ball98}. 
As before, the quantum corrections may not be expected to vanish 
trivially in the limit ``$\hbar \to 0$'' since 
the Moyal terms may contain implicit dependence on $\hbar$.

%


\begin{thebibliography}{99}
\addcontentsline{toc}{chapter}{Bibliography}


\bibitem{Ammann98}
H. Ammann, R. Gray, I. Shvarchuck, and N. Christensen, Phys. Rev. 
Lett. {\bf 80}, 4111 (1998). 
\bibitem{Arnold}
V.I. Arnold, {\it Mathematical Methods of Classical Mechanics} 
(Springer-Verlag, New York, 1978).
\bibitem{BallQM}
L.E. Ballentine, Rev. Mod. Phys. {\bf 42}, 358 (1970).
\bibitem{B91a}
L. E. Ballentine, Phys.\ Rev.\ A {\bf 44}, 4126 (1991). 
\bibitem{B91b}
L. E. Ballentine, Phys.\ Rev.\ A {\bf 44}, 4133 (1991). 
\bibitem{B93}
L. E. Ballentine, Phys.\ Rev.\ A {\bf 47}, 2592 (1993).
\bibitem{Ball94}
L.E. Ballentine, Y. Yang, and J.P. Zibin, Phys.\ Rev.\ A {\bf 50}, 2854 (1994).
\bibitem{Ball95}
L.E. Ballentine in {\it Fundamental Problems in Quantum Physics}, 
Eds. M. Ferrero and A. van der Merwe (Kluwer Academic Publishers, 1995).
\bibitem{Ball96}
L.E. Ballentine, {\it Quantum Mechanics} (World Scientific, Singapore, 1996).
\bibitem{BZ96}
L.E. Ballentine and J. Zibin, Phys. Rev. A {\bf 54}, 3813 (1996). 
\bibitem{Ball98}
L.E. Ballentine and S.M. McRae, Phys.\ Rev.\ A {\bf 58}, 1799 (1998).
\bibitem{Ball00}
L.E.\ Ballentine, Phys.\ Rev.\ A {\bf 63}, 024101 (2001). 
\bibitem{Ballprivate}
L.E. Ballentine, private communication. 
\bibitem{BayKoch74}
J.E. Bayfield and P.M. Koch, Phys. Rev. Lett. {\bf 33}, 258 (1974). 
\bibitem{BZ78}
G.P. Berman and G.M. Zaslavsky, Physica  {\bf 91A}, 450 (1978).
\bibitem{Berry79a}
M.V. Berry, N.L. Balasz, M. Tabor, and A. Voros, Ann. Phys. {\bf 122}, 
26 (1979).
\bibitem{Berry79b}
M.V. Berry and N.L. Balasz, J. Phys. A: Math. Gen. {\bf 12}, 625 (1979).
\bibitem{BGLR96}
L. Bonci, P. Grigolini, Phys.\ Rev.\ A {\bf 54}, 112 (1996).
\bibitem{BGI98}
F. Borgonovi, I. Guarneri, F.M. Izrailev, Phys.\ Rev.\ E {\bf 57}, 5291 (1998).
\bibitem{Hyperion94}
P. Boyd, G. Midlin, R. Gilmore, and H. Solari, 
The Astrophysical Journal {\bf 431}, 425-431 (1994).
\bibitem{Haake96}
A. Braun, P. Gerwinski, F. Haake, H. Schomerus, 
Z.\ Phys. B {\bf 100}, 115 (1996).
\bibitem{CC}
G. Casati and B.V. Chirikov, 
in {\it Quantum Chaos: Between Order and Disorder} 
(Cambridge University Press, Cambridge, 1995).
\bibitem{CC95}
G. Casati and B.V. Chirikov, Phys. Rev. Letters {\bf 75}, 350 (1995).
\bibitem{CC81}
B.V. Chirikov, F.M. Izrailev, and D.L. 
Shepelyansky,  Sov. Sci. Rev. C2, 209 (1981).
\bibitem{Chi88}
B.V. Chirikov, F.M. Israilev, and D.L. Shepelyansky, Physica D {\bf
33}, 77 (1988).
\bibitem{Cirac95}
J.I. Cirac and P. Zoller, Phys Rev Lett {\bf 74}, 4091 (1995).
\bibitem{Coryprivate}
D. Cory and Y. Weinstein, private communication. 
\bibitem{deGroot}
S.R. Degroot, {\it La Transformation de Weyl et la Fonction de Wigner: 
une Forme Alternative de la Mecanique Quantique} (Les Presses 
Universitaires de Montreal, Montreal, 1974).
\bibitem{DV95}
D.P. DiVincenzo, Phys. Rev. A {\bf 51}, 1015 (1995). 
\bibitem{Dorfman}
J.R.\ Dorfman, {\it An Introduction to Chaos in NonEquilibrium Statistical Mechanics} 
(Cambridge University Press, Cambridge, 1999).
\bibitem{Eckhardt}
B. Eckhardt, S. Fishman, J. Keating, O. Agam, J. Main, and K. Muller, 
Phys.\ Rev.\ E {\bf 52}, 5893 (1995). 
\bibitem{Edmonds}
A.R. Edmonds, {\it Angular Momentum in Quantum Mechanics} 
(Princeton University Press, Princeton, 1957).
\bibitem{Ehrenfest}
P. Ehrenfest, Z. Physik {\bf 45}, 455 (1927).
\bibitem{EPR}
A. Einstein, B. Podolsky, and N. Rosen, Phys. Rev. {\bf 47}, 777 (1935). 
\bibitem{EB00a}
J.\ Emerson and L.E.\ Ballentine, Phys.\ Rev.\ A {\bf 63}, 052103 (2001); 
quant-ph/0011020.
\bibitem{EB00b}
J.\ Emerson and L.E.\ Ballentine, Phys. Rev. E {\bf 64}, 026217 (2001); 
quant-ph/0103050.
\bibitem{Farini}
A. Farini, S. Boccaletti, and F.T. Arecchi, Phys. 
Rev. E. {\bf 53}, 4447 (1996).
\bibitem{FP83}
M. Feingold and A. Peres, Physica {\bf 9D} 433 (1983).
\bibitem{Peres84b}
M.\ Feingold, N.\ Moiseyev, and A.\ Peres, Phys. Rev. A {\bf 30}, 509 (1984).
\bibitem{FMP85}
M.\ Feingold, N.\ Moiseyev, and A.\ Peres , Chem.\ Phys.\ Lett.\ 
\bibitem{FP86}
M.\ Feingold and A.\ Peres, Phys.\ Rev.\ A {\bf 34}, 591 (1986).
{\bf 117}, 344 (1985).
\bibitem{Ford91}
J. Ford, G. Mantica, G. Ristow, Physica D {\bf 50}, 493-520, (1991).
\bibitem{Ford92a}
J. Ford and M. Ilg, Phys. Rev. A {\bf 45}, 6165, 1992.
\bibitem{Ford92b}
J. Ford and G. Mantica, Am. J. Phys 60 (12), 1086, (1992).
\bibitem{Fox90}
R.F. Fox and B.L. Lan, Phys.\ Rev.\ A {\bf 41}, 2952 (1990).   
\bibitem{Fox94b}
R.F. Fox and T.C. Elston, Phys.\ Rev.\ E {\bf 50}, 2553 (1994).
\bibitem{Fox94a}
R.F. Fox and T.C. Elston, Phys.\ Rev.\ E {\bf 49}, 3683 (1994).
\bibitem{Frahm85}
H. Frahm and H.J. Mikeska, Z.\ Phys. B {\bf 60}, 117 (1985).
\bibitem{Gaspard}
P. Gaspard, {\it Chaos, Scattering and Statistical Mechanics} (Cambridge 
University Press, Cambridge, 1998).
\bibitem{Dec96}
D.\ Giulini, E. Joos, C. Kiefer, J. Kupsch, I. Stamatescu, and H. D. Zeh,
{\it Decoherence and the Appearance of a Classical World in 
Quantum Theory} (Springer, Berlin, 1996).
\bibitem{Goldstein}
H. Goldstein, {\it Classical Mechanics}, 2nd ed. 
(Addison-Wesley, Reading, 1980). 
\bibitem{Gong99}
J. Gong and P. Brumer, Phys.\ Rev.\ E {\bf 60}, 1643 (1999).
\bibitem{Gutzwiller}
M.C. Gutzwiller, {Chaos in Classical and Quantum Mechanics} 
(Springer-Verlag, New York, 1990).
\bibitem{Haake87}
F. Haake, M. Kus and R. Scharf, Z.\ Phys. B {\bf 65}, 361 (1987).
\bibitem{Haake91}
F. Haake, {\it Quantum Signatures of Chaos} (Springer-Verlag, New York, 1991).
\bibitem{Zurek98a}
S. Habib, K. Shizume and W.H. Zurek, Phys. Rev. Lett. {\bf 80}, 4361 (1998).
\bibitem{HB94}
B.S. Helmkamp and D.A. Browne, Phys.\ Rev.\ E {\bf 49}, 1831 (1994).
\bibitem{HB96}
B.S. Helmkamp and D.A. Browne, Phys.\ Rev.\ Lett.\ {\bf 76}, 3691 (1996).
\bibitem{Amiet}
Ph. Jacquod and J.-P. Amiet, J. Phys. A {\bf 30}, 2963 (1997).
\bibitem{Jensen}
R.V. Jensen, Nature {\bf 355}, 311 (1992).
\bibitem{Joos85}
E. Joos and H.D. Zeh, Condensed Matter {\bf 59}, 223-243 (1985).
\bibitem{Kol96}
A.R. Kolovsky, Phys. Rev. Letters {\bf 76}, 340 (1996).
\bibitem{Lan91}
B.L. Lan and R.F. Fox, Phys.\ Rev.\ A {\bf 43}, 646 (1991).       
\bibitem{LP78}
J.G. Leopold and I.C. Percival, Phys. Rev. Lett. {\bf 41}, 944 (1978).
\bibitem{Liboff}
R.L. Liboff, {\it Introductory Quantum Mechanics} 
(Addison-Wesley, Reading, 1989). 
\bibitem{LL}
A.J.\ Lichtenberg and M.A.\ Lieberman, {\it Regular and Chaotic Motion} 
(Springer-Verlag, New York, 1992).
\bibitem{Messiah}
A. Messiah, {\it Quantum Mechanics} (Wiley, New York, 1966).
\bibitem{Merzbacher}
E. Merzbacher, {\it Quantum Mechanics} (Wiley, New York, 1970).
\bibitem{Mil99}
G.J. Milburn, S. Schneider, and D.F.V. James, {\it Ion Trap Quantum 
Computing with Warm Ions}, Fortschritte der Physik, 801-810, 
{\bf 48} (2000); quant-ph/9908037. 
\bibitem{Sark99}
P. Miller and S. Sarkar, Phys.\ Rev.\ E {\bf 60}, 1542 (1999).
\bibitem{Moyal}
J.E. Moyal, Proc. Cambridge Phys. Soc. {\bf 45}, 99 (1949).
\bibitem{cs}
A.\ Perelomov,  {\it Generalized Coherent States and Their Applications}, 
(Springer-Verlag, New York , 1986).
\bibitem{Peres84a}
A. Peres, Phys. Rev. A {\bf 30}, 504 (1984).
\bibitem{Peres84}
A. Peres, Phys. Rev. A {\bf 30}, 1610 (1984).
\bibitem{Peres93}
A. Peres, {\it Quantum Theory: Concepts and Methods} (Kluwer, Dordrecht, 1993).
\bibitem{Peres96} 
A. Peres, Phys. Rev. E {\bf 53}, 4524 (1996). 
\bibitem{Cory99}
M. Price, S. Somaroo, A. Dunlop, T. Havel, and D. Cory, 
Phys. Rev. A {\bf 60}, 2777 (1999). 
\bibitem{RR98}
D. T. Robb and L. E. Reichl, Phys.\ Rev.\ E {\bf 57}, 2458 (1998).  
\bibitem{RBWG95}
R. Roncaglia, L. Bonci, B.J. West, and P. Grigolini, Phys.\ Rev.\ E {\bf 51}, 5524 (1995).
\bibitem{Rose}
M.E. Rose, {\it Elementary Theory of Angular Momentum} (Wiley, New York, 1957).
\bibitem{Sakurai}
J.J.\ Sakurai, {\it Modern Quantum Mechanics} 
(Benjamin-Cummings, Menlo Park Calif., 1985).
\bibitem{SC92}
R. Schack and C. Caves, Phys. Rev. Lett. {\bf 69}, 3413 (1992).
\bibitem{SC93}
R. Schack and C. Caves, Phys. Rev. Lett. {\bf 71}, 525 (1993).
\bibitem{SC94}
R. Schack, G. D'Ariano, and C. Caves, Phys. Rev. E {\bf 50}, 972(1994).
\bibitem{SC96}
R. Schack and C. Caves, Phys. Rev. E {\bf 53}, 3257 (1996).
\bibitem{SC98}
R. Schack and C. Caves, quant-ph/9810050.
\bibitem{Schack98}
R. Schack, Phys. Rev. A {\bf 57}, (1998); quant-ph/9705016.
\bibitem{Schiff}
L.I. Schiff, {\it Quantum Mechanics}(McGraw-Hill, New York, 1968).
\bibitem{Schn}
A.I. Schnirelman, in {\it KAM Theory and 
Semiclassical Approximations to Eigenfunctions} 
(Spinger-Verlag, New York, 1993).
\bibitem{Tabor}
M. Tabor, {\it Chaos and Integrability in Nonlinear Dynamics} 
(Wiley, New York, 1989).
\bibitem{Tak89}
K. Takahashi, Prog. Theor. Phys. Supp. 98 (1989). 
\bibitem{TH93}
S. Tomsovic and E.J. Heller, Phys.\ Rev.\ E {\bf 47}, 282 (1993).
\bibitem{Wigner}
E.P. Wigner, Phys. Rev. 40, 749 (1932).
\bibitem{Hyperion84}
J. Wisdom, S. Peale, and F. Mignard, Icarus {\bf 58}, 137 (1984)
\bibitem{Zas81}
G.M. Zaslavsky, Phy. Rep. {\bf 80}, 157 (1981).
\bibitem{ZP94}
W.H. Zurek and J.P. Paz, Phys. Rev. Lett. {\bf 72}, 2508 (1994).
\bibitem{ZP95a}
W.H. Zurek and J.P. Paz, Phys. Rev. Letters {\bf 75}, 351 (1995).
\bibitem{ZP95b}
W.H. Zurek and J.P. Paz, Physica D {\bf 83}, 300 (1995).
\bibitem{Zurek98b}
W.H. Zurek, Physica Scripta {\bf T76}, 186 (1998).











\end{thebibliography}
\end{document}